\documentclass{aa}  
\usepackage{graphicx}
\usepackage{txfonts}
\usepackage[utf8]{inputenc}
\usepackage{adjustbox}
\usepackage[usenames, dvipsnames]{color}
\usepackage{todonotes}
\usepackage{natbib}
\usepackage{paralist}
\usepackage{tabularx}
\newcolumntype{L}[1]{>{\raggedright\arraybackslash}m{#1}}
\newcolumntype{C}[1]{>{\centering\arraybackslash}m{#1}}
\newcolumntype{R}[1]{>{\raggedleft\arraybackslash}m{#1}}
\usepackage{soul}

\usepackage{subfig}

\usepackage{lscape}
\usepackage{multicol}
\usepackage{graphicx}

\usepackage{tablefootnote}
\usepackage{threeparttable}

\usepackage{multirow}

\usepackage{changes}

\usepackage{color}

\usepackage{url}

\def \vsini {$\textit{v sin i}$}

\def \teff {$T_{\mathrm{eff}}$}

\usepackage[bookmarksnumbered=true]{hyperref} 
\renewcommand\thefootnote{\textcolor{green}{\arabic{footnote}}}

\hypersetup{
     colorlinks = true,
     linkcolor = blue,
     anchorcolor = blue,
     citecolor = blue,
     filecolor = blue,
     urlcolor = blue
     }

\begin{document} 

   \title{The correlation between photometric variability and radial velocity jitter \thanks{Based on observations collected at the La Silla Observatory,
ESO(Chile), with the HARPS spectrograph at the 3.6-m telescope. 
See the acknowledgements for the list of specific programs.}} 

   \subtitle{based on TESS and HARPS observations}

   \author{S. Hojjatpanah\inst{\ref{IA}, \ref{U-Porto}}
   \and
    M. Oshagh \inst{\ref{iac}, \ref{gottingen}, \ref{IA}}
        \and
        P. Figueira\inst{\ref{ESO}, \ref{IA}}
            \and
    N.C. Santos  \inst{\ref{IA}, \ref{U-Porto}} 
        \and
        E. M. Amazo-G\'omez \inst{\ref{max_eliana},\ref{gottingen}}
           \and
    S. G. Sousa \inst{\ref{IA}}
    \and
       V. Adibekyan\inst{\ref{IA}, \ref{U-Porto}}
    \and
    B. Akinsanmi  \inst{\ref{IA}, \ref{U-Porto}, \ref{Akin2}} 
    \and
        O. Demangeon \inst{\ref{IA}}
    \and
    J. Faria \inst{\ref{IA}}
        \and
    J. Gomes da Silva \inst{\ref{IA}}
    \and
    N. Meunier \inst{\ref{Grenoble}}
    }
      
   \institute{Instituto de Astrof\'isica e Ci\^{e}ncias do Espa\c co, Universidade do Porto, CAUP, Rua das Estrelas, 4150-762 Porto, Portugal\label{IA}
  \email{Saeed.Hojjatpanah@astro.up.pt}     
\and
  Departamento de Fisica e Astronomia, Faculdade de Ci\^{e}ncias, Universidade do Porto, Rua Campo Alegre, 4169-007 Porto, Portugal\label{U-Porto}
\and
Instituto de Astrof\'isica de Canarias (IAC), E-38200 La Laguna, Tenerife, Spain \label{iac}
\and
    Institut f\"ur Astrophysik, Georg-August-Universit\"at, Friedrich-Hund-Platz 1, 37077 G\"ottingen, Germany \label{gottingen}
\and
  European Southern Observatory, Alonso de Cordova 3107, Vitacura, Santiago, Chile \label{ESO}
\and
    Max-Planck-Institut fur Sonnensystemforschung, G\"ottingen, Germany \label{max_eliana}
 \and
 National Space Research and Development Agency. Airport Road, Abuja, Nigeria \label{Akin2}
 \and
Univ. Grenoble Alpes, CNRS, IPAG, F-38000 Grenoble, France \label{Grenoble}
}

   \date{Received ----, ----; accepted --- , ----}

 
  \abstract
  {Characterizing the relation between stellar photometric variability and radial velocity (RV) jitter can help us to better understand the physics behind these phenomena. The current and upcoming high precision photometric surveys such as TESS, CHEOPS, and PLATO will provide the community with thousands of new exoplanet candidates. As a consequence, the presence of such a correlation is crucial in selecting the targets with the lowest RV jitter for efficient RV follow-up of exoplanetary candidates. Studies of this type are also crucial to design optimized observational strategies to mitigate RV jitter when searching for Earth-mass exoplanets.}
  {Our goal is to assess the correlation between high-precision photometric variability measurements and high-precision RV jitter over different time scales.} 
  {We analyze 171 G, K, and M stars with available TESS high precision photometric time-series and HARPS precise RVs. We derived the stellar parameters for the stars in our sample and measured the RV jitter and photometric variability. We also estimated chromospheric Ca II H $\&$ K activity indicator $log(R' _{HK})$, $\textit{v sin i}$, and the stellar rotational period. Finally, we evaluate how different stellar parameters and an RV sampling subset can have an impact on the potential correlations.}
  {We find a varying correlation between the photometric variability and RV jitter as function of time intervals between the TESS photometric observation and HARPS RV. As the time intervals of the observations considered for the analysis increases, the correlation value and significance becomes smaller and weaker, to the point that it becomes negligible. We also find that for stars with a photometric variability above 6.5 ppt the correlation is significantly stronger. We show that such a result can be due to the transition between the spot-dominated and the faculae-dominated regime. We quantified the correlations and updated the relationship between chromospheric Ca II H $\&$ K activity indicator $log(R' _{HK})$ and RV jitter.}
  {}


  \maketitle
%
\section{Introduction}

Detecting and accurately characterizing low-mass exoplanets is a challenge for observational astronomy. Besides instrumental limitations, one of the main remaining obstacles is the intrinsic stellar variability, often called "jitter" in the exoplanet's field. The two main detection methods, radial velocity (RV) and transit photometry, struggle with the impact of stellar activity induced signal \citep[e.g.,][]{2018ASSP...49..239O}, which is generated by active regions, such as spots and plages, associated to strong magnetic fields. The presence of active regions on a rotating star not only generates RV jitter but also induces photometric modulations in transit observations \citep[e.g.,][]{2009A&A...505.1277C,2018ASSP...49..239O}.

Stellar activity can both mimic RV planetary signals (similar to planetary signals) and hide a real planetary signal by adding significant stellar activity induced signal. \citep[e.g.,][]{1998ApJ...498L.153S,2011A&A...525A.140D, 2011A&A...527A..82D, 2001A&A...379..279Q,2002A&A...392..215S,2010A&A...511A..54S, 2011IAUS..273..281B,2014A&A...566A..35S,2016A&A...585A.134D}. Moreover, it can degrade the precision of the planetary parameters derived through RV \citep[e.g.,][]{2018AJ....156...82C,2016PASP..128f6001F} or transit measurements \citep{2009A&A...505.1277C, 2013A&A...549A..35O,2013A&A...556A..19O,2014A&A...569A..74B,2015EPJWC.10105003O}. 

\citet{2008A&A...489L...9H} and \citet{2010A&A...513L...8F} demonstrated that the stellar spots can even reproduce the RV variations as high as those produced by giant planets on short orbits. Additionally, \citet{2014Sci...345..440R} showed that the activity can masquerade as a planet located inside the habitable zone of M dwarfs. 

It has further been shown that the stellar photometric variability is correlated with the RV jitter \citep[e.g.,][]{2012MNRAS.419.3147A,bastien-2014-a,oshagh-2017-a}. Defining such relations as precisely as possible is crucial in optimizing the telescope time invested in the RV follow-up of upcoming planetary candidates from ongoing and upcoming missions such as TESS \citep{2014SPIE.9143E..20R} and PLATO \citep{2014ExA....38..249R} with high-resolution spectrographs, such as HARPS \citep{HARPS} and ESPRESSO \citep{pepe2014espresso}. Several studies have attempted to assess this potential correlation. \citet{bastien-2014-a} compared the variability in Kepler light-curves and the root-mean-square (RMS) of RV time series of twelve G and K stars. They did not find any strong correlation between the RV-RMS and the amplitude of photometric variability. Instead, they found a correlation between the number of significant peaks in the Lomb-Scargle periodograms of the light-curves and the RMS of stellar RV.\\

Later, \citet{cegla-2014-a} studied a sample of 900 stars using GALEX \citep{2005ApJ...619L...1M} and Kepler surveys \citep{borucki2010kepler}. They estimated the chromospheric activity index of each target using the UV flux measured by GALEX \citep{2011AJ....142...23F} and used it to calculate the RV jitter by using empirical relations reported in several studies \citep{2000A&A...361..265S,2005PASP..117..657W,2003csss...12..694S}. For magnetically quiet stars, they found a strong correlation between photometric variability and the estimated RV jitter. They also evaluated the correlation between RV jitter and $F_{8}$ \footnote{F$_{8}$ is the RMS of photometric light-curve on timescale shorter than 8 hr, and has been shown to also correlate strongly with the stellar
log g.} \citep{2013Natur.500..427B} and additionally with the number of zero crossings (X0) in photometric observation. They found that the correlation between RV jitter and F$_{8}$ is sensitive to the effective temperature of the stars.
 
\citet{oshagh-2017-a} performed an intensive study on a sample of nine stars with different levels of activity. They obtained simultaneous K2 photometric light-curves and HARPS RV measurements. From these observations the photometric variability, $F_{8}$, RV jitter, spectroscopic activity indicators BIS, FWHM, and $log(R' _{HK})$ were derived. The authors found that a strong correlation exists between RV jitter and photometric variation for highly active stars ($log(R' _{HK})$ $\geq$ -4.66). Moreover they found that for highly active stars strong correlations exist between RV jitter and variability of activity indicators, such as BIS, FWHM, and $log(R' _{HK})$. The authors pointed out that all of these correlations become weaker for stars with low-amplitude photometric variations and for slowly-rotating stars. The same study also showed that even though $F_{8}$ has a strong correlation with the RV jitter, it does not display a strong correlation with rotationally-induced photometric variability. They also confirmed the evidence of the two spot-dominated and plage-dominated regimes presented in several other studies \citep{1997ApJ...485..789L,1998ApJS..118..239R,2016A&A...589A..46S}.

Recently, \citet{2019ApJ...883..195T} investigated the relationship between stellar properties (such as $\log g$, temperatures, and metallicities) and convective granulation using $F_{8}$. They used a large sample of dwarfs, subgiants and red giants (2465 stars in total) observed by Kepler and APOGEE surveys. With their study, they updated the relationship between convective granulation and stellar properties. Using the empirical Flicker-jitter relations presented by \citep{bastien-2014-a,oshagh-2017-a}, they predicted the granulation-driven RV jitter amplitudes for a sample of stars observed by TESS.

The main objective of the current study is to extend these studies to explore, in unprecedented detail, the existence of correlations between RV jitter and photometric variability measurements. This study is performed for 171 G, K and M dwarfs observed with the high-resolution and stable spectrograph HARPS and with the TESS mission, which both provide some of the best RV and photometric precision available.

This paper is organized as following: we explain the target selection and data reduction in Sect.~\ref{targets_sam}. The derivation of stellar parameters is described in Sect.~\ref{st_param}. In Sect.~\ref{cor_result}, we analyze the correlation between RV jitter and photometric peak-to-peak variation. The application of our obtained result on TESS planet candidate host stars is presented in Sect.~\ref{tess_app} by identifying the potential targets which will have the lowest RV jitter based on their light curve photometric variability. Finally, conclusions are drawn in Sect.~{\ref{con}}.

\section{Target selection and data reduction \label{targets_sam}}
\subsection{Target selection}

We selected our targets by applying the following criteria: 

\begin{enumerate}[I.]
\item Cross matching the stars in the HARPS archive with those observed by TESS during its first year (covering 13 observation sectors in the Southern Ecliptic Hemisphere) \footnote{\url{https://tess.mit.edu/observations/}}. All targets have their RV and photometric observations publicly available in the ESO archive and Mikulski Archive for Space Telescopes (MAST)\footnote{\url{https://archive.stsci.edu}}, respectively.

\item Selecting only the stars with RV measurements taken after the fiber change of HARPS on June 2015 \citep{2015Msngr.162....9L} to avoid offset in RV measurements and to make the RV measurements as homogeneous as possible.\footnote{Including the data before the fiber change would add few points per star and one free parameter, the offset between the two, which brings only a little gain.}

\item Removing the targets with known exoplanets to eliminate RV variations due to the known companions \footnote{\url{https://exoplanetarchive.ipac.caltech.edu} \url{exoplanet.eu}}.

\item Selecting stars with measurements on at least 5 nights over a minimum period of one week. \footnote{We have applied this criteria on the reduced final RV measurements which we explained in Sect.~\ref{rv_reduction}}
This criterion aims at producing a minimum sampling of the rotation modulation of G, K and M dwarfs \citep{2013A&A...557L..10N,2020arXiv200108214R}.

\end{enumerate}

\begin{figure*}
   \centering
   \includegraphics[width=14cm]{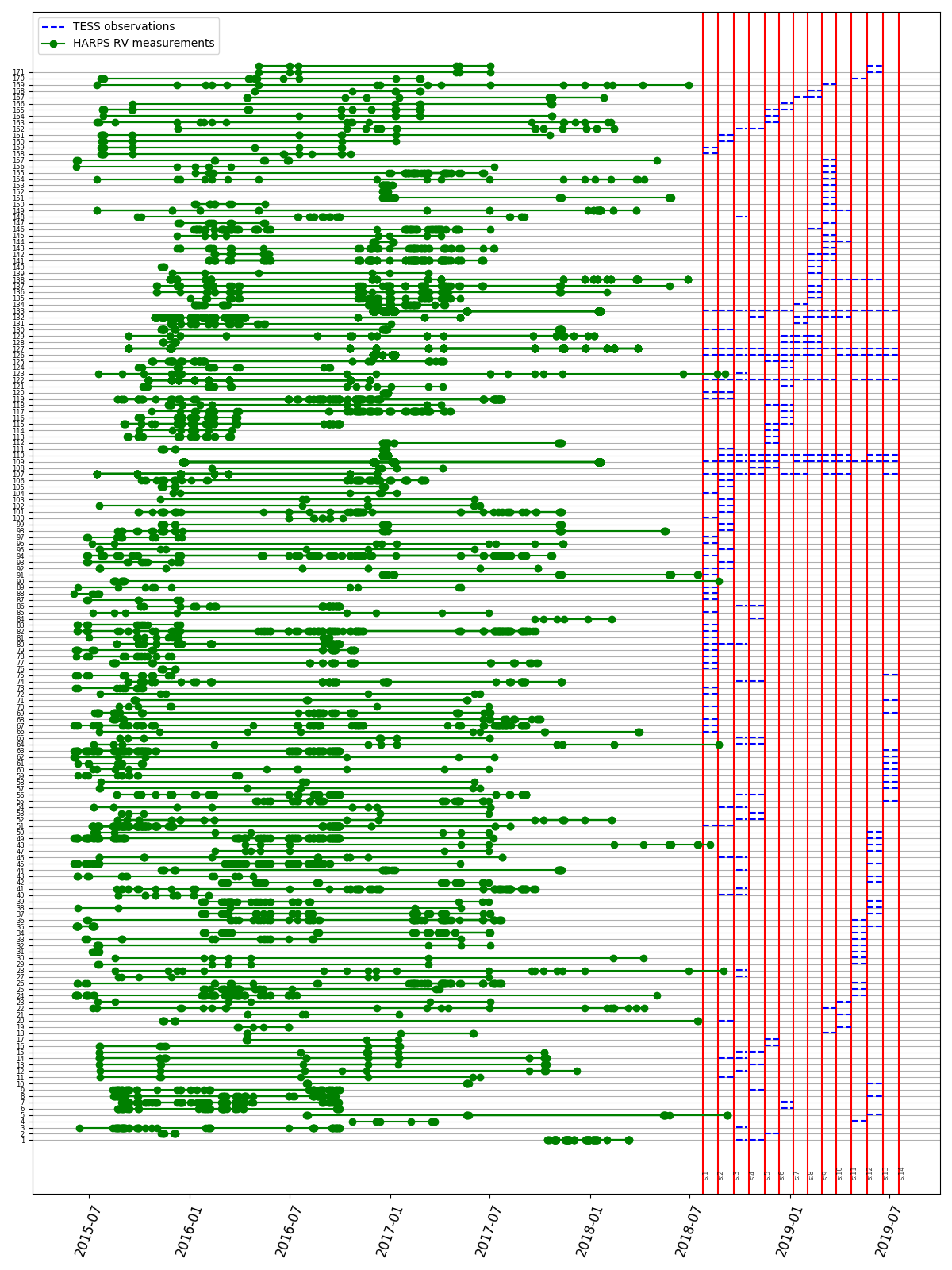}

      \caption{Total RV measurements and TESS observation windows. The green points show the RV measurements dates. Blue dash and red vertical lines show the TESS observations and sectors, respectively. 
      The name of each star is presented in Table.~\ref{stars_param_table} using index number in y-axis.}
         \label{windows}
   \end{figure*}

A total of 171 stars satisfy all of the above criteria. They are all listed in Table.~\ref{stars_param_table}. In Fig.~\ref{windows}, we illustrated the TESS observation windows (sectors) and RV measurement epochs of each target.

\subsection{Light curve: TESS \label{lc_analays}}

\begin{figure}
   \centering
   \includegraphics[width=9.4cm]{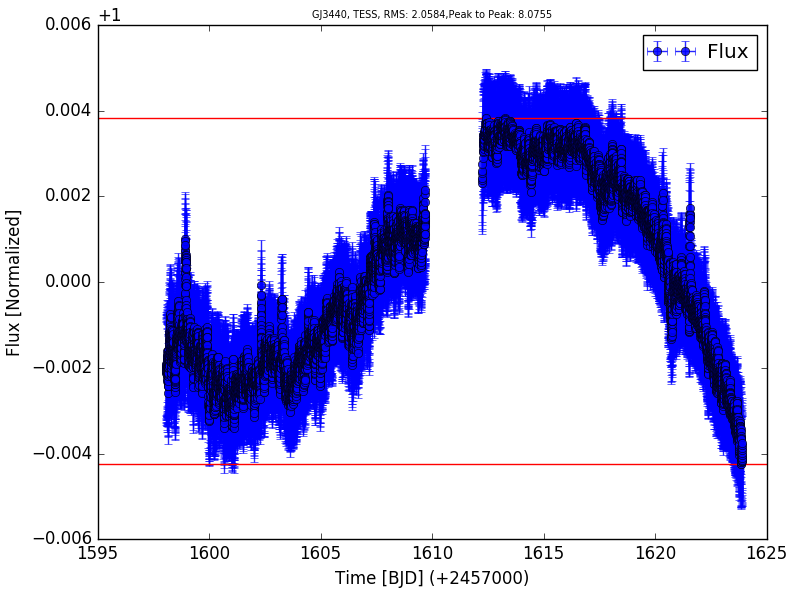}

      \caption{Example of light curve of \object{GJ3440} during one TESS sector (blue points) and the red lines presenting the peak-to-peak of light curve.}
         \label{lc_sample}
\end{figure}

We obtained the light-curves of all 171 stars from MAST. MAST contains simple aperture photometry (SAP$\_$flux) \citep{2017ksci.rept....6M} as well as presearch data conditioning (PDCSAP$\_$flux). Most of the targets (96 $\%$) were observed in only one TESS sector. For stars with light-curves in two or more consecutive sectors, we merged all available  light-curves. We used quality-flag\footnote{$101010111111$: \url{https://outerspace.stsci.edu/display/TESS/2.0+-+Data+Product+Overview}} as suggested by the TESS Data Product review, also recently used and tested in \citep{2019ApJ...884..160V}. We used SAP$\_$flux which optimizes the aperture for the best signal-to-noise for the each target \citep{2017ksci.rept....3B} and also the calibrated pixels in order to perform a summation of the flux. The SAP light curves provided by the TESS pipeline are also background corrected. We removed outlier flux points using a sigma$\-$clipping of three standard deviations and smoothed the fluxes using Savitzky-Golay filter within 15 data point windows ($\sim$ ~30 minutes) to reduce the effect of the short-time scale photometric variability. Since we are interested in variability at stellar rotation timescales, this smoothing does not affect our results. We then normalized the flux by the median flux values. We derived the ratio between the peak-to-peak of light curve variability of SAP$\_$flux and the peak-to-peak of light curve variability of PDCSAP$\_$flux. If this ratio was larger than 3.0, we checked the light curves visually to ensure if there was any evidence for systematic errors in SAP$\_$flux. For more than 90 $\%$ of the stars, we used SAP$\_$flux, and for the rest we assumed that the light-curves are dominated by systematic errors therefore we used PDCSAP$\_$FLUX. In Fig.~\ref{lc_sample}, you can see a sample of reduced light curve in one sector for one star.

\subsection{HARPS RVs \label{rv_reduction}}

\begin{figure}
   \centering
   \includegraphics[width=9.2cm]{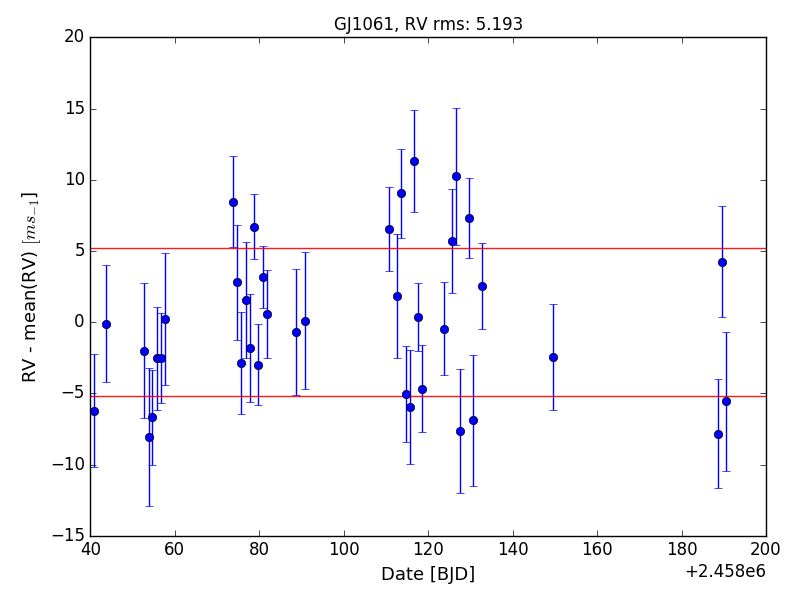}
      \caption{Example of RV measurements for \object{GJ1061} (blue dots), and the red lines presenting the $\pm$ RV-RMS values.}
         \label{rv-sample}
   \end{figure}

For the 171 stars selected in Section \ref{targets_sam} we collected the available HARPS data through ESO archive\footnote{\url{https://www.eso.org/sci/observing/phase3/data_streams.html}}. The data were reduced using the official HARPS pipeline (DRS), that delivers the RV and full width at half maximum (FWHM) measured on each Cross-Correlation Function (CCF). We determined the significant outliers of the RV time-series by applying a sigma\-clipping of three standard deviations, and removed the few measurements that show an RV error larger than 9 $ms^{-1}$ as an admittedly arbitrary value. This threshold was set for being significantly larger than the median RV error ($\sim$ 0.80 $ms^{-1}$) for the whole sample. After the nightly binning of the data, in total, over 3971 spectra were collected and 81 RV measurements ($\sim$ 2\%) were removed based on the previous outlier and photon noise criteria. For each star, we subtracted the mean value of the RV from the RV measurements. Moreover, we corrected for linear long term RV trends using a linear fit for 27 stars of our sample \footnote{GJ3440, HD021175, HD107094, HD1388, HD16548, HD185283, HD19230, HD19641, HD200133, HD202917, HD205536, HD207129, HD211415, HD213042, HD218860, HD221638, HD222335, HD28471, HD3074A, HD40397, HD42936, HD6107, HD76849, HD88218, HD89839, HD94771, HD96700}. This linear trend removal was done to ensure eliminating the signal from any possible unknown long period companions around our targets. It is worth mentioning that unknown planets on short orbits could also affect our RV-RMS measurements, however, if undiscovered and passing the previous vetting criteria, it means that they must be low mass planets and induce low RV semi-amplitude, and thus they do not affect our study significantly. In Fig.~\ref{rv-sample}, we present an example of final RV time-series of one star as well as its associated RV-RMS measurement. Minimum, maximum, and the average of the time spans of the RV observations are $\sim$ 7 days, $\sim$ 3 years, and 1.5 years, respectively. The histogram of the RV time spans is presented in Fig.~\ref{rv-span-hist}. \\

 \subsection{Subsamples \label{sub_sampleing}}
 
 The HARPS measurement are not contemporaneous with TESS observations. Since stellar activity patterns are expected to evolve with time, we divided our whole sample into 4 subsamples as a function of time lag between TESS and HARPS. For the first subsample we considered only the stars with RVs measured up to one year prior to the TESS observations, hereafter named as ``year 1''. The second subsample contains the stars that have RV measurements in a time interval from one year to two years before TESS observations, hereafter named as ``year 2''; the same criteria were used to create the third and the fourth subsamples hereafter named as ``year 3'' and ``year 4'', respectively. Thus, it is important to emphasize that we divided all the RV measurements into four subsamples, and some stars are present in more than one subsample. According to our criteria for dividing the whole RV measurements into the four subsamples, few stars were not categorized in any of these four subsamples, and are only considered in the whole sample. Despite the ability for a star to appear in multiple subsamples, the RV-RMS for a star in a given subsample only uses the RVs taken during that subsample's time-frame. The samples in which targets are present is listed in Table.~\ref{stars_param_table}; stars which are used only in whole sample are labeled as ``w''.
 
\section{Stellar parameters \label{st_param}}

Accurate stellar parameters are necessary for the characterization of exoplanets and their hosts \citep[][]{mayor2004coralie,fischer2005planet,reffert2015precise,2018ASSP...49..225A}. Stellar parameters rule the physics behind the generation of stellar activity regions, and are thus of interest in this analysis. Therefore, we derived the main spectroscopic stellar parameters: effective temperature (\teff), metallicity ([Fe/H]), surface gravity ($\log g$) and stellar rotation rate for the stars in our sample. The stellar parameters for G and K dwarfs are determined with the procedure described in \citet{Sousa-14}. For the M dwarfs in the sample, we used a modified version of the method and software developed by \citet{neves2014metallicity}\footnote{\url{http://www.astro.up.pt/resources/mcal/Site/MCAL.html}}.

\begin{figure}
   \centering
   \includegraphics[width=9.3cm]{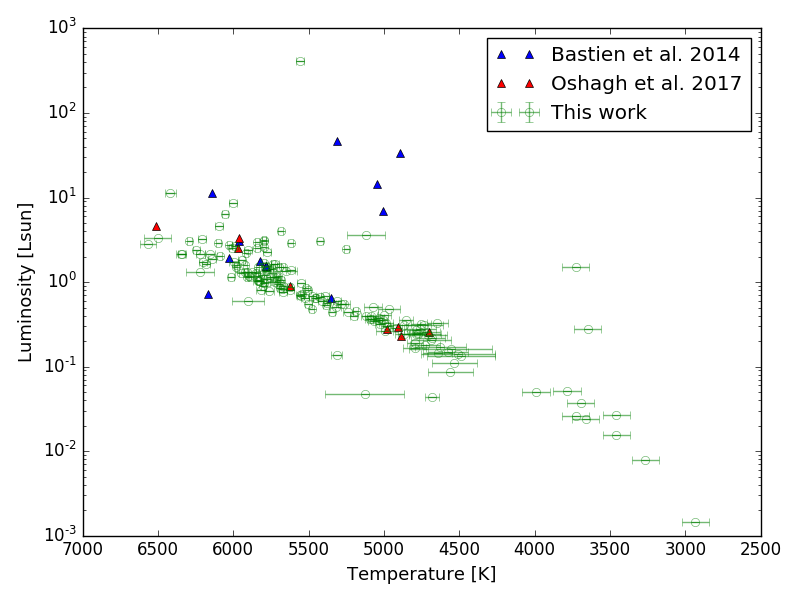}

      \caption{HR diagram for the sample of stars used in this work.}
         \label{teff-lum}
   \end{figure}

Fig.~\ref{teff-lum} presents the Hertzsprung–Russell diagram (HR), for the stars in our sample on which we over-plotted the samples that were used in the two similar previous studies \citep{oshagh-2017-a,bastien-2014-a}. Our sample size is one order of magnitude larger than the previous studies, and covers a wider range of effective temperatures for G and K dwarfs; however, the number of M-dwarfs in our sample is relatively small. 

\subsection{Chromospheric activity indicator $log(R' _{HK})$ and $\textit{v sin i}$         \label{ste-vsini-rhk}}

We derived values for the S-index of each star to evaluate the chromospheric magnetic activity level and its evolution.
We used the open source package $ACTIN$ \footnote{\url{https://pypi.org/project/actin/}, \url{https://github.com/gomesdasilva/ACTIN}} \citep{2018JOSS....3..667G} to calculate the $S_{CaII}$ index \citep{1978PASP...90..267V,1991ApJS...76..383D} homogeneously for all individual spectra following the instructions in \citep[][and references therein]{2011A&A...534A..30G,2019A&A...629A..80H}. We then calibrated $S_{CaII}$ to the Mt. Wilson scale using open source package $pyrhk$ \footnote{\url{https://github.com/gomesdasilva/pyrhk}} and calculated the $R' _{HK}$ chromospheric emission ratio as defined in \citet{1984ApJ...279..763N}.

\begin{figure}
   \centering
   \includegraphics[width=9.4cm]{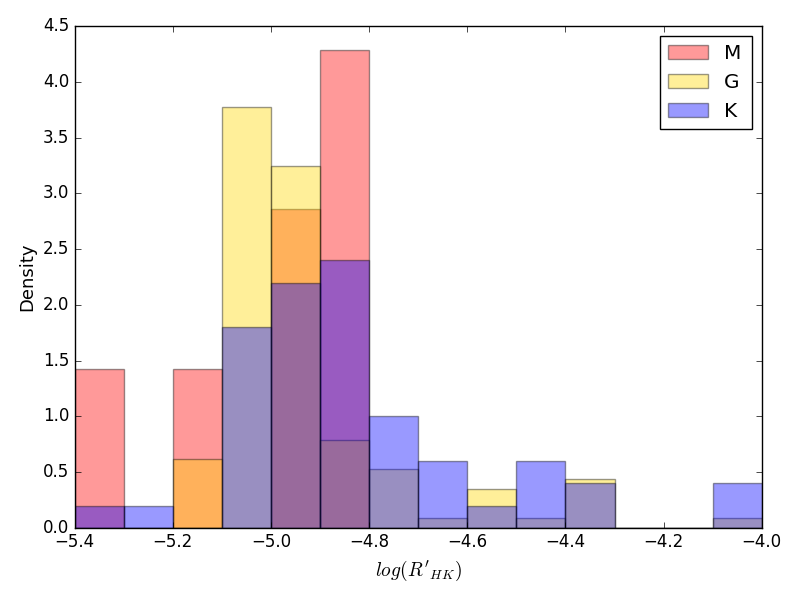}
      \caption{Histogram of mean $log(R' _{HK})$  for the sample of stars used in this work. Red, gold and blue correspond to the spectral type of the star, M, G and K respectively.}
         \label{logrhk-hist}
\end{figure}

For each star we created a $log(R' _{HK})$ time-series and calculated the mean and peak-to-peak of variation of $log(R' _{HK})$ to evaluate the variation of chromospheric activity during the observation span. In Fig.~\ref{logrhk-hist} we present the distribution of mean $log(R' _{HK})$ as a function of spectral types. This figure shows that the majority of our stars are magnetically relativity similar to the solar average value of $log(R' _{HK})$ $\sim$ -4.90 \citep{2008ApJ...687.1264M}, which is a consequence of a bias in the target selection of stars for precise RV searches with HARPS.

 It has been demonstrated that the RV jitter depends strongly on the projected rotational velocity ($\textit{v sin i}$) \citep[e.g.,][]{1997ApJ...485..319S,2007A&A...473..983D,boisse2011disentangling,2015MNRAS.448.3038K}. Thus, we expect that $\textit{v sin i}$ should be one of the influential parameters in this investigation. Therefore, we estimated $\textit{v sin i}$ of all our stars based on the measured FWHM of the CCF. We used the procedure described in \citet[][and references therein]{2002A&A...392..215S,maldonado2017hades,2019A&A...629A..80H}. 

\subsection{Stellar rotation period \label{st-rot-d}}

Photometric contrast differences associated to magnetic features (e.g., dark spots and bright faculae) generate traceable signatures of stellar rotation periods on light curves. 

We analyze the presence of a periodic modulation signal from stellar rotation on the TESS photometric time-series using the gradient of the power spectra (GPS)~\citep[see][]{paperI,Eliana1}. We successfully recover the rotation period for 71 out of 171 stars of the sample. We report the estimated rotation period from the GPS method in Table.~\ref{stars_param_table}. The rotation period from GPS is determined from the enhanced profile of the high-frequency tail of the power spectrum. In particular, we identify the point where the gradient of the power spectrum GPS in log-log scale reaches its maximum value. Such a point corresponds to the high frequency inflection point (HFIP), that is, where the concavity of the power spectrum plotted in the log-log scale changes sign. The position of inflection point is related to the rotation period of star by the calibration factor $\alpha_{Sun}$, for Sun-like stars.

For the calculations presented in this project we adopt a solar-like calibration factor $\alpha_{\rm Sun}\pm~2\sigma=0.158\pm0.014$, and 2~sigma uncertainty~\citet[for more details see,][]{paperI,Eliana1}.

We also estimated the faculae to spot ratio for 29 of the 71 stars. Following~\cite{paperI,Eliana1,Eliana2}, the light curve is faculae dominated when the ratio between HFIP and the independent rotation period ranges between [0.11-0.16], and spot is dominated when the value falls between [0.16-1.24].

\section{Correlation between RV jitter and photometric peak-to-peak \label{cor_result}}
\subsection{Correlation as a function of time \label{cor1}}

\begin{figure*}
    \centering
    \subfloat[year 1]{{\includegraphics[width=8.6cm]{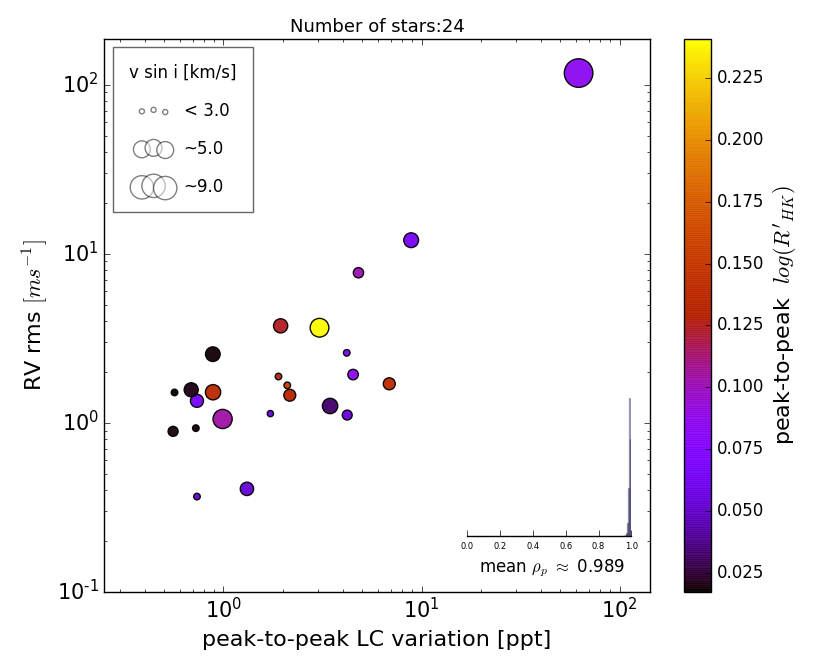}}}
    \qquad
    \subfloat[year 2]{{\includegraphics[width=8.6cm]{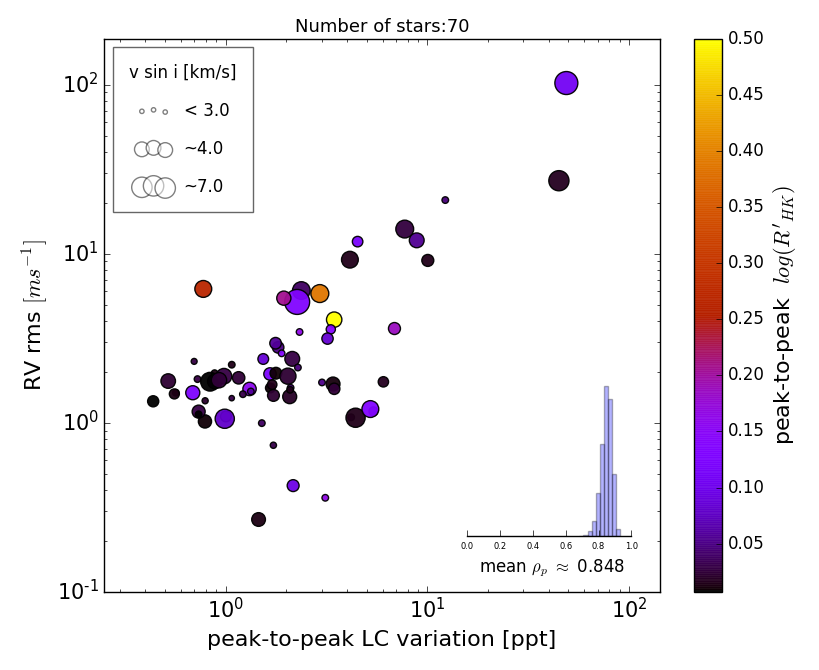}}}

    \subfloat[year 3]{{\includegraphics[width=8.6cm]{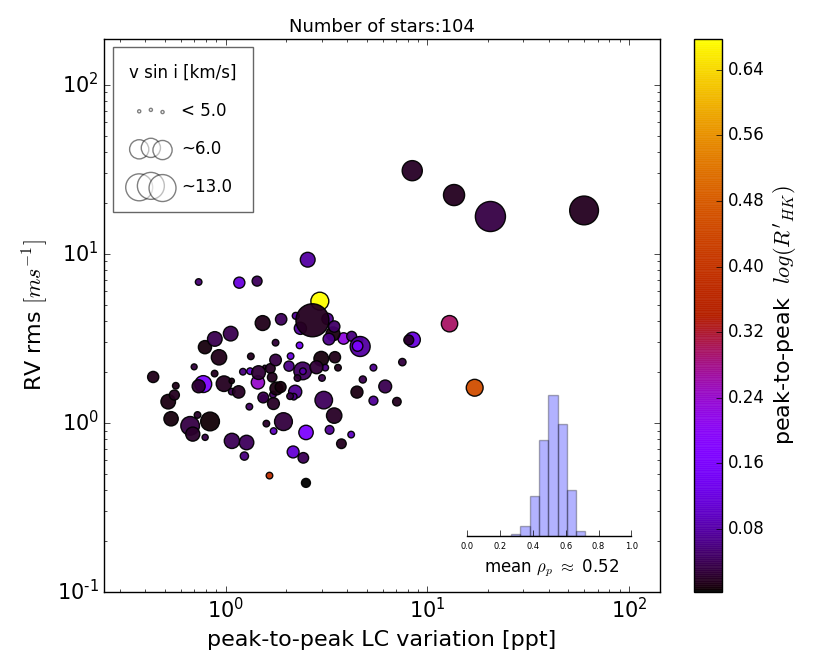} }}
    \qquad
    \subfloat[year 4]{{\includegraphics[width=8.6cm]{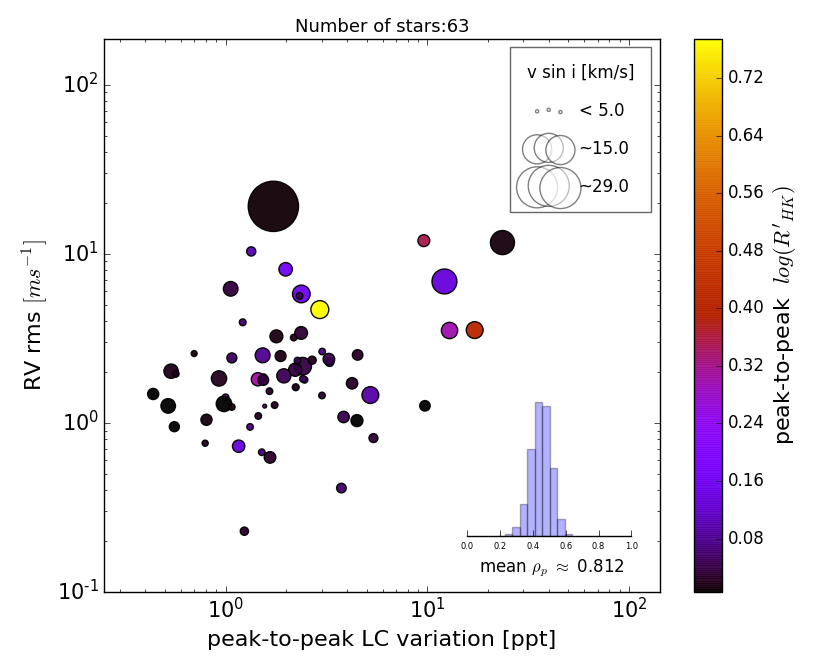} }}
    
    \caption{RV-RMS versus photometric variability (peak-to-peak of light curve) for the year 1 (a), year 2 (b), year 3 (c) and year 4 (d) RV measurements sampling. Color bars present the peak-to-peak variation of $log(R' _{HK})$ during the whole HARPS observations and the size-scale presents the estimation of the average $\textit{v sin i}$ derived from the CCF FWHM described in Sec.~\ref{ste-vsini-rhk}. The calculated value of correlation coefficient $\rho_{sp}$ and its posterior distribution is presented in each panel.}
    \label{cor-plot-1234}
\end{figure*}

In Fig.~\ref{cor-plot-1234} we present the RV-RMS versus peak-to-peak of the photometric time series for the four different subsamples. To evaluate the possible influence of stellar rotation velocity and its magnetic activity level variation, we code these variables into marker color and size. 

In order to evaluate the strength of the correlation between RV-RMS and peak-to-peak of light curve, we derived two correlation coefficients. We used the Pearson's correlation coefficient ($\rho_{pe}$) that evaluates the presence of a linear relationship between data pairs, as well as Spearman's correlation coefficient ($\rho_{sp}$) that, working on ranked data, assesses the presence of a monotonic relationship. Both correlation coefficients were derived using the code described in \citet{2016OLEB...46..385F} which uses the Bayesian framework to evaluate the significance of the presence of a correlation.

 \begin{table*}
 \centering  
\begin{tabular}{cccccc}
\hline
\multirow{2}{*}{Time interval with TESS} & \multicolumn{2}{c}{$\rho$} & \multicolumn{2}{c}{95$\%$} & \multirow{2}{*}{sample size} \\
 & $\rho_{pe}$ & $\rho_{sp}$ & $\rho_{pe}$ & $\rho_{sp}$ & \\
\hline
 year 1 & 0.989 $\pm$ 0.005 & 0.550 $\pm$ 0.133   & [0.979 0.996]&[0.287 0.791] & 24\\

 year 2 & 0.848 $\pm$ 0.033 & 0.394 $\pm$ 0.099   &[0.782 0.908] &[0.196 0.578] & 70\\

 year 3 & 0.520 $\pm$ 0.071 & 0.286 $\pm$ 0.087   &[0.384 0.656] &[0.120 0.456] & 104\\

 year 4 & 0.331 $\pm$ 0.109 & 0.265 $\pm$ 0.113 & [0.111 0.531] &[0.043 0.486] & 63\\

 whole sample & 0.812 $\pm$ 0.026 & 0.444 $\pm$ 0.061  & [0.764 0.862] & [0.319 0.555] & 171\\
 
\hline \hline
\end{tabular}
\caption{Result for Pearson's correlation coefficient ($\rho_{pe}$) and Spearman's rank-order correlation ($\rho_{sp}$) coefficient on ranked data, \citep{2016OLEB...46..385F} with $\rho$: correlation coefficient and its standard deviation and 95$\%$ highest posterior density (HPD) Credible Interval}
\label{cor_table}
\end{table*}

In each panel of Fig.~\ref{cor-plot-1234} we also report the value of the $\rho_{pe}$ correlation coefficient and its posterior distribution. For data obtained within year 1 time span, we obtained the correlation coefficient $\rho_{pe}$ =  0.989 $\pm$ 0.005 ($\rho_{sp}$ =  0.550 $\pm$ 0.133) for the correlation between photometric variability and RV jitter. As the time lag between RV measurement and TESS observation increases (subsample year 2, year 3 and year 4), the correlation becomes less strong and less significant. In the Table.~\ref{cor_table}, we presented the values of the $\rho_{pe}$ and $\rho_{sp}$ \citep{2016OLEB...46..385F}.  We also calculated the standard deviation and 95$\%$ Highest Posterior Density (HPD) Credible Interval as described in \citet{2016OLEB...46..385F} for the both $\rho_{pe}$ and $\rho_{sp}$ values. As can be read on Table.~\ref{cor_table} and Fig.~\ref{cor-plot-1234}, the correlation coefficients decrease to $\rho_{pe}$ = 0.848 $\pm$ 0.033 ($\rho_{sp}$ = 0.394 $\pm$ 0.099) for year 2 data, down to $\rho_{pe}$ = 0.520 $\pm$ 0.071 ($\rho_{sp}$ = 0.286 $\pm$ 0.087) and $\rho_{pe}$ = 0.331 $\pm$ 0.109 ($\rho_{sp}$ = 0.256 $\pm$ 0.113) for year 3 and year 4 data, respectively. $\rho_{pe}$ is always larger than $\rho_{sp}$ because of the extreme values in the subsets.\\

Therefore, our result shows that a correlation exists and decreases significantly when the time interval between the RV observations and photometric observation increases. This can be interpreted as a consequence of the magnetic activity cycles which lead to departure from periodic variability to a quasi-periodic variability, and thus result in weakening the correlation over time. This has an impact on the predictive ability of using one dataset to estimate the other: since several studies aimed at estimating the RV jitter using photometric measurements \citep[e.g.,][]{oshagh-2017-a,bastien-2014-a}, here we note that such a prediction after more than 2 years of photometric observation can be imprecise since the correlation between RV jitter and peak-to-peak light curve variability decreases ( $\rho_{pe}$ $\sim$ 0.8 at year 2 to $\rho_{pe}$ $\sim$ 0.5 at year 3).

\subsection{Overall correlation using the whole sample}

\begin{figure}
   \centering
   \includegraphics[width=0.5 \textwidth, height=6 cm]{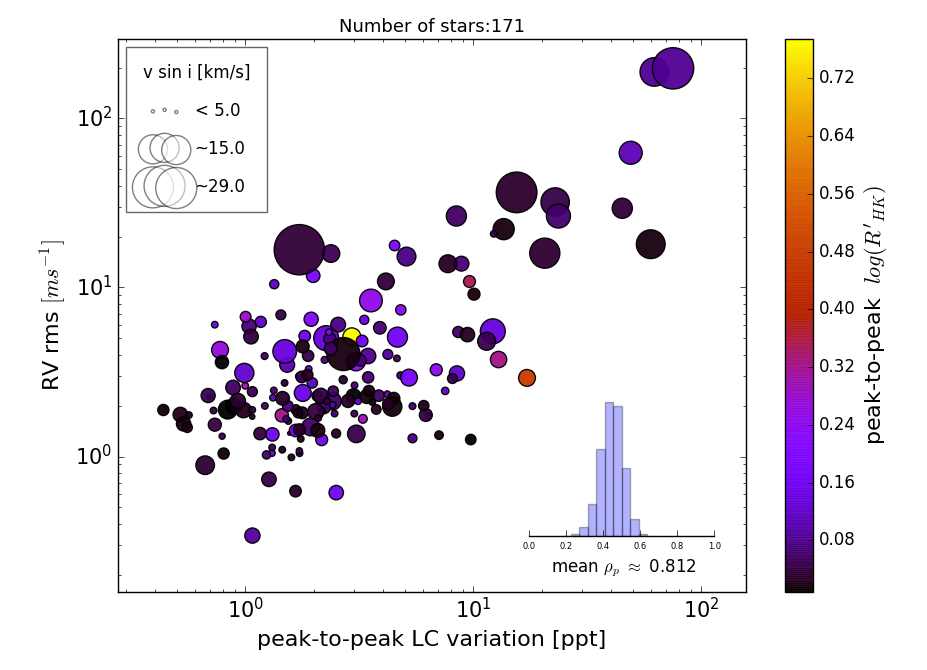}
      \caption{Correlation between RV-RMS and peak-to-peak of light curve variation using the all years RV sampling (whole sample). Size of the points indicate the \vsini~and color represent and peak-to-peak of $log(R' _{HK})$ within the RV measurements. The points above 6 ppt have larger marker sizes and they filled up more space, thus this might cause difficulties in clearly recognizing the double law feature.}
         \label{all_harps}
   \end{figure}

The overall correlation between RV-RMS and peak-to-peak of light curve for the whole sample within the whole 4 years is presented in Fig.~\ref{all_harps}. The correlation is significant; The mean value of $\rho_{pe}$ is 0.812 $\pm$ 0.026 and that of $\rho_{sp}$ is 0.444 $\pm$ 0.061. In this comparison it is important to mention that while the RV jitter is derived using relatively long-term observations (see Fig.~\ref{rv-span-hist}), the photometric peak-to-peak is measured during TESS observations, of roughly one month.

\subsubsection{Functional form of the dependence \label{Functional}}

Careful inspection of the shape of the RV-RMS vs peak-to-peak photometry dependence in Fig.~\ref{all_harps} (see also Fig.~\ref{cor-plot-1234}, especially the year 2 subplot), shows hints of a double behavior, a region where the dependence is weak or plateau (peak-to-peak of light curve variation < 6 ppt) and a region where the dependence is steeper, being associated to a stronger correlation coefficient (peak-to-peak of light curve variation > 6 ppt).  \\

\begin{figure}
  \centering
   \includegraphics[width=9.1cm]{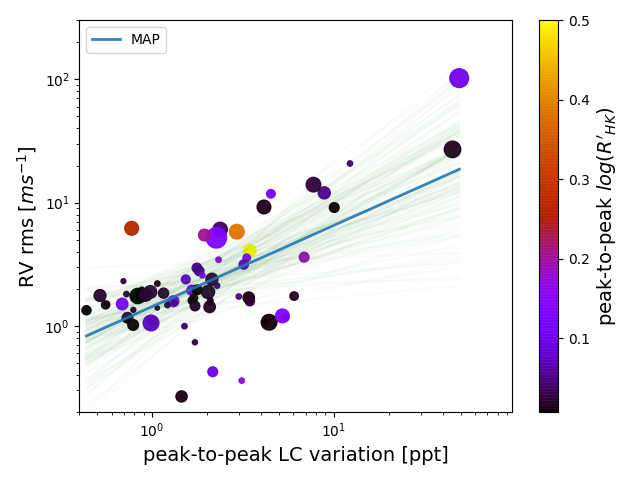}
  \includegraphics[width=9.1cm]{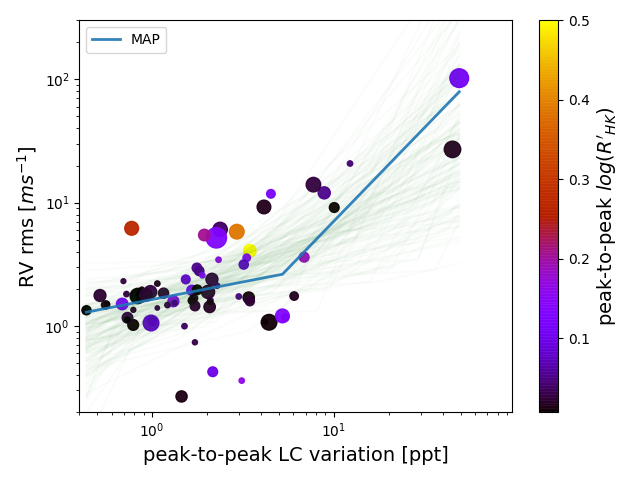}
      \caption{Year 2 sample with single power law function fit (top plot) and a piecewise function with two power law function fit (bottom plot); In both panels the best fit indicated with the Maximum a Posteriori (MAP).}
         \label{second_year_fit}
  \end{figure}

\begin{figure}
  \centering
   \includegraphics[width=9.1cm]{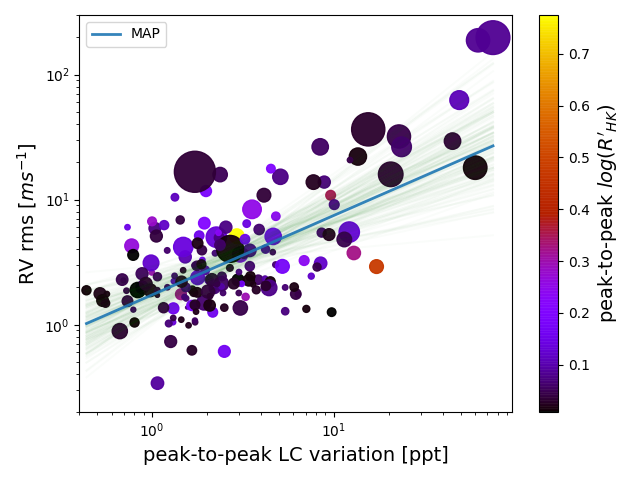}
  \includegraphics[width=9.1cm]{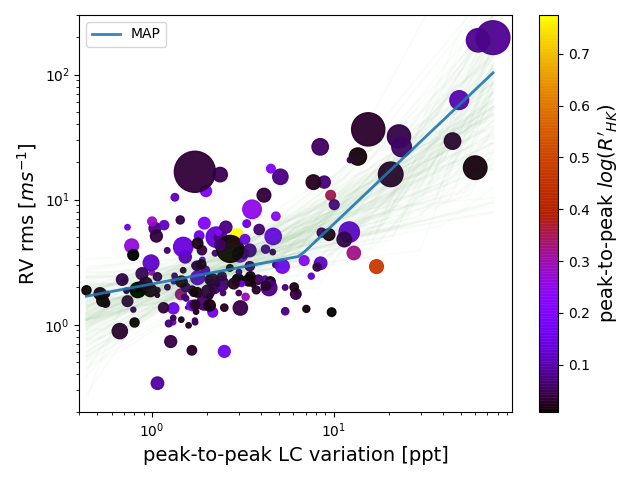}
        \caption{Top plot: the same with single power law function fit, bottom plot: all RV sampling group with two piecewise power law fitting. In both panels the best fit indicated with the Maximum a Posteriori (MAP).}
         \label{cor_all}
  \end{figure}

Here we aim to examine different functional forms that can properly describe the whole sample (Fig.~\ref{all_harps}) and individual subsamples (Fig.~\ref{cor-plot-1234}). To test this, we decided to fit RV jitter and photometric peak-to-peak variability using two types of functional models, a) a power law function and b) piecewise power law function (which is composed of two power law functions with a knee). We used Markov-chain Monte Carlo (MCMC) sampling using the library
$emcee$ \citep{2013PASP..125..306F} to perform the fit and obtain the uncertainties of the fitted parameters; for details refer to Appendix.~\ref{mcmc}. The best fits are overplotted in Fig.~\ref{second_year_fit} for year 2 which is closest subsampling group to the TESS observations with fairly sufficient number of data points. The best fits are also overplotted in Fig.~\ref{cor_all} for the all RV sampling. All fitted parameters are reported in the Table.~\ref{mcmc_table}. The best fit indicated with the Maximum a Posteriori (MAP).

We compare the best fitted models using RMS of the residual ($RMS_{resi}$) and Bayesian information criterion (BIC). For all the subsamples and also for the  whole sample the BIC values are similar between two models. However, since the piecewise function is penalized for the two extra parameters, BIC values for the single power law function has lower BIC than piecewise function, which suggests that the single power law function as a better model, but with a low significant evidence. We considered BIC difference to be significant if $\Delta$BIC > 10  \citep{2007MNRAS.377L..74L,2017EPJC...77..565A}. On the other hand, for the subsample and also for the whole sample, the fits using piecewise model have lower $RMS_{resi}$, which means it can describe the data more precisely. For instance for the whole sample ($RMS_{resi}$ $\sim$ 0.35 vs 0.29) for single and piecewise model, respectively. If we consider the piecewise function as the best model for describing the data, the knee is located at a value of $\sim$ 6.5 ppt peak-to-peak of light curve variation.

\section{Correlations with activity level and rotational period}

The large data-set presented here also allows us to explore other correlations that may be relevant for the estimation of RV jitter in different stars.

\subsection{Correlation with $log(R' _{HK})$}

\begin{figure}
   \centering
   \includegraphics[width=9.4cm]{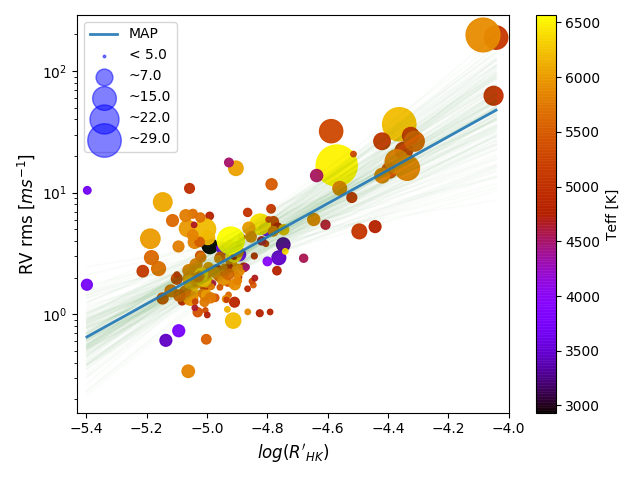}
      \caption{$log(R' _{HK})$ vs RV-RMS of light curve with the best fit. The same method in MCMC analysis explained in Sec.\ref{mcmc}, log10(y) = ax + b : $\mathrm{a} = 1.382_{-0.344}^{0.341}$,
$\mathrm{b} = 7.271_{-1.683}^{1.670}$.
}
         \label{RV_rms_rhk}
   \end{figure}

\begin{figure}
   \centering
   \includegraphics[width=9.4cm]{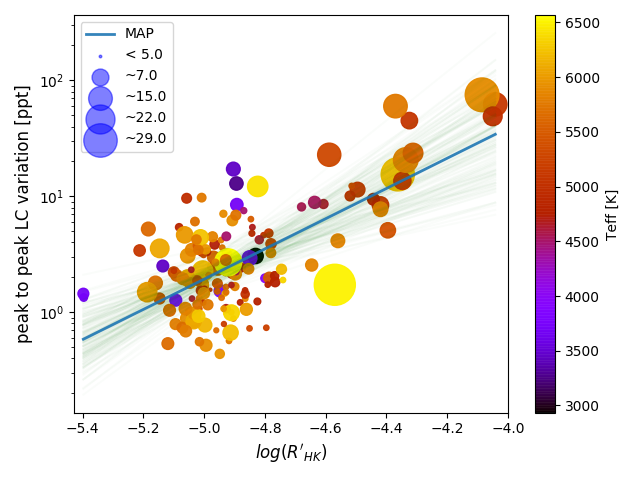}
      \caption{$log(R' _{HK})$ vs peak-to-peak of light curve variation with the best fit. The same method in MCMC analysis explained in Sec.\ref{mcmc}, log10(y) = ax + b : $\mathrm{a} = 1.304_{-0.335}^{0.338}$,
$\mathrm{b} = 6.804_{-1.641}^{1.659}$.
}
         \label{lc_pp_rhk}
   \end{figure}

In the Fig.~\ref{RV_rms_rhk} and Fig.~\ref{lc_pp_rhk}, we present the mean $log(R' _{HK})$ values versus RV-RMS and peak-to-peak photometric variability for the whole sample. We also fitted the power law function and use MCMC to find the error in order to provide a relationship between the RV-RMS, peak-to-peak variability and $log(R' _{HK})$ given by:
\begin{equation}
  \label{log_rhk_eq}
RV_{RMS} \propto (R'_{HK})^{a},~\mathrm{a} = 1.382_{-0.344}^{0.341}.
\end{equation}
Our result is compatible with \citet{2003csss...12..694S} which reported $a = 1.1$ however less compatible with \citet{2000A&A...361..265S} which reported $a = 0.55$. We obtained using the sample of stars in this project. If we assume that the same functional dependence exists between the $log(R' _{HK})$ and peak-to-peak photometric we can write it as
\begin{equation}
  \label{log_rhk_lc_eq}
LC_{pp} \propto (R'_{HK})^{a},~\mathrm{a} = 1.304_{-0.335}^{0.338},
\end{equation}
where $LC_{pp}$ is peak-to-peak of light curve variation.

\subsection{Correlation with stellar rotation period}

\begin{figure}
  \centering
   \includegraphics[width=9.1cm]{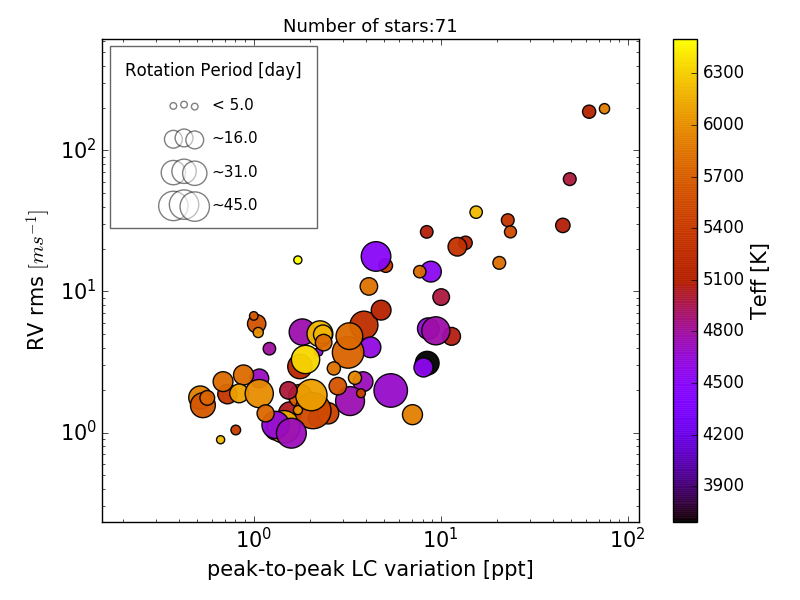}
      \includegraphics[width=9.1cm]{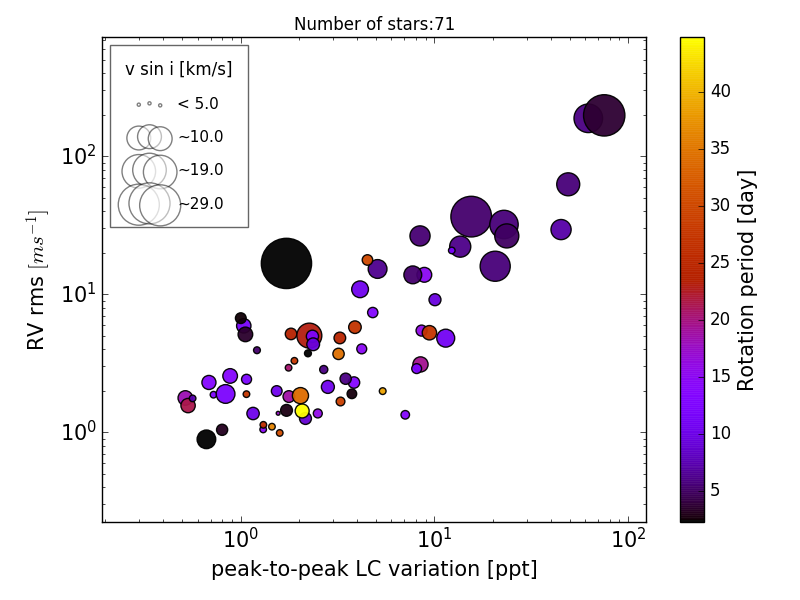}
      \caption{RV-RMS and peak-to-peak of light curve variation for the subsample of 71 stars. Top plot: Circle sizes represents the value of rotation period found by GPS method, see Table.~\ref{stars_param_table}. Color bar indicates the stellar effective temperature. Bottom plot: Similar than top panel but, color bar indicates the rotation period value, and circle size the \vsini~obtained spectroscopically}
         \label{Prot_1}
  \end{figure}

In Top panel in Fig.~\ref{Prot_1}, we present RV-RMS and the peak-to-peak light curve variation for the subsample of 71 stars with measured rotation periods (coded with marker size). The color bar represents the effective temperature. One can easily notice that stars with large RV-RMS and a large peak-to-peak photometric variability are mostly fast rotating stars (less than 13 days) and there is a hint of temperature dependency. Bottom panel in Fig.~\ref{Prot_1} shows the same 71 stars but color bar indicates the rotation period value, and circle size the \vsini~obtained spectroscopically, which again confirms the previous result we found. 

\begin{figure}
  \centering
      \includegraphics[width=9.1cm]{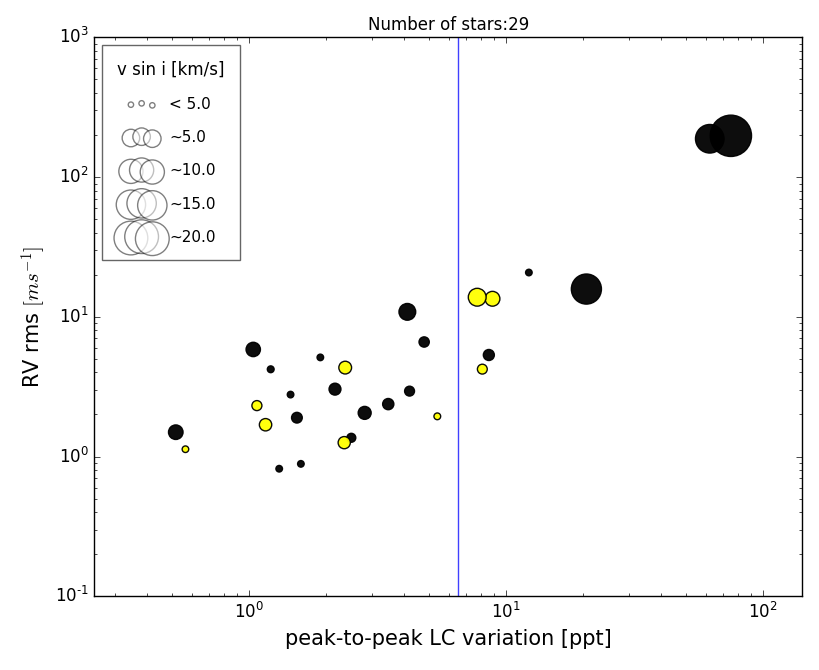}
      \caption{RV-RMS and the peak-to-peak of light curve variation for the subsample of 20 stars with spot dominance and 9 stars with faculae dominance in their light curves. Circle sizes represents the \vsini~value determined spectroscopically. Black color indicates spot dominated and yellow indicates faculae dominated. The Blue vertical line shows the knee point presented in Fig. \ref{cor_all} in peak-to-peak light curve variation at 6.5 ppt.}
         \label{Prot_spot_fac}
  \end{figure}

In Fig.~\ref{Prot_spot_fac}, we present a similar plot to Fig.~\ref{Prot_1} but this time for 29 stars where we could identify faculae or spot dominated patterns using the method described in Sec.~\ref{st-rot-d}. We found 9 faculae dominated stars (which were also slow rotators as is expected for faculae dominated stars), and 20 spot-dominated stars. We show faculae and spot dominated stars in yellow and black, respectively. In this sample, 20 stars can be classified as fast rotators (rotation period < 15 days), and large fraction of them (13 out 20) are spot dominated. This result is in strong agreement with \citet{2017ApJ...851..116M}, where they reported 15 days as the threshold in rotation period for separating spot-faculae dominated regimes. Moreover, faculae-dominated stars tend to have low photometric peak-to-peak variability, due to the low contrast of faculae's region, and therefore are mostly below the 6.5 ppt limit. Thus, the 6.5 ppt limit can be also interpreted as the photometric variability transition between the spot-dominated and the faculae-dominated regime. However, the sample of 29 stars is too small to generalize.

\section{Application of our result on the TESS planet candidates' host stars \label{tess_app}}

It is difficult to define the criteria for the most suitable targets for RV follow up of transiting planet candidates. In Sect.~\ref{Functional} we saw that the dependence of RV jitter on photometric peak-to-peak variation shows a hint of knee at 6.5 ppt. We use this value as an upper threshold to identify the targets on the TESS object of interest (TOI) catalog \footnote{\url{https://tess.mit.edu/toi-releases/}} with the lowest RV jitter. 
We selected TOI catalog targets with publicly available light curves in the archive. Since we are mostly interested in the cooler stars, we selected the stars with \teff~lower than 6800 K. Although in our sample we have filtered out the evolved stars already, we selected the stars with the radius less than 2.0 $R_{sun}$ to narrow our analysis and select stars similar to our the Sun. In this sample, we used \teff, $R_{sun}$ and TESS magnitudes as reported in the TIC \citep{2018AJ....156..102S}.
\\
We then removed all the flux points within the transit reported by the TOI catalog. After eliminating the transit light-curves, we applied the same procedures explained in the section \ref{lc_analays} and estimated the peak-to-peak flux variation for each TOI host star.

   \begin{figure}
   \centering
   \includegraphics[width=9.4cm]{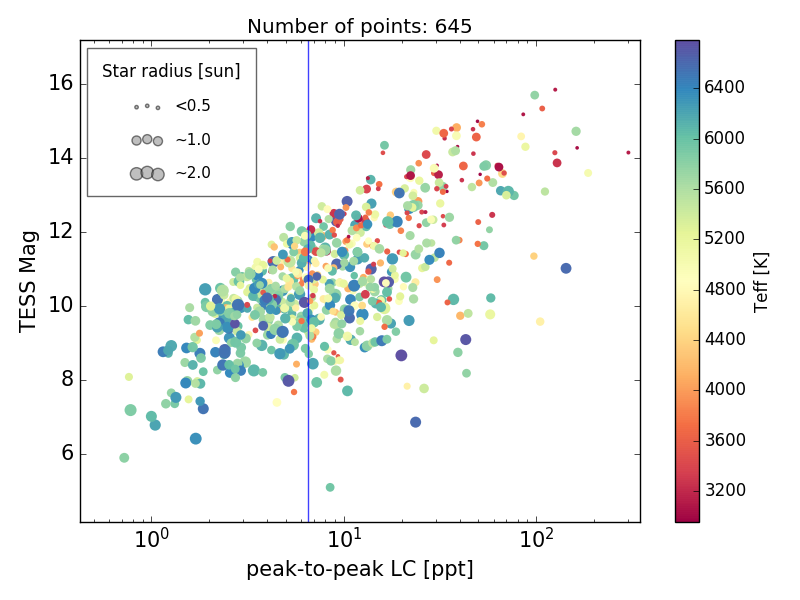}
   \caption{Peak-to-peak of light curve versus TESS mag. Color bar shows the stellar effective temperature and size shows the stellar radius. The Blue vertical line shows the knee point presented in Fig. \ref{cor_all} in peak-to-peak light curve variation at 6.5 ppt.}
         \label{mag_p_tess}
   \end{figure}

In the Fig.~\ref{mag_p_tess}, we present the results of the peak-to-peak variation versus TESS magnitude. Stars with different $R_{sun}$ values are denoted by points with different sizes. The color denotes the effective temperature of the star. The vertical line shows where we believe a critical knee point in Fig.~\ref{cor_all} is. The amplitude of RV jitter above this limit of photometric variability (6.5 ppt) can dramatically increase. 43 $\%$ of the stars in this sample have light curve peak-to-peak variations less than 6.5 ppt (279 out of 645 stars). Mean, median, and interquartile range of RV-RMS of the stars with photometric peak-to-peak variability below 6.5 ppt in our sample are 3.62 $ms^{-1}$, 2.77 $ms^{-1}$, and 2.54 $ms^{-1}$, respectively. We note that in this figure we can also clearly see a correlation between the TESS magnitude and peak-to-peak variation as well as \teff, which can be part of the instrumental as well as the astrophysical effects. The correlation of the peak-to-peak light curve variation with star magnitudes can be explained by the TESS photon noise \citep[see more Fig. 8 in][]{Ricker_2014}. We tested the impact of photon noise by seeking for any correlation between the RV-RMS and stars' magnitude, as well as between the photometric peak-to-peak variation and stars' magnitude in our sample, and could not find any significant correlation.

\section{Discussion and conclusion \label{con}}
  
We present a study on the correlation between RV jitter and photometric variability. We used a sample of 171 starts that have been observed by both the HARPS spectrograph and TESS space telescope. We derived the RV-RMS and peak-to-peak of light curve variability as well as the stellar parameters. \\

We found a strong correlation between the RV jitter and photometric flux variability. We divided our sample into four subsamples in order to investigate the impact of time interval between RV measurements and light curves. We found that the correlation becomes weaker as the time lags between the RV measurements and light curves increase. Within year 1 and year 2, the correlation is strong but the year 3 and year 4 show relatively weaker correlation. We can conclude the maximum time difference in which the light curve peak-to-peak can be used to predict the RV jitter is two years. After that the information content of photometric viability is lost, probably the consequence of the evolution of the stellar magnetic cycle. This could imply that majority of these stars with different spectral types should have a cycle shorter than 6 years, assuming that half an activity cycle is sufficient to bring the observables out of phase and diminishes the correlation (for instance if the RVs were taken at activity maximum and photometry were obtained at activity minimum or vice-versa).
\\
We also examined different functional forms which could represent better the observed correlations. We compare the best fitted model using $RMS_{resi}$ and BIC. For the whole sample the fit of piecewise power law functions, with a knee at 6.5 ppt in peak-to-peak of light curve variation, describes the data more precisely with lower $RMS_{resi}$ without considering the cost of the two extra parameters. We also fitted the two models on the year 1, year 2, year 3, and year 4 subsampling (see Table.~\ref{mcmc_table}). The results show that for the year 2, which is the closest subsampling group to the TESS observations with fairly sufficient number of data points as well as a strong correlation, the piecewise power law also describes the data better than the single power law with lower $RMS_{resi}$, however, the BIC is larger for the piecewise model because of the two extra parameters in this model.
The correlation between RV-RMS and peak-to-peak of light curve variability of showed two regions separating with a knee point where the strength of the correlation changes. This might be due to the transition from the spot-dominated regime to the faculae-dominated regime. These two regimes have already been discussed in several studies \citep{1997ApJ...485..789L,1998ApJS..118..239R,2016A&A...589A..46S,oshagh-2017-a,2020A&A...633A..32S}
\\

\citet{2019A&A...629A..42M,2019A&A...632A..81M} simulated the RVs and photometric time series and derived the flux variability as well as the global RV jitter for F to early K stars. They showed that there is a relationship between the photometric variability and RV jitter considering the dependency of stellar parameters such as rotation period and \teff~as well as inclination. These simulations showed presence of a knee point in the correlation between photometric variability and RV jitter, however, the knee point and also the range of values are in a different domain than our findings. There could be several reason for these disagreements such as: a) our observational sample of stars could have more complex and random spot or faculae than the ones simulated in  \citet{2019A&A...629A..42M}. b) our observational RV sample were not simultaneous with the TESS light curve, while in \citet{2019A&A...629A..42M} the simulation were done for simultaneous observations. c) the observation sample could have unknown companions which can produce RV variation in time series. However, the dispersion of the simulations for RV-RMS and photometric variability \citep[see Fig.~9 in][]{2019A&A...629A..42M} are compatible with the observed dispersion of the whole sample.\\
We also found that there is a strong correlation between the chromospheric activity index ($log(R' _{HK})$) and RV jitter and also with photometric variability, which are both in agreement with previous studies \citep{2000A&A...361..265S,2005PASP..117..657W,2003csss...12..694S,2019A&A...632A..81M}. Also the observed correlation between $log(R' _{HK})$ and peak-to-peak of photometric variation is compatible to the simulation's result within the range -5.0 < $log(R' _{HK})$ < -4.6 \citep[see Fig.~5 in][]{2019A&A...629A..42M}. We provided an updated version of the relation between $log(R' _{HK})$ and RV jitter and also with peak-to-peak of light curve variability.\\

We were able to estimate the rotation period of 71 stars, out of 171 stars in our sample, using the TESS light curve. Then we investigated the effect of this parameter on the correlation between RV-RMS and peak-to-peak of light curve variability. Our result demonstrated that slow rotating stars (which are the ones also we found to be faculae dominated) create lower RV jitter as well as lower peak-to-peak photometric variability, and on the other hand fast rotating star, $P_{rot}$ $\leq$ 5 day (which are the ones also we found to be spot dominated) generate much larger RV jitter and photometric variability. 
\\
We have also looked into the correlation of the other parameters which might affect the peak-to-peak of light curve vs RV-RMS correlation such as \teff, $[Fe/H]$ and the number of nights that the star has been observed in RV. We could not find any significant trend caused by these parameters on the peak-to-peak of light curve vs RV-RMS correlation (Appendix~\ref{extra_p}). 
We also examined our results by using just the first sector for the stars which have more than one TESS sector observations and we did not see any significant changes in the results.
Finally, we attempted to select the best targets, in terms of the lowest RV jitter, in TESS Objects of Interest catalog. To do that, we used our estimated knee point at 6.5 ppt in photometric variability. We estimated that 42$\%$ of the stars in a specific sample of the TOI catalog have the peak-to-peak photometric variability of less than 6.5 ppt and, therefore, will have a small amplitude RV jitter (less than $\sim$ 3 $ms^{-1}$) which can be essential for the confirmation of Earth-mass exoplanets at short orbits around low-mass stars in TESS objects of interest. (see Table.~\ref{list_tic}).

\begin{acknowledgements}
We would like to thank the anonymous referee for insightful and 
constructive comments, which added significantly to the clarity of this paper. This work was supported by FCT - Funda\c{c}\~ao para a Ci\^encia e a Tecnologia through national funds and by FEDER through COMPETE2020 - Programa Operacional Competitividade e Internacionaliza\c{c}\~ao by these grants: UID/FIS/04434/2019; UIDB/04434/2020 \& UIDP/04434/2020; PTDC/FIS-AST/32113/2017 \& POCI-01-0145-FEDER-032113; PTDC/FIS-AST/28953/2017 \& POCI-01-0145-FEDER-028953.

S.H. and B.A. acknowledge support by the fellowships PD/BD/128119/2016 and PD/BD/135226/2017 funded by FCT (Portugal) and POCH/FSE (EC).

S.S., V.A., O.D.S.D. and J.P.F. acknowledge support from FCT through work contracts nºs IF/00028/2014/CP1215/CT0002, IF/00650/2015/CP1273, DL 57/2016/CP1364/CT0004, DL 57/2016/CP1364/CT0005.

M.O. acknowledges the support of the Deutsche
Forschungsgemeinschft (DFG) priority program SPP 1992 “Exploring the Diversity of Extrasolar Planets (RE 1664/17-1)”. M.O., E.A.G also acknowledge the support of the FCT/DAAD bilateral grant 2019 (DAAD ID: 57453096). M.O. acknowledges research funding from the Deutsche Forschungsgemeinschft (DFG, German Research Foundation) - OS 508/1-1.
This research made use of Astropy, a
community-developed core Python package for Astronomy \citep{astropy:2018,2013A&A...558A..33A}, and the NumPy, SciPy, Matplotlib, tesscut, celerite Python modules \citep{van2011numpy,scipy,Hunter:2007,2019ascl.soft05007B,celerite} and \textit{pandas} \citep{mckinney2010data,mckinney2011pandas}. This research made use of \textit{Lightkurve}, a Python package for Kepler and TESS data analysis \citep{2018ascl.soft12013L}. This paper includes data collected with the TESS mission, obtained from the MAST data archive at the Space Telescope Science Institute (STScI). Funding for the TESS mission is provided by the NASA Explorer Program. STScI is operated by the Association of Universities for Research in Astronomy, Inc., under NASA contract NAS 5–26555. This work is based on observations collected at the European Organization for Astronomical Research in the Southern
Hemisphere under ESO programs: '0100.C-0097(A)','0100.C-0487(A)','0100.C-0836(A)','0100.D-0444(A)','0101.C-0379(A)','0101.D-0494(A)','0101.D-0494(B)','060.A-9036(A)','089.C-0732(A)','090.C-0421(A)','093.C-0062(A)','095.C-0040(A)','095.C-0551(A)','095.C-0799','095.C-0799(A)','096.C-0053(A)','096.C-0460(A)','096.C-0499(A)','096.C-0876(A)','097.C-0021(A)','097.C-0090(A)','097.C-0390(B)','097.C-0571(A)','098-C-0518(A)','098.C-0269(A)','098.C-0269(B)','098.C-0366(A)','098.C-0518(A)','098.C-0739(A)','099.C-0205(A)','099.C-0458(A)','099.C-0798(A)','183.C-0437(A)','188.C-0265(O)','188.C-0265(P)','188.C-0265(Q)','188.C-0265(R)','190.C-0027(A)','191.C-0873(A)','192.C-0224','192.C-0224(C)','192.C-0852(A)','196.C-0042','196.C-0042(D)','196.C-0042(E)','196.C-1006(A)','198.C-0836(A)','198.C-0838(A)','60.A-9036(A)','Lagrange'. 
      
\end{acknowledgements}

\bibliographystyle{aa}
\bibliography{bt}

\begin{thebibliography}{86}
\expandafter\ifx\csname natexlab\endcsname\relax\def\natexlab#1{#1}\fi

\bibitem[{{Adibekyan} {et~al.}(2018){Adibekyan}, {Sousa}, \&
  {Santos}}]{2018ASSP...49..225A}
{Adibekyan}, V., {Sousa}, S.~G., \& {Santos}, N.~C. 2018, in Asteroseismology
  and Exoplanets: Listening to the Stars and Searching for New Worlds, ed.
  T.~L. {Campante}, N.~C. {Santos}, \& M.~J.~P.~F.~G. {Monteiro}, Vol.~49, 225

\bibitem[{{Aigrain} {et~al.}(2012){Aigrain}, {Pont}, \&
  {Zucker}}]{2012MNRAS.419.3147A}
{Aigrain}, S., {Pont}, F., \& {Zucker}, S. 2012, \mnras, 419, 3147

\bibitem[{{Amazo-Gomez, E.M.} {et~al.}(2020){Amazo-Gomez, E.M.}, {Shapiro,
  A.I.}, {Solanki, S.K.}, {Krivova}, {Kopp, N.K.}, {Reinhold}, \&
  {Oshagh}}]{Eliana2}
{Amazo-Gomez, E.M.}, {Shapiro, A.I.}, {Solanki, S.K.}, {et~al.} 2020, \aap,
  submitted

\bibitem[{Amazo-Gómez {et~al.}(2020)Amazo-Gómez, Shapiro, Solanki, Krivova,
  Kopp, Reinhold, Oshagh, \& Reiners}]{Eliana1}
Amazo-Gómez, E.~M., Shapiro, A.~I., Solanki, S.~K., {et~al.} 2020, Rotation
  periods from the inflection point in the power spectrum of stellar brightness
  variations: II. The Sun

\bibitem[{{Arevalo} {et~al.}(2017){Arevalo}, {Cid}, \&
  {Moya}}]{2017EPJC...77..565A}
{Arevalo}, F., {Cid}, A., \& {Moya}, J. 2017, European Physical Journal C, 77,
  565

\bibitem[{{Astropy Collaboration} {et~al.}(2013){Astropy Collaboration},
  {Robitaille}, {Tollerud}, {Greenfield}, {Droettboom}, {Bray}, {Aldcroft},
  {Davis}, {Ginsburg}, {Price-Whelan}, {Kerzendorf}, {Conley}, {Crighton},
  {Barbary}, {Muna}, {Ferguson}, {Grollier}, {Parikh}, {Nair}, {Unther},
  {Deil}, {Woillez}, {Conseil}, {Kramer}, {Turner}, {Singer}, {Fox}, {Weaver},
  {Zabalza}, {Edwards}, {Azalee Bostroem}, {Burke}, {Casey}, {Crawford},
  {Dencheva}, {Ely}, {Jenness}, {Labrie}, {Lim}, {Pierfederici}, {Pontzen},
  {Ptak}, {Refsdal}, {Servillat}, \& {Streicher}}]{2013A&A...558A..33A}
{Astropy Collaboration}, {Robitaille}, T.~P., {Tollerud}, E.~J., {et~al.} 2013,
  \aap, 558, A33

\bibitem[{{Barros} {et~al.}(2014){Barros}, {Almenara}, {Deleuil}, {Diaz},
  {Csizmadia}, {Cabrera}, {Chaintreuil}, {Collier Cameron}, {Hatzes},
  {Haywood}, {Lanza}, {Aigrain}, {Alonso}, {Bord{\'e}}, {Bouchy}, {Deeg},
  {Erikson}, {Fridlund}, {Grziwa}, {Gandolfi}, {Guillot}, {Guenther}, {Leger},
  {Moutou}, {Ollivier}, {Pasternacki}, {P{\"a}tzold}, {Rauer}, {Rouan},
  {Santerne}, {Schneider}, \& {Wuchterl}}]{2014A&A...569A..74B}
{Barros}, S.~C.~C., {Almenara}, J.~M., {Deleuil}, M., {et~al.} 2014, \aap, 569,
  A74

\bibitem[{{Bastien} {et~al.}(2013){Bastien}, {Stassun}, {Basri}, \&
  {Pepper}}]{2013Natur.500..427B}
{Bastien}, F.~A., {Stassun}, K.~G., {Basri}, G., \& {Pepper}, J. 2013, \nat,
  500, 427

\bibitem[{{Bastien} {et~al.}(2014){Bastien}, {Stassun}, {Pepper}, {Wright},
  {Aigrain}, {Basri}, {Johnson}, {Howard}, \& {Walkowicz}}]{bastien-2014-a}
{Bastien}, F.~A., {Stassun}, K.~G., {Pepper}, J., {et~al.} 2014, \aj, 147, 29

\bibitem[{{Boisse} {et~al.}(2011){Boisse}, {Bouchy}, {H{\'e}brard}, {Bonfils},
  {Santos}, \& {Vauclair}}]{2011IAUS..273..281B}
{Boisse}, I., {Bouchy}, F., {H{\'e}brard}, G., {et~al.} 2011, in IAU Symposium,
  Vol. 273, Physics of Sun and Star Spots, ed. D.~{Prasad Choudhary} \& K.~G.
  {Strassmeier}, 281--285

\bibitem[{Boisse {et~al.}(2011)Boisse, Bouchy, H{\'e}brard, Bonfils, Santos, \&
  Vauclair}]{boisse2011disentangling}
Boisse, I., Bouchy, F., H{\'e}brard, G., {et~al.} 2011, \aap, 528, A4

\bibitem[{Borucki {et~al.}(2010)Borucki, Koch, Basri, Batalha, Brown, Caldwell,
  Caldwell, Christensen-Dalsgaard, Cochran, DeVore,
  {et~al.}}]{borucki2010kepler}
Borucki, W.~J., Koch, D., Basri, G., {et~al.} 2010, Science, 327, 977

\bibitem[{{Brasseur} {et~al.}(2019){Brasseur}, {Phillip}, {Fleming},
  {Mullally}, \& {White}}]{2019ascl.soft05007B}
{Brasseur}, C.~E., {Phillip}, C., {Fleming}, S.~W., {Mullally}, S.~E., \&
  {White}, R.~L. 2019, {Astrocut: Tools for creating cutouts of TESS images}

\bibitem[{{Bryson} {et~al.}(2017){Bryson}, {Jenkins}, {Klaus}, {Cote},
  {Quintana}, {Campbell}, {Zamudio}, {Chandrasekaran}, {Caldwell}, {Van Cleve},
  \& {Haas}}]{2017ksci.rept....3B}
{Bryson}, S.~T., {Jenkins}, J.~M., {Klaus}, T.~C., {et~al.} 2017, {Kepler Data
  Processing Handbook: Target and Aperture Definitions: Selecting Pixels for
  Kepler Downlink}, Tech. rep.

\bibitem[{{Cegla} {et~al.}(2014){Cegla}, {Stassun}, {Watson}, {Bastien}, \&
  {Pepper}}]{cegla-2014-a}
{Cegla}, H.~M., {Stassun}, K.~G., {Watson}, C.~A., {Bastien}, F.~A., \&
  {Pepper}, J. 2014, \apj, 780, 104

\bibitem[{{Cloutier} {et~al.}(2018){Cloutier}, {Doyon}, {Bouchy}, \&
  {H{\'e}brard}}]{2018AJ....156...82C}
{Cloutier}, R., {Doyon}, R., {Bouchy}, F., \& {H{\'e}brard}, G. 2018, \aj, 156,
  82

\bibitem[{{Czesla} {et~al.}(2009){Czesla}, {Huber}, {Wolter}, {Schr{\"o}ter},
  \& {Schmitt}}]{2009A&A...505.1277C}
{Czesla}, S., {Huber}, K.~F., {Wolter}, U., {Schr{\"o}ter}, S., \& {Schmitt},
  J.~H.~M.~M. 2009, \aap, 505, 1277

\bibitem[{{Desort} {et~al.}(2007){Desort}, {Lagrange}, {Galland}, {Udry}, \&
  {Mayor}}]{2007A&A...473..983D}
{Desort}, M., {Lagrange}, A.~M., {Galland}, F., {Udry}, S., \& {Mayor}, M.
  2007, \aap, 473, 983

\bibitem[{{D{\'\i}az} {et~al.}(2016){D{\'\i}az}, {S{\'e}gransan}, {Udry},
  {Lovis}, {Pepe}, {Dumusque}, {Marmier}, {Alonso}, {Benz}, {Bouchy},
  {Coffinet}, {Collier Cameron}, {Deleuil}, {Figueira}, {Gillon}, {Lo Curto},
  {Mayor}, {Mordasini}, {Motalebi}, {Moutou}, {Pollacco}, {Pompei}, {Queloz},
  {Santos}, \& {Wyttenbach}}]{2016A&A...585A.134D}
{D{\'\i}az}, R.~F., {S{\'e}gransan}, D., {Udry}, S., {et~al.} 2016, \aap, 585,
  A134

\bibitem[{{Dumusque} {et~al.}(2011{\natexlab{a}}){Dumusque}, {Santos}, {Udry},
  {Lovis}, \& {Bonfils}}]{2011A&A...527A..82D}
{Dumusque}, X., {Santos}, N.~C., {Udry}, S., {Lovis}, C., \& {Bonfils}, X.
  2011{\natexlab{a}}, \aap, 527, A82

\bibitem[{{Dumusque} {et~al.}(2011{\natexlab{b}}){Dumusque}, {Udry}, {Lovis},
  {Santos}, \& {Monteiro}}]{2011A&A...525A.140D}
{Dumusque}, X., {Udry}, S., {Lovis}, C., {Santos}, N.~C., \& {Monteiro},
  M.~J.~P.~F.~G. 2011{\natexlab{b}}, \aap, 525, A140

\bibitem[{{Duncan} {et~al.}(1991){Duncan}, {Vaughan}, {Wilson}, {Preston},
  {Frazer}, {Lanning}, {Misch}, {Mueller}, {Soyumer}, {Woodard}, {Baliunas},
  {Noyes}, {Hartmann}, {Porter}, {Zwaan}, {Middelkoop}, {Rutten}, \&
  {Mihalas}}]{1991ApJS...76..383D}
{Duncan}, D.~K., {Vaughan}, A.~H., {Wilson}, O.~C., {et~al.} 1991, \apjs, 76,
  383

\bibitem[{{Figueira} {et~al.}(2016){Figueira}, {Faria}, {Adibekyan}, {Oshagh},
  \& {Santos}}]{2016OLEB...46..385F}
{Figueira}, P., {Faria}, J.~P., {Adibekyan}, V.~Z., {Oshagh}, M., \& {Santos},
  N.~C. 2016, Origins of Life and Evolution of the Biosphere, 46, 385

\bibitem[{{Figueira} {et~al.}(2010){Figueira}, {Marmier}, {Bonfils}, {di
  Folco}, {Udry}, {Santos}, {Lovis}, {M{\'e}gevand}, {Melo}, {Pepe}, {Queloz},
  {S{\'e}gransan}, {Triaud}, \& {Viana Almeida}}]{2010A&A...513L...8F}
{Figueira}, P., {Marmier}, M., {Bonfils}, X., {et~al.} 2010, \aap, 513, L8

\bibitem[{{Findeisen} {et~al.}(2011){Findeisen}, {Hillenbrand}, \&
  {Soderblom}}]{2011AJ....142...23F}
{Findeisen}, K., {Hillenbrand}, L., \& {Soderblom}, D. 2011, \aj, 142, 23

\bibitem[{{Fischer} {et~al.}(2016){Fischer}, {Anglada-Escude}, {Arriagada},
  {Baluev}, {Bean}, {Bouchy}, {Buchhave}, {Carroll}, {Chakraborty}, {Crepp},
  {Dawson}, {Diddams}, {Dumusque}, {Eastman}, {Endl}, {Figueira}, {Ford},
  {Foreman-Mackey}, {Fournier}, {F{\H{u}}r{\'e}sz}, {Gaudi}, {Gregory},
  {Grundahl}, {Hatzes}, {H{\'e}brard}, {Herrero}, {Hogg}, {Howard}, {Johnson},
  {Jorden}, {Jurgenson}, {Latham}, {Laughlin}, {Loredo}, {Lovis}, {Mahadevan},
  {McCracken}, {Pepe}, {Perez}, {Phillips}, {Plavchan}, {Prato}, {Quirrenbach},
  {Reiners}, {Robertson}, {Santos}, {Sawyer}, {Segransan}, {Sozzetti},
  {Steinmetz}, {Szentgyorgyi}, {Udry}, {Valenti}, {Wang}, {Wittenmyer}, \&
  {Wright}}]{2016PASP..128f6001F}
{Fischer}, D.~A., {Anglada-Escude}, G., {Arriagada}, P., {et~al.} 2016, \pasp,
  128, 066001

\bibitem[{Fischer \& Valenti(2005)}]{fischer2005planet}
Fischer, D.~A. \& Valenti, J. 2005, The Astrophysical Journal, 622, 1102

\bibitem[{{Foreman-Mackey} {et~al.}(2017){Foreman-Mackey}, {Agol}, {Angus}, \&
  {Ambikasaran}}]{celerite}
{Foreman-Mackey}, D., {Agol}, E., {Angus}, R., \& {Ambikasaran}, S. 2017, ArXiv

\bibitem[{{Foreman-Mackey} {et~al.}(2013){Foreman-Mackey}, {Hogg}, {Lang}, \&
  {Goodman}}]{2013PASP..125..306F}
{Foreman-Mackey}, D., {Hogg}, D.~W., {Lang}, D., \& {Goodman}, J. 2013, \pasp,
  125, 306

\bibitem[{{Gomes da Silva} {et~al.}(2018){Gomes da Silva}, {Figueira},
  {Santos}, \& {Faria}}]{2018JOSS....3..667G}
{Gomes da Silva}, J., {Figueira}, P., {Santos}, N., \& {Faria}, J. 2018, The
  Journal of Open Source Software, 3, 667

\bibitem[{{Gomes da Silva} {et~al.}(2011){Gomes da Silva}, {Santos}, {Bonfils},
  {Delfosse}, {Forveille}, \& {Udry}}]{2011A&A...534A..30G}
{Gomes da Silva}, J., {Santos}, N.~C., {Bonfils}, X., {et~al.} 2011, \aap, 534,
  A30

\bibitem[{{Hojjatpanah} {et~al.}(2019){Hojjatpanah}, {Figueira}, {Santos},
  {Adibekyan}, {Sousa}, {Delgado-Mena}, {Alibert}, {Cristiani}, {Gonz{\'a}lez
  Hern{\'a}ndez}, {Lanza}, {Di Marcantonio}, {Martins}, {Micela}, {Molaro},
  {Neves}, {Oshagh}, {Pepe}, {Poretti}, {Rojas-Ayala}, {Rebolo}, {Su{\'a}rez
  Mascare{\~n}o}, \& {Zapatero Osorio}}]{2019A&A...629A..80H}
{Hojjatpanah}, S., {Figueira}, P., {Santos}, N.~C., {et~al.} 2019, \aap, 629,
  A80

\bibitem[{{Hu{\'e}lamo} {et~al.}(2008){Hu{\'e}lamo}, {Figueira}, {Bonfils},
  {Santos}, {Pepe}, {Gillon}, {Azevedo}, {Barman}, {Fern{\'a}ndez}, {di Folco},
  {Guenther}, {Lovis}, {Melo}, {Queloz}, \& {Udry}}]{2008A&A...489L...9H}
{Hu{\'e}lamo}, N., {Figueira}, P., {Bonfils}, X., {et~al.} 2008, \aap, 489, L9

\bibitem[{Hunter(2007)}]{Hunter:2007}
Hunter, J.~D. 2007, Computing in Science \& Engineering, 9, 90

\bibitem[{Jones {et~al.}(2001--)Jones, Oliphant, Peterson, {et~al.}}]{scipy}
Jones, E., Oliphant, T., Peterson, P., {et~al.} 2001--, {SciPy}: Open source
  scientific tools for {Python}, [Online; accessed <today>]

\bibitem[{{Korhonen} {et~al.}(2015){Korhonen}, {Andersen}, {Piskunov},
  {Hackman}, {Juncher}, {J{\"a}rvinen}, \&
  {J{\o}rgensen}}]{2015MNRAS.448.3038K}
{Korhonen}, H., {Andersen}, J.~M., {Piskunov}, N., {et~al.} 2015, \mnras, 448,
  3038

\bibitem[{{Liddle}(2007)}]{2007MNRAS.377L..74L}
{Liddle}, A.~R. 2007, \mnras, 377, L74

\bibitem[{{Lightkurve Collaboration} {et~al.}(2018){Lightkurve Collaboration},
  {Cardoso}, {Hedges}, {Gully-Santiago}, {Saunders}, {Cody}, {Barclay}, {Hall},
  {Sagear}, {Turtelboom}, {Zhang}, {Tzanidakis}, {Mighell}, {Coughlin}, {Bell},
  {Berta-Thompson}, {Williams}, {Dotson}, \& {Barentsen}}]{2018ascl.soft12013L}
{Lightkurve Collaboration}, {Cardoso}, J.~V.~d.~M., {Hedges}, C., {et~al.}
  2018, {Lightkurve: Kepler and TESS time series analysis in Python},
  Astrophysics Source Code Library

\bibitem[{{Lo Curto} {et~al.}(2015){Lo Curto}, {Pepe}, {Avila}, {Boffin},
  {Bovay}, {Chazelas}, {Coffinet}, {Fleury}, {Hughes}, {Lovis}, {Maire},
  {Manescau}, {Pasquini}, {Rihs}, {Sinclaire}, \& {Udry}}]{2015Msngr.162....9L}
{Lo Curto}, G., {Pepe}, F., {Avila}, G., {et~al.} 2015, The Messenger, 162, 9

\bibitem[{{Lockwood} {et~al.}(1997){Lockwood}, {Skiff}, \&
  {Radick}}]{1997ApJ...485..789L}
{Lockwood}, G.~W., {Skiff}, B.~A., \& {Radick}, R.~R. 1997, \apj, 485, 789

\bibitem[{Maldonado {et~al.}(2017)Maldonado, Scandariato, Stelzer, Biazzo,
  Lanza, Maggio, Micela, Gonz{\'a}lez-{\'A}lvarez, Affer, Claudi,
  {et~al.}}]{maldonado2017hades}
Maldonado, J., Scandariato, G., Stelzer, B., {et~al.} 2017, Astronomy \&
  Astrophysics, 598, A27

\bibitem[{{Mamajek} \& {Hillenbrand}(2008)}]{2008ApJ...687.1264M}
{Mamajek}, E.~E. \& {Hillenbrand}, L.~A. 2008, \apj, 687, 1264

\bibitem[{{Martin} {et~al.}(2005){Martin}, {Fanson}, {Schiminovich},
  {Morrissey}, {Friedman}, {Barlow}, {Conrow}, {Grange}, {Jelinsky},
  {Milliard}, {Siegmund}, {Bianchi}, {Byun}, {Donas}, {Forster}, {Heckman},
  {Lee}, {Madore}, {Malina}, {Neff}, {Rich}, {Small}, {Surber}, {Szalay},
  {Welsh}, \& {Wyder}}]{2005ApJ...619L...1M}
{Martin}, D.~C., {Fanson}, J., {Schiminovich}, D., {et~al.} 2005, \apjl, 619,
  L1

\bibitem[{{Mayor} {et~al.}(2003){Mayor}, {Pepe}, {Queloz}, {Bouchy},
  {Rupprecht}, {Lo Curto}, {Avila}, {Benz}, {Bertaux}, {Bonfils}, {Dall},
  {Dekker}, {Delabre}, {Eckert}, {Fleury}, {Gilliotte}, {Gojak}, {Guzman},
  {Kohler}, {Lizon}, {Longinotti}, {Lovis}, {Megevand}, {Pasquini}, {Reyes},
  {Sivan}, {Sosnowska}, {Soto}, {Udry}, {van Kesteren}, {Weber}, \&
  {Weilenmann}}]{HARPS}
{Mayor}, M., {Pepe}, F., {Queloz}, D., {et~al.} 2003, The Messenger, 114, 20

\bibitem[{Mayor {et~al.}(2004)Mayor, Udry, Naef, Pepe, Queloz, Santos, \&
  Burnet}]{mayor2004coralie}
Mayor, M., Udry, S., Naef, D., {et~al.} 2004, Astronomy \& Astrophysics, 415,
  391

\bibitem[{McKinney(2010)}]{mckinney2010data}
McKinney, W. 2010, in Proceedings of the 9th Python in Science Conference, Vol.
  445, Austin, TX, 51--56

\bibitem[{McKinney(2011)}]{mckinney2011pandas}
McKinney, W. 2011, Python for High Performance and Scientific Computing, 14

\bibitem[{{Meunier} \& {Lagrange}(2019)}]{2019A&A...629A..42M}
{Meunier}, N. \& {Lagrange}, A.~M. 2019, \aap, 629, A42

\bibitem[{{Meunier} {et~al.}(2019){Meunier}, {Lagrange}, \&
  {Cuzacq}}]{2019A&A...632A..81M}
{Meunier}, N., {Lagrange}, A.~M., \& {Cuzacq}, S. 2019, \aap, 632, A81

\bibitem[{{Montet} {et~al.}(2017){Montet}, {Tovar}, \&
  {Foreman-Mackey}}]{2017ApJ...851..116M}
{Montet}, B.~T., {Tovar}, G., \& {Foreman-Mackey}, D. 2017, \apj, 851, 116

\bibitem[{{Morris} {et~al.}(2017){Morris}, {Twicken}, {Smith}, {Clarke},
  {Jenkins}, {Bryson}, {Girouard}, \& {Klaus}}]{2017ksci.rept....6M}
{Morris}, R.~L., {Twicken}, J.~D., {Smith}, J.~C., {et~al.} 2017, {Kepler Data
  Processing Handbook: Photometric Analysis}, Tech. rep.

\bibitem[{Neves {et~al.}(2014)Neves, Bonfils, Santos, Delfosse, Forveille,
  Allard, \& Udry}]{neves2014metallicity}
Neves, V., Bonfils, X., Santos, N., {et~al.} 2014, Astronomy \& Astrophysics,
  568, A121

\bibitem[{{Nielsen} {et~al.}(2013){Nielsen}, {Gizon}, {Schunker}, \&
  {Karoff}}]{2013A&A...557L..10N}
{Nielsen}, M.~B., {Gizon}, L., {Schunker}, H., \& {Karoff}, C. 2013, \aap, 557,
  L10

\bibitem[{{Noyes} {et~al.}(1984){Noyes}, {Hartmann}, {Baliunas}, {Duncan}, \&
  {Vaughan}}]{1984ApJ...279..763N}
{Noyes}, R.~W., {Hartmann}, L.~W., {Baliunas}, S.~L., {Duncan}, D.~K., \&
  {Vaughan}, A.~H. 1984, \apj, 279, 763

\bibitem[{{Oshagh}(2018)}]{2018ASSP...49..239O}
{Oshagh}, M. 2018, in Asteroseismology and Exoplanets: Listening to the Stars
  and Searching for New Worlds, Vol.~49, 239

\bibitem[{{Oshagh} {et~al.}(2013{\natexlab{a}}){Oshagh}, {Boisse}, {Bou{\'e}},
  {Montalto}, {Santos}, {Bonfils}, \& {Haghighipour}}]{2013A&A...549A..35O}
{Oshagh}, M., {Boisse}, I., {Bou{\'e}}, G., {et~al.} 2013{\natexlab{a}}, \aap,
  549, A35

\bibitem[{{Oshagh} {et~al.}(2015){Oshagh}, {Santos}, {Boisse}, {Bou{\'e}},
  {Ehrenreich}, {Haghighipour}, {Figueira}, {Santerne}, \&
  {Dumusque}}]{2015EPJWC.10105003O}
{Oshagh}, M., {Santos}, N.~C., {Boisse}, I., {et~al.} 2015, in European
  Physical Journal Web of Conferences, Vol. 101, 05003

\bibitem[{{Oshagh} {et~al.}(2013{\natexlab{b}}){Oshagh}, {Santos}, {Boisse},
  {Bou{\'e}}, {Montalto}, {Dumusque}, \& {Haghighipour}}]{2013A&A...556A..19O}
{Oshagh}, M., {Santos}, N.~C., {Boisse}, I., {et~al.} 2013{\natexlab{b}}, \aap,
  556, A19

\bibitem[{{Oshagh} {et~al.}(2017){Oshagh}, {Santos}, {Figueira}, {Barros},
  {Donati}, {Adibekyan}, {Faria}, {Watson}, {Cegla}, {Dumusque}, {H{\'e}brard},
  {Demangeon}, {Dreizler}, {Boisse}, {Deleuil}, {Bonfils}, {Pepe}, \&
  {Udry}}]{oshagh-2017-a}
{Oshagh}, M., {Santos}, N.~C., {Figueira}, P., {et~al.} 2017, \aap, 606, A107

\bibitem[{Pepe {et~al.}(2014)Pepe, Molaro, Cristiani, Rebolo, Santos, Dekker,
  M{\'e}gevand, Zerbi, Cabral, Di~Marcantonio, {et~al.}}]{pepe2014espresso}
Pepe, F., Molaro, P., Cristiani, S., {et~al.} 2014, Astronomische Nachrichten,
  335, 8

\bibitem[{{Price-Whelan} {et~al.}(2018){Price-Whelan}, {Sip{\H{o}}cz},
  {G{\"u}nther}, {Lim}, {Crawford}, {Conseil}, {Shupe}, {Craig}, {Dencheva},
  {Ginsburg}, {VanderPlas}, {Bradley}, {P{\'e}rez-Su{\'a}rez}, {de Val-Borro},
  {Paper Contributors}, {Aldcroft}, {Cruz}, {Robitaille}, {Tollerud},
  {Coordination Committee}, {Ardelean}, {Babej}, {Bach}, {Bachetti}, {Bakanov},
  {Bamford}, {Barentsen}, {Barmby}, {Baumbach}, {Berry}, {Biscani}, {Boquien},
  {Bostroem}, {Bouma}, {Brammer}, {Bray}, {Breytenbach}, {Buddelmeijer},
  {Burke}, {Calderone}, {Cano Rodr{\'\i}guez}, {Cara}, {Cardoso}, {Cheedella},
  {Copin}, {Corrales}, {Crichton}, {D{\textquoteright}Avella}, {Deil},
  {Depagne}, {Dietrich}, {Donath}, {Droettboom}, {Earl}, {Erben}, {Fabbro},
  {Ferreira}, {Finethy}, {Fox}, {Garrison}, {Gibbons}, {Goldstein}, {Gommers},
  {Greco}, {Greenfield}, {Groener}, {Grollier}, {Hagen}, {Hirst}, {Homeier},
  {Horton}, {Hosseinzadeh}, {Hu}, {Hunkeler}, {Ivezi{\'c}}, {Jain}, {Jenness},
  {Kanarek}, {Kendrew}, {Kern}, {Kerzendorf}, {Khvalko}, {King}, {Kirkby},
  {Kulkarni}, {Kumar}, {Lee}, {Lenz}, {Littlefair}, {Ma}, {Macleod},
  {Mastropietro}, {McCully}, {Montagnac}, {Morris}, {Mueller}, {Mumford},
  {Muna}, {Murphy}, {Nelson}, {Nguyen}, {Ninan}, {N{\"o}the}, {Ogaz}, {Oh},
  {Parejko}, {Parley}, {Pascual}, {Patil}, {Patil}, {Plunkett}, {Prochaska},
  {Rastogi}, {Reddy Janga}, {Sabater}, {Sakurikar}, {Seifert}, {Sherbert},
  {Sherwood-Taylor}, {Shih}, {Sick}, {Silbiger}, {Singanamalla}, {Singer},
  {Sladen}, {Sooley}, {Sornarajah}, {Streicher}, {Teuben}, {Thomas},
  {Tremblay}, {Turner}, {Terr{\'o}n}, {van Kerkwijk}, {de la Vega}, {Watkins},
  {Weaver}, {Whitmore}, {Woillez}, {Zabalza}, \& {Contributors}}]{astropy:2018}
{Price-Whelan}, A.~M., {Sip{\H{o}}cz}, B.~M., {G{\"u}nther}, H.~M., {et~al.}
  2018, \aj, 156, 123

\bibitem[{{Queloz} {et~al.}(2001){Queloz}, {Henry}, {Sivan}, {Baliunas},
  {Beuzit}, {Donahue}, {Mayor}, {Naef}, {Perrier}, \&
  {Udry}}]{2001A&A...379..279Q}
{Queloz}, D., {Henry}, G.~W., {Sivan}, J.~P., {et~al.} 2001, \aap, 379, 279

\bibitem[{{Radick} {et~al.}(1998){Radick}, {Lockwood}, {Skiff}, \&
  {Baliunas}}]{1998ApJS..118..239R}
{Radick}, R.~R., {Lockwood}, G.~W., {Skiff}, B.~A., \& {Baliunas}, S.~L. 1998,
  \apjs, 118, 239

\bibitem[{{Rauer} {et~al.}(2014){Rauer}, {Catala}, {Aerts}, {Appourchaux},
  {Benz}, {Brandeker}, {Christensen-Dalsgaard}, {Deleuil}, {Gizon}, {Goupil},
  {G{\"u}del}, {Janot-Pacheco}, {Mas-Hesse}, {Pagano}, {Piotto}, {Pollacco},
  {Santos}, {Smith}, {Su{\'a}rez}, {Szab{\'o}}, {Udry}, {Adibekyan}, {Alibert},
  {Almenara}, {Amaro-Seoane}, {Eiff}, {Asplund}, {Antonello}, {Barnes},
  {Baudin}, {Belkacem}, {Bergemann}, {Bihain}, {Birch}, {Bonfils}, {Boisse},
  {Bonomo}, {Borsa}, {Brand {\~a}o}, {Brocato}, {Brun}, {Burleigh}, {Burston},
  {Cabrera}, {Cassisi}, {Chaplin}, {Charpinet}, {Chiappini}, {Church},
  {Csizmadia}, {Cunha}, {Damasso}, {Davies}, {Deeg}, {D{\'\i}az}, {Dreizler},
  {Dreyer}, {Eggenberger}, {Ehrenreich}, {Eigm{\"u}ller}, {Erikson}, {Farmer},
  {Feltzing}, {de Oliveira Fialho}, {Figueira}, {Forveille}, {Fridlund},
  {Garc{\'\i}a}, {Giommi}, {Giuffrida}, {Godolt}, {Gomes da Silva}, {Granzer},
  {Grenfell}, {Grotsch-Noels}, {G{\"u}nther}, {Haswell}, {Hatzes},
  {H{\'e}brard}, {Hekker}, {Helled}, {Heng}, {Jenkins}, {Johansen},
  {Khodachenko}, {Kislyakova}, {Kley}, {Kolb}, {Krivova}, {Kupka}, {Lammer},
  {Lanza}, {Lebreton}, {Magrin}, {Marcos-Arenal}, {Marrese}, {Marques},
  {Martins}, {Mathis}, {Mathur}, {Messina}, {Miglio}, {Montalban}, {Montalto},
  {Monteiro}, {Moradi}, {Moravveji}, {Mordasini}, {Morel}, {Mortier},
  {Nascimbeni}, {Nelson}, {Nielsen}, {Noack}, {Norton}, {Ofir}, {Oshagh},
  {Ouazzani}, {P{\'a}pics}, {Parro}, {Petit}, {Plez}, {Poretti}, {Quirrenbach},
  {Ragazzoni}, {Raimondo}, {Rainer}, {Reese}, {Redmer}, {Reffert},
  {Rojas-Ayala}, {Roxburgh}, {Salmon}, {Santerne}, {Schneider}, {Schou},
  {Schuh}, {Schunker}, {Silva-Valio}, {Silvotti}, {Skillen}, {Snellen}, {Sohl},
  {Sousa}, {Sozzetti}, {Stello}, {Strassmeier}, {{\v{S}}vanda}, {Szab{\'o}},
  {Tkachenko}, {Valencia}, {Van Grootel}, {Vauclair}, {Ventura}, {Wagner},
  {Walton}, {Weingrill}, {Werner}, {Wheatley}, \&
  {Zwintz}}]{2014ExA....38..249R}
{Rauer}, H., {Catala}, C., {Aerts}, C., {et~al.} 2014, Experimental Astronomy,
  38, 249

\bibitem[{Reffert {et~al.}(2015)Reffert, Bergmann, Quirrenbach, Trifonov, \&
  K{\"u}nstler}]{reffert2015precise}
Reffert, S., Bergmann, C., Quirrenbach, A., Trifonov, T., \& K{\"u}nstler, A.
  2015, Astronomy \& Astrophysics, 574, A116

\bibitem[{{Reinhold} \& {Hekker}(2020)}]{2020arXiv200108214R}
{Reinhold}, T. \& {Hekker}, S. 2020, arXiv e-prints, arXiv:2001.08214

\bibitem[{{Ricker} {et~al.}(2014){Ricker}, {Winn}, {Vanderspek}, {Latham},
  {Bakos}, {Bean}, {Berta-Thompson}, {Brown}, {Buchhave}, {Butler}, {Butler},
  {Chaplin}, {Charbonneau}, {Christensen-Dalsgaard}, {Clampin}, {Deming},
  {Doty}, {De Lee}, {Dressing}, {Dunham}, {Endl}, {Fressin}, {Ge}, {Henning},
  {Holman}, {Howard}, {Ida}, {Jenkins}, {Jernigan}, {Johnson}, {Kaltenegger},
  {Kawai}, {Kjeldsen}, {Laughlin}, {Levine}, {Lin}, {Lissauer}, {MacQueen},
  {Marcy}, {McCullough}, {Morton}, {Narita}, {Paegert}, {Palle}, {Pepe},
  {Pepper}, {Quirrenbach}, {Rinehart}, {Sasselov}, {Sato}, {Seager},
  {Sozzetti}, {Stassun}, {Sullivan}, {Szentgyorgyi}, {Torres}, {Udry}, \&
  {Villasenor}}]{2014SPIE.9143E..20R}
{Ricker}, G.~R., {Winn}, J.~N., {Vanderspek}, R., {et~al.} 2014, in Society of
  Photo-Optical Instrumentation Engineers (SPIE) Conference Series, Vol. 9143,
  \procspie, 914320

\bibitem[{Ricker {et~al.}(2014)Ricker, Winn, Vanderspek, Latham, Bakos, Bean,
  Berta-Thompson, Brown, Buchhave, Butler, \& et~al.}]{Ricker_2014}
Ricker, G.~R., Winn, J.~N., Vanderspek, R., {et~al.} 2014, Journal of
  Astronomical Telescopes, Instruments, and Systems, 1, 014003

\bibitem[{{Robertson} {et~al.}(2014){Robertson}, {Mahadevan}, {Endl}, \&
  {Roy}}]{2014Sci...345..440R}
{Robertson}, P., {Mahadevan}, S., {Endl}, M., \& {Roy}, A. 2014, Science, 345,
  440

\bibitem[{{Saar} {et~al.}(1998){Saar}, {Butler}, \&
  {Marcy}}]{1998ApJ...498L.153S}
{Saar}, S.~H., {Butler}, R.~P., \& {Marcy}, G.~W. 1998, \apjl, 498, L153

\bibitem[{{Saar} \& {Donahue}(1997)}]{1997ApJ...485..319S}
{Saar}, S.~H. \& {Donahue}, R.~A. 1997, \apj, 485, 319

\bibitem[{{Saar} {et~al.}(2003){Saar}, {Hatzes}, {Cochran}, \&
  {Paulson}}]{2003csss...12..694S}
{Saar}, S.~H., {Hatzes}, A., {Cochran}, W., \& {Paulson}, D. 2003, in Cambridge
  Workshop on Cool Stars, Stellar Systems, and the Sun, Vol.~12, The Future of
  Cool-Star Astrophysics: 12th Cambridge Workshop on Cool Stars, Stellar
  Systems, and the Sun, ed. A.~{Brown}, G.~M. {Harper}, \& T.~R. {Ayres},
  694--698

\bibitem[{{Santos} {et~al.}(2010){Santos}, {Gomes da Silva}, {Lovis}, \&
  {Melo}}]{2010A&A...511A..54S}
{Santos}, N.~C., {Gomes da Silva}, J., {Lovis}, C., \& {Melo}, C. 2010, \aap,
  511, A54

\bibitem[{{Santos} {et~al.}(2000){Santos}, {Mayor}, {Naef}, {Pepe}, {Queloz},
  {Udry}, \& {Blecha}}]{2000A&A...361..265S}
{Santos}, N.~C., {Mayor}, M., {Naef}, D., {et~al.} 2000, \aap, 361, 265

\bibitem[{{Santos} {et~al.}(2002){Santos}, {Mayor}, {Naef}, {Pepe}, {Queloz},
  {Udry}, {Burnet}, {Clausen}, {Helt}, {Olsen}, \&
  {Pritchard}}]{2002A&A...392..215S}
{Santos}, N.~C., {Mayor}, M., {Naef}, D., {et~al.} 2002, \aap, 392, 215

\bibitem[{{Santos} {et~al.}(2014){Santos}, {Mortier}, {Faria}, {Dumusque},
  {Adibekyan}, {Delgado-Mena}, {Figueira}, {Benamati}, {Boisse}, {Cunha},
  {Gomes da Silva}, {Lo Curto}, {Lovis}, {Martins}, {Mayor}, {Melo}, {Oshagh},
  {Pepe}, {Queloz}, {Santerne}, {S{\'e}gransan}, {Sozzetti}, {Sousa}, \&
  {Udry}}]{2014A&A...566A..35S}
{Santos}, N.~C., {Mortier}, A., {Faria}, J.~P., {et~al.} 2014, \aap, 566, A35

\bibitem[{{Shapiro} {et~al.}(2020){Shapiro}, {Amazo-G{\'o}mez}, {Krivova}, \&
  {Solanki}}]{2020A&A...633A..32S}
{Shapiro}, A.~I., {Amazo-G{\'o}mez}, E.~M., {Krivova}, N.~A., \& {Solanki},
  S.~K. 2020, \aap, 633, A32

\bibitem[{{Shapiro} {et~al.}(2016){Shapiro}, {Solanki}, {Krivova}, {Yeo}, \&
  {Schmutz}}]{2016A&A...589A..46S}
{Shapiro}, A.~I., {Solanki}, S.~K., {Krivova}, N.~A., {Yeo}, K.~L., \&
  {Schmutz}, W.~K. 2016, \aap, 589, A46

\bibitem[{{Shapiro, A. I.} {et~al.}(2020){Shapiro, A. I.}, {Amazo-G\'omez, E.
  M.}, {Krivova, N. A.}, \& {Solanki, S. K.}}]{paperI}
{Shapiro, A. I.}, {Amazo-G\'omez, E. M.}, {Krivova, N. A.}, \& {Solanki, S. K.}
  2020, A\&A, 633, A32

\bibitem[{{Sousa}(2014)}]{Sousa-14}
{Sousa}, S.~G. 2014, [arXiv:1407.5817] [\eprint[arXiv]{1407.5817}]

\bibitem[{{Stassun} {et~al.}(2018){Stassun}, {Oelkers}, {Pepper}, {Paegert},
  {De Lee}, {Torres}, {Latham}, {Charpinet}, {Dressing}, {Huber}, {Kane},
  {L{\'e}pine}, {Mann}, {Muirhead}, {Rojas-Ayala}, {Silvotti}, {Fleming},
  {Levine}, \& {Plavchan}}]{2018AJ....156..102S}
{Stassun}, K.~G., {Oelkers}, R.~J., {Pepper}, J., {et~al.} 2018, \aj, 156, 102

\bibitem[{{Tayar} {et~al.}(2019){Tayar}, {Stassun}, \&
  {Corsaro}}]{2019ApJ...883..195T}
{Tayar}, J., {Stassun}, K.~G., \& {Corsaro}, E. 2019, \apj, 883, 195

\bibitem[{Van Der~Walt {et~al.}(2011)Van Der~Walt, Colbert, \&
  Varoquaux}]{van2011numpy}
Van Der~Walt, S., Colbert, S.~C., \& Varoquaux, G. 2011, Computing in Science
  \& Engineering, 13, 22

\bibitem[{{Vaughan} {et~al.}(1978){Vaughan}, {Preston}, \&
  {Wilson}}]{1978PASP...90..267V}
{Vaughan}, A.~H., {Preston}, G.~W., \& {Wilson}, O.~C. 1978, \pasp, 90, 267

\bibitem[{{Vida} {et~al.}(2019){Vida}, {Ol{\'a}h}, {K{\H{o}}v{\'a}ri}, {van
  Driel-Gesztelyi}, {Mo{\'o}r}, \& {P{\'a}l}}]{2019ApJ...884..160V}
{Vida}, K., {Ol{\'a}h}, K., {K{\H{o}}v{\'a}ri}, Z., {et~al.} 2019, \apj, 884,
  160

\bibitem[{{Wright}(2005)}]{2005PASP..117..657W}
{Wright}, J.~T. 2005, \pasp, 117, 657

\end{thebibliography}

\begin{appendix}

\section{Correlation and uncertainties estimation \label{mcmc}}

We compare the piecewise two power law function fit and single power law function fit in the peak-to-peak of photometric variability and RV jitter correlation plot. We used Markov-chain Monte Carlo (MCMC) sampling using the library
$emcee$ \citep{2013PASP..125..306F} to estimate the uncertainties of the parameters in both models.
\\
For the piecewise function we used
\begin{equation}
  \label{peac-eq}
log10 (y) = \left\{ \begin{array}{cc} 
                ~k_{1}log10(x) + y_{0} - k_{1}x_{0} & \hspace{5mm} log(x) < x_{0} \\
                ~k_{2}log10(x) + y_{0} - k_{2}x_{0} & \hspace{5mm} log(x) \geqslant x_{0} \\
                \end{array} \right.
\end{equation}
\\
which here {($10^{x_{0}}$,$10^{y_{0}}$)} is the knee point coordination in the plots. We also fit a single power law function fit as,
  \begin{equation}
  \label{lin-eq}
log10(y) = a~log10(x) + b.
\end{equation}

We derived $RMS_{resi}$ and BIC value in order to compare the significance of the two models. All best fitted parameters are presented in Table.~\ref{mcmc_table}.

\begin{table*}
\small
\centering
\begin{tabular}{ccccccccccc}
\hline \hline
\multirow{3}{*}{} & \multicolumn{6}{c}{Posterior}                                                                                                                                   & \multicolumn{2}{c}{$RMS_{resi}$}                     & \multicolumn{2}{c}{BIC}                              \\
                  & \multicolumn{2}{c}{Single}                          & \multicolumn{4}{c}{Piecewise}                                                                             & \multirow{2}{*}{Single} & \multirow{2}{*}{Piecewise} & \multirow{2}{*}{Single} & \multirow{2}{*}{Piecewise} \\
                  & a                        & b                        & $x_{0}$                  & $y_{0}$                  & $k_{1}$                  & $k_{2}$                 &                         &                            &                         &                            \\
\hline
year 1            & $0.825_{-0.399}^{0.417}$ & $0.038_{-0.225}^{0.233}$ & $0.610_{-0.788}^{0.573}$ & $0.338_{-0.470}^{0.664}$ & $0.392_{-0.458}^{0.635}$ & $1.450_{-0.833}^{1.396}$ & 0.30                    & 0.25                       & 8.6                     & 14.3                       \\
year 2            & $0.659_{-0.285}^{0.289}$ & $0.156_{-0.148}^{0.144}$ & $0.718_{-0.875}^{0.570}$ & $0.419_{-0.384}^{0.514}$ & $0.285_{-0.347}^{0.537}$ & $1.523_{-0.690}^{1.427}$ & 0.32                    & 0.28                       & 15.7                    & 22.7                       \\
year 3            & $0.374_{-0.222}^{0.249}$ & $0.185_{-0.126}^{0.120}$ & $0.733_{-1.031}^{0.527}$ & $0.343_{-0.262}^{0.367}$ & $0.159_{-0.252}^{0.420}$ & $1.004_{-0.647}^{1.554}$ & 0.29                    & 0.27                       & 18.1                    & 26.6                       \\
year 4            & $0.304_{-0.241}^{0.307}$ & $0.214_{-0.154}^{0.148}$ & $0.733_{-0.864}^{0.473}$ & $0.320_{-0.255}^{0.298}$ & $0.100_{-0.328}^{0.983}$ & $0.939_{-0.522}^{1.621}$ & 0.33                    & 0.32                       & 15.3                    & 23.1                       \\
Whole sample        & $0.636_{-0.183}^{0.179}$ & $0.238_{-0.108}^{0.107}$ & $0.813_{-0.792}^{0.484}$ & $0.548_{-0.366}^{0.437}$ & $0.273_{-0.268}^{0.306}$ & $1.381_{-0.735}^{1.305}$ & 0.35                    & 0.29                       & 28.3                    & 35.7                      
\end{tabular}
\caption{Maximum a Posteriori (MAP) with error for parameters which described in the Eq.~\ref{peac-eq} and Eq.~\ref{lin-eq} as well as the Bayesian information criterion (BIC) and RMS of the residual ($RMS_{resi}$) between the model and data for the two models: single power law function and piecewise power laws functions. Prior for b, $x_{0}$ and $y_{0}$ is a uniform prior with lower and upper limits (e.g., : for $x_{0}$: $\mathcal{U}(min(x),max(x))$) and for a, $k_{1}$ and $k_{2}$ is $\mathcal{U}(0,4)$.}

\label{mcmc_table}
\end{table*}

\section{Correlation with other observable \label{extra_p}}
 
\begin{figure}
   \centering
   \includegraphics[width=9.4cm]{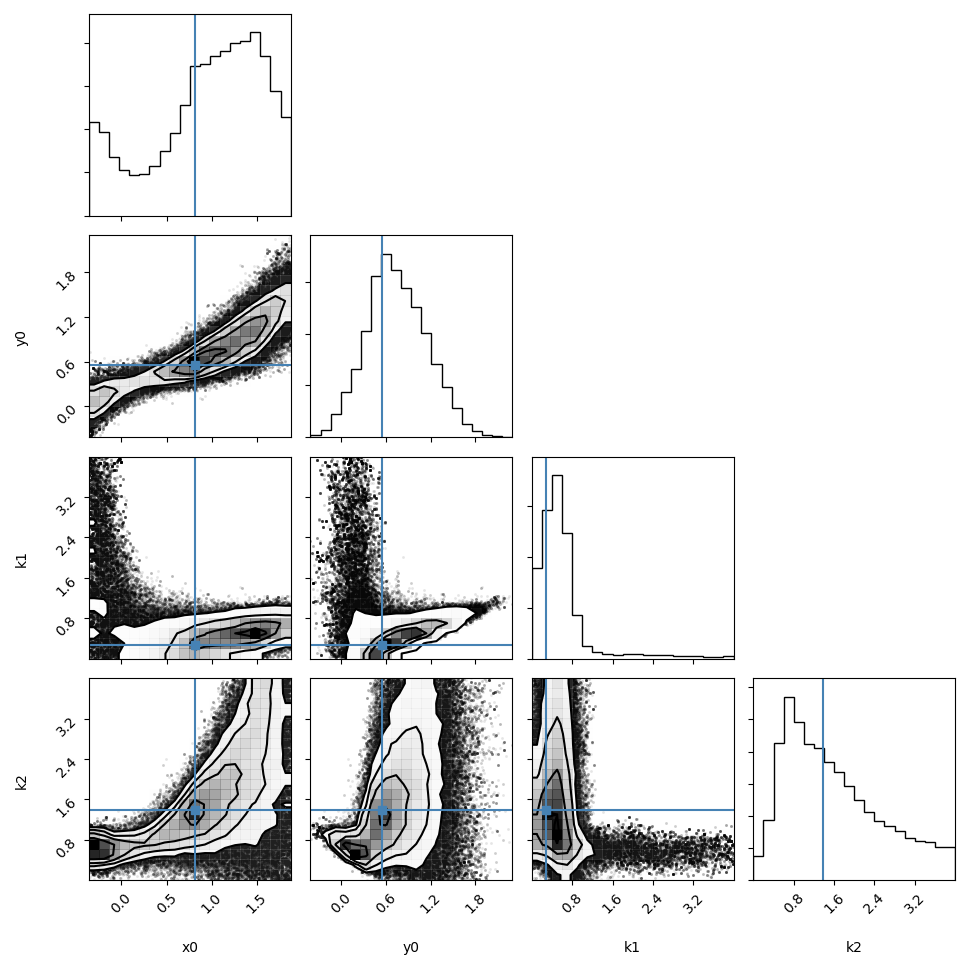}
      \caption{Posterior distribution for parameters $x_{0}$, $y_{0}$, $k_{1}$, $k_{2}$ in Eq.~\ref{peac-eq}}
         \label{mcmc_1}
   \end{figure}
 
\begin{figure}
   \centering
   \includegraphics[width=9.4cm]{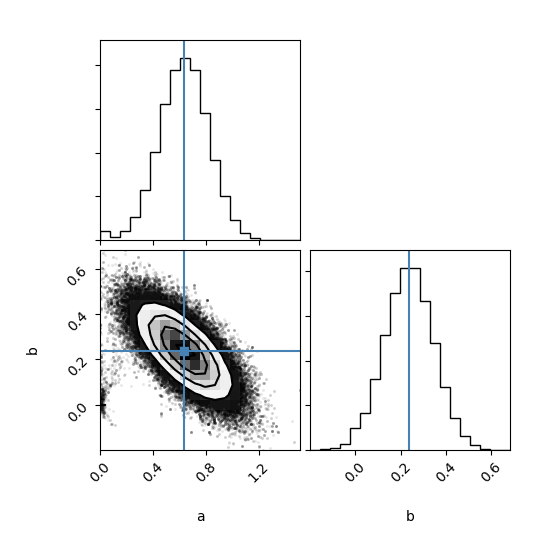}
      \caption{Posterior distribution for parameters a, b in Eq.~\ref{lin-eq}}
         \label{mcmc_lin}
   \end{figure}

   \begin{figure}
   \centering
  \includegraphics[width=9.5cm]{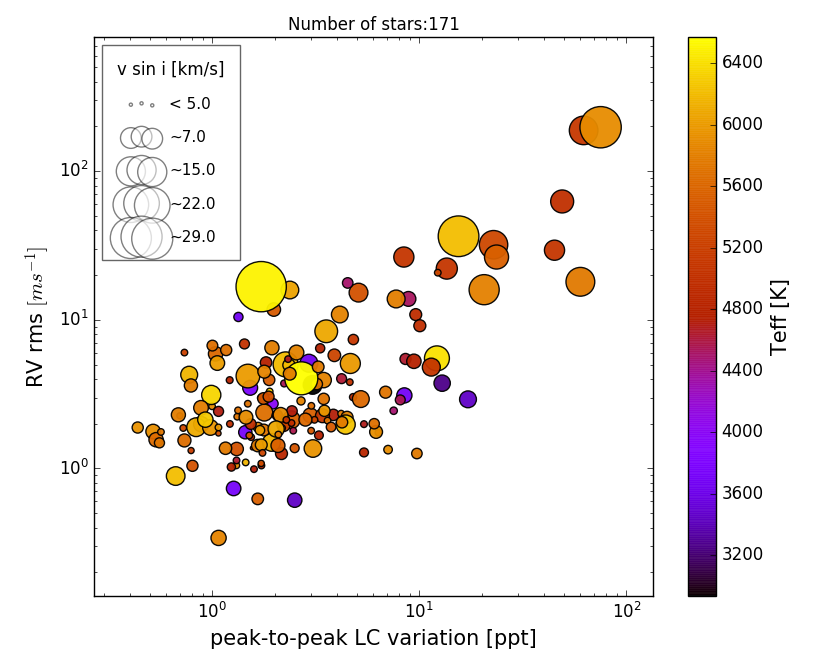}
      \caption{Same plots as Fig.~\ref{all_harps} with color bar \teff }
         \label{fwhm-night-teff}
   \end{figure}

    \begin{figure}
   \centering
  \includegraphics[width=9.5cm]{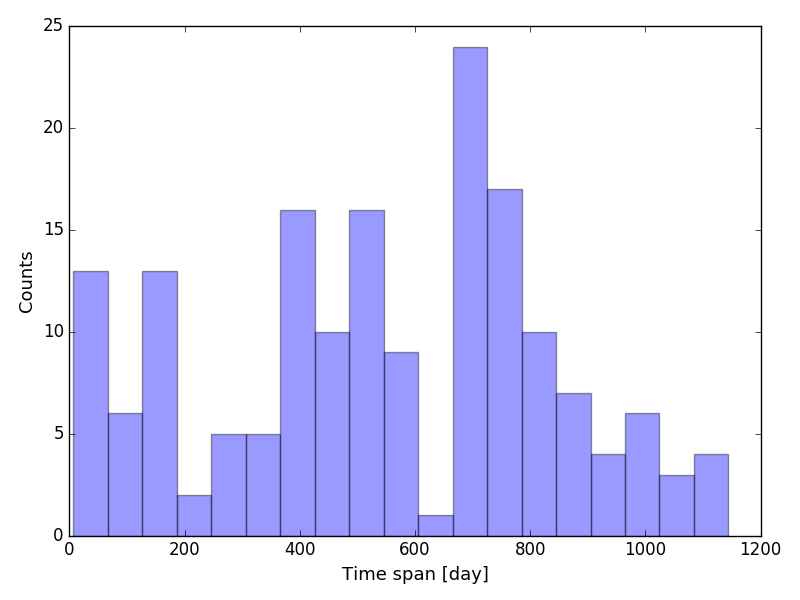}
      \caption{Time spans histogram of RV measurements in the sample.}
         \label{rv-span-hist}
   \end{figure}

 By having the effective temperature estimated, we investigated the correlation between \teff~and RV-RMS versus peak-to-peak photometric variability. In Fig.~\ref{fwhm-night-teff}, we used \teff~for the color bar. We cannot confirm a significant dependency of the plots on \teff.
\\
We also inspected if the correlation between RV and photometric jitter has any dependency on the $[Fe/H]$, and also the number of nights that the star has been observed in RV (see in Fig.~\ref{rv-span-hist}). We did not find significant correlation between these parameters.

 \section{Table: stellar parameters \label{all_param}}

\renewcommand{\thefootnote}{\fnsymbol{footnote}}
\longtab[1]{
 \tiny

\begin{longtable}{ccccccccccc}

\caption{ \label{stars_param_table} Derived parameters for the stars} \\
\hline\hline

Index & Star & \teff & RV-RMS \footnotemark[1] & pp Light curves & \vsini \footnotemark[2] & mean($log(R' _{HK})$) \footnotemark[3] & $\log g$  & $[Fe/H]$ & $P_{rot}$ & Sample \footnotemark[4] \\
 &  & [K] & [$m \;s^{-1}$] & [ppt] & [$km \;s^{-1}$] &  & [cm $s^{-2}$] &  & [day] & \\
 
 \hline
\endfirsthead
\caption{continued.}\\
\hline\hline
Index & Star & \teff & RV-RMS & pp Light curves & \vsini  & mean($log(R' _{HK})$)  & $\log g$  & $[Fe/H]$ & $P_{rot}$ & Sample \\
\hline
\endhead

\hline
\endfoot

1 & GJ1061 & 2934.0 $\pm$ 90 & 3.65 & 3.055 & 4.0 & -4.74 & -- & -0.08 $\pm$ 0.05 & -- & 1 \\
2 & GJ180 & 3662.0 $\pm$ 90 & 10.456 & 1.339 & 1.0 & -5.4 & -- & -0.24 $\pm$ 0.05 & -- & 4 \\
3 & GJ3135 & 3649.0 $\pm$ 90 & 5.105 & 2.931 & 3.7 & -4.95 & -- & -0.34 $\pm$ 0.05 & -- & 2,3,4 \\
4 & GJ3440 & 3693.0 $\pm$ 90 & 3.101 & 8.47 & 2.7 & -4.89 & -- & -- & 19.5 $\pm$ 1.2 & 3 \\
5 & GJ588 & 3459.0 $\pm$ 90 & 0.612 & 2.503 & 2.4 & -5.14 & -- & 0.06 $\pm$ 0.05 & -- & 3 \\
6 & Gl213 & 3266.0 $\pm$ 90 & 3.748 & 12.893 & 3.1 & -4.94 & -- & -0.16 $\pm$ 0.05 & -- & 3,4 \\
7 & Gl229 & 3785.0 $\pm$ 90 & 3.488 & 1.525 & 2.6 & -4.95 & -- & -0.02 $\pm$ 0.05 & -- & 3,4 \\
8 & Gl693 & 3457.0 $\pm$ 90 & 2.919 & 17.185 & 3.3 & -4.86 & -- & -0.275 $\pm$ 0.05 & -- & 3,4 \\
9 & Gl87 & 3726.0 $\pm$ 90 & 1.752 & 1.445 & 2.0 & -5.4 & -- & -0.35 $\pm$ 0.05 & -- & 3,4 \\
10 & HD 149189 & 3726.0 $\pm$ 90 & 0.733 & 1.269 & 2.4 & -5.09 & -- & 0.19 $\pm$ 0.05 & -- & 3 \\
11 & HD000105 & 6221.0 $\pm$ 94 & 36.513 & 15.471 & 19.1 & -4.36 & 4.71 $\pm$ 0.09 & -0.13 $\pm$ 0.068 & 5.3 $\pm$ 1.9 & w \\
12 & HD007661 & 5410.0 $\pm$ 30 & 15.244 & 5.088 & 4.1 & -4.4 & 4.38 $\pm$ 0.05 & -0.01 $\pm$ 0.022 & 6.3 $\pm$ 1.1 & w \\
13 & HD017925 & 5086.0 $\pm$ 41 & 29.416 & 44.961 & 4.7 & -4.32 & 4.29 $\pm$ 0.09 & 0.075 $\pm$ 0.023 & 7.2 $\pm$ 2.6 & 2 \\
14 & HD018599 & 5077.0 $\pm$ 38 & 26.463 & 8.423 & 4.7 & -4.42 & 4.34 $\pm$ 0.08 & -0.01 $\pm$ 0.022 & 5.4 $\pm$ 0.7 & 3 \\
15 & HD021175 & 5263.0 $\pm$ 39 & 5.773 & 3.889 & 1.8 & -4.78 & 4.33 $\pm$ 0.07 & 0.12 $\pm$ 0.026 & 27.1 $\pm$ 2.3 & w \\
16 & HD024916 & 4644.0 $\pm$ 97 & 5.454 & 8.58 & 1.4 & -4.61 & 4.14 $\pm$ 0.36 & -0.097 $\pm$ 0.042 & 16.0 $\pm$ 3.9 & w \\
17 & HD037572 & 5314.0 $\pm$ 36 & 32.029 & 22.83 & 9.3 & -4.59 & 4.4 $\pm$ 0.07 & -0.344 $\pm$ 0.023 & 5.6 $\pm$ 3.0 & w \\
18 & HD092945 & 5091.0 $\pm$ 38 & 22.13 & 13.576 & 5.1 & -4.35 & 4.39 $\pm$ 0.08 & -0.04 $\pm$ 0.022 & 6.2 $\pm$ 11.1 & 3 \\
19 & HD102438 & 5547.0 $\pm$ 12 & 2.277 & 2.98 & 2.5 & -4.97 & 4.42 $\pm$ 0.03 & -0.29 $\pm$ 0.01 & -- & 3 \\
20 & HD10360 & 5026.0 $\pm$ 31 & 3.808 & 4.619 & 0.5 & -4.8 & 4.37 $\pm$ 0.07 & -0.205 $\pm$ 0.017 & -- & 3 \\
21 & HD103743 & 5763.0 $\pm$ 33 & 18.031 & 59.935 & 9.5 & -4.37 & 4.55 $\pm$ 0.04 & -0.02 $\pm$ 0.025 & -- & 3 \\
22 & HD103891 & 6053.0 $\pm$ 20 & 15.874 & 2.373 & 3.6 & -4.9 & 3.99 $\pm$ 0.03 & -0.2 $\pm$ 0.016 & -- & 24 \\
23 & HD107094 & 5522.0 $\pm$ 27 & 1.742 & 6.351 & 0.5 & -4.85 & 4.43 $\pm$ 0.03 & -0.51 $\pm$ 0.021 & -- & w \\
24 & HD109200 & 5040.0 $\pm$ 30 & 1.373 & 1.562 & 0.2 & -4.98 & 4.27 $\pm$ 0.07 & -0.285 $\pm$ 0.018 & 17.4 $\pm$ 0.7 & 3,4 \\
25 & HD111777 & 5622.0 $\pm$ 22 & 1.446 & 1.649 & 0.5 & -4.93 & 4.37 $\pm$ 0.04 & -0.7 $\pm$ 0.018 & -- & 3,4 \\
26 & HD114853 & 5668.0 $\pm$ 13 & 2.262 & 0.698 & 0.4 & -4.96 & 4.36 $\pm$ 0.02 & -0.23 $\pm$ 0.01 & -- & 2,3,4 \\
27 & HD11507 & 3991.0 $\pm$ 90 & 2.725 & 1.965 & 1.3 & -4.8 & -- & -0.21 $\pm$ 0.05 & -- & w \\
28 & HD11608 & 4864.0 $\pm$ 106 & 1.259 & 2.161 & 1.6 & -4.91 & 4.08 $\pm$ 0.23 & 0.22 $\pm$ 0.057 & 8.8 $\pm$ 2.1 & 1,2,3 \\
29 & HD118466 & 4967.0 $\pm$ 72 & 10.849 & 9.613 & 1.6 & -5.06 & 4.1 $\pm$ 0.19 & 0.2 $\pm$ 0.038 & -- & 4 \\
30 & HD118563 & 5502.0 $\pm$ 22 & 11.73 & 1.984 & 2.1 & -4.79 & 4.39 $\pm$ 0.05 & -0.04 $\pm$ 0.017 & -- & 4 \\
31 & HD121004 & 5672.0 $\pm$ 27 & 2.639 & 3.012 & 0.5 & -4.93 & 4.39 $\pm$ 0.03 & -0.71 $\pm$ 0.021 & -- & 4 \\
32 & HD123651 & 5877.0 $\pm$ 27 & 1.048 & 1.308 & 0.5 & -4.86 & 4.4 $\pm$ 0.03 & -0.48 $\pm$ 0.022 & 14.7 $\pm$ 4.9 & w \\
33 & HD125072 & 4731.0 $\pm$ 126 & 2.293 & 3.845 & 1.5 & -4.97 & 3.95 $\pm$ 0.3 & 0.23 $\pm$ 0.058 & 13.6 $\pm$ 2.8 & 3,4 \\
34 & HD126803 & 5455.0 $\pm$ 19 & 1.794 & 3.005 & 0.5 & -4.93 & 4.42 $\pm$ 0.04 & -0.63 $\pm$ 0.016 & -- & 2,3,4 \\
35 & HD128571 & 6156.0 $\pm$ 32 & 1.098 & 1.451 & 0.5 & -4.93 & 4.26 $\pm$ 0.04 & -0.37 $\pm$ 0.021 & 36.3 $\pm$ 6.5 & 4 \\
36 & HD129642 & 4851.0 $\pm$ 46 & 1.282 & 5.408 & 0.9 & -5.08 & 4.18 $\pm$ 0.14 & -0.09 $\pm$ 0.026 & -- & 2,3,4 \\
37 & HD133633 & 5557.0 $\pm$ 24 & 2.128 & 2.409 & 3.6 & -4.95 & 4.38 $\pm$ 0.03 & -0.45 $\pm$ 0.019 & -- & 3,4 \\
38 & HD134606 & 5619.0 $\pm$ 30 & 5.906 & 1.039 & 2.4 & -5.11 & 4.28 $\pm$ 0.07 & 0.27 $\pm$ 0.023 & 11.6 $\pm$ 2.0 & w \\
39 & HD137676 & 5249.0 $\pm$ 16 & 1.961 & 2.212 & 2.0 & -5.1 & 3.88 $\pm$ 0.04 & -0.53 $\pm$ 0.013 & -- & 3,4 \\
40 & HD13808 & 5034.0 $\pm$ 36 & 6.884 & 1.432 & 1.1 & -4.87 & 4.34 $\pm$ 0.08 & -0.2 $\pm$ 0.02 & -- & 3 \\
41 & HD1388 & 5920.0 $\pm$ 13 & 1.772 & 0.519 & 2.4 & -4.99 & 4.33 $\pm$ 0.02 & -0.01 $\pm$ 0.011 & 18.3 $\pm$ 2.6 & 2,3,4 \\
42 & HD141624 & 5887.0 $\pm$ 18 & 2.14 & 2.27 & 0.5 & -4.94 & 4.4 $\pm$ 0.04 & -0.38 $\pm$ 0.014 & -- & 3,4 \\
43 & HD142709 & 4573.0 $\pm$ 132 & 1.795 & 2.463 & 0.5 & -4.98 & 4.44 $\pm$ 0.34 & -0.32 $\pm$ 0.031 & -- & 4 \\
44 & HD14412 & 5344.0 $\pm$ 21 & 1.871 & 0.725 & 0.5 & -4.85 & 4.42 $\pm$ 0.05 & -0.47 $\pm$ 0.017 & 13.0 $\pm$ 0.3 & 1,2,3 \\
45 & HD144628 & 4979.0 $\pm$ 42 & 2.021 & 2.416 & 0.5 & -4.89 & 4.34 $\pm$ 0.08 & -0.36 $\pm$ 0.024 & -- & 3,4 \\
46 & HD14745 & 6182.0 $\pm$ 27 & 4.274 & 0.774 & 3.3 & -5.0 & 4.42 $\pm$ 0.02 & -0.14 $\pm$ 0.019 & -- & 2,3 \\
47 & HD150139 & 5904.0 $\pm$ 26 & 2.62 & 0.999 & 0.5 & -4.91 & 4.16 $\pm$ 0.03 & -0.51 $\pm$ 0.02 & -- & 4 \\
48 & HD150474 & 5423.0 $\pm$ 18 & 1.352 & 1.313 & 2.1 & -5.15 & 3.93 $\pm$ 0.03 & 0.01 $\pm$ 0.015 & -- & 1,2 \\
49 & HD154577 & 4868.0 $\pm$ 36 & 1.623 & 1.543 & 0.5 & -4.87 & 4.44 $\pm$ 0.09 & -0.65 $\pm$ 0.018 & -- & 3,4 \\
50 & HD156098 & 6418.0 $\pm$ 37 & 5.506 & 12.165 & 7.2 & -4.82 & 3.86 $\pm$ 0.05 & 0.12 $\pm$ 0.026 & -- & 4 \\
51 & HD1581 & 5946.0 $\pm$ 17 & 1.885 & 0.437 & 1.4 & -4.95 & 4.44 $\pm$ 0.03 & -0.2 $\pm$ 0.013 & -- & 2,3,4 \\
52 & HD16280 & 4625.0 $\pm$ 168 & 4.02 & 4.216 & 1.1 & -4.82 & 4.22 $\pm$ 0.41 & -0.19 $\pm$ 0.103 & 14.8 $\pm$ 3.9 & 13 \\
53 & HD16548 & 5685.0 $\pm$ 15 & 2.377 & 1.785 & 3.3 & -5.16 & 3.95 $\pm$ 0.04 & 0.15 $\pm$ 0.012 & -- & w \\
54 & HD16905 & 4760.0 $\pm$ 104 & 5.159 & 1.821 & 1.5 & -4.77 & 4.14 $\pm$ 0.27 & 0.15 $\pm$ 0.06 & 23.8 $\pm$ 0.2 & 2 \\
55 & HD172568 & 5690.0 $\pm$ 18 & 1.865 & 2.176 & 0.5 & -4.93 & 4.41 $\pm$ 0.03 & -0.37 $\pm$ 0.014 & -- & 3,4 \\
56 & HD17970 & 5005.0 $\pm$ 30 & 1.994 & 1.218 & 0.5 & -5.0 & 4.34 $\pm$ 0.08 & -0.45 $\pm$ 0.019 & -- & 2,3 \\
57 & HD181327 & 6502.0 $\pm$ 90 & 16.722 & 1.724 & 29.1 & -4.57 & -- & 0.055 $\pm$ 0.05 & 2.3 $\pm$ 4.6 & 4 \\
58 & HD183414 & 5816.0 $\pm$ 33 & 15.956 & 20.565 & 10.4 & -4.34 & 4.51 $\pm$ 0.04 & -0.165 $\pm$ 0.026 & 5.6 $\pm$ 1.0 & 3 \\
59 & HD183783 & 4796.0 $\pm$ 77 & 3.93 & 1.215 & 0.5 & -4.88 & 4.14 $\pm$ 0.47 & -0.21 $\pm$ 0.031 & 5.4 $\pm$ 2.2 & 4 \\
60 & HD185283 & 4754.0 $\pm$ 65 & 1.671 & 3.277 & 0.9 & -5.01 & 4.24 $\pm$ 0.16 & -0.06 $\pm$ 0.03 & 28.9 $\pm$ 0.6 & 3,4 \\
61 & HD187456 & 4715.0 $\pm$ 90 & 2.42 & 1.073 & 1.2 & -4.93 & 4.09 $\pm$ 0.24 & 0.02 $\pm$ 0.054 & 12.1 $\pm$ 1.6 & 4 \\
62 & HD189567 & 5710.0 $\pm$ 12 & 2.726 & 1.486 & 0.5 & -4.93 & 4.37 $\pm$ 0.02 & -0.24 $\pm$ 0.01 & -- & w \\
63 & HD190248 & 5653.0 $\pm$ 77 & 1.561 & 0.536 & 2.4 & -5.12 & 4.6 $\pm$ 0.13 & 0.33 $\pm$ 0.056 & 21.4 $\pm$ 9.3 & 3,4 \\
64 & HD19230 & 5118.0 $\pm$ 31 & 7.372 & 4.805 & 1.2 & -4.79 & 4.37 $\pm$ 0.06 & -0.57 $\pm$ 0.021 & 13.5 $\pm$ 4.1 & 1 \\
65 & HD19641 & 5811.0 $\pm$ 12 & 1.426 & 1.659 & 1.8 & -5.01 & 4.38 $\pm$ 0.03 & -0.01 $\pm$ 0.01 & -- & 2 \\
66 & HD197481 & 5129.0 $\pm$ 262 & 188.286 & 62.137 & 9.4 & -4.04 & 4.36 $\pm$ 0.5 & 0.1 $\pm$ 0.172 & 6.0 $\pm$ 1.1 & 1 \\
67 & HD199288 & 5756.0 $\pm$ 17 & 1.76 & 0.566 & 0.5 & -4.92 & 4.47 $\pm$ 0.03 & -0.63 $\pm$ 0.014 & 7.5 $\pm$ 1.3 & 1,2,3,4 \\
68 & HD199604 & 5785.0 $\pm$ 34 & 2.338 & 4.192 & 0.5 & -4.94 & 4.27 $\pm$ 0.03 & -0.62 $\pm$ 0.027 & -- & 1,2,3 \\
69 & HD199847 & 5722.0 $\pm$ 27 & 2.23 & 1.321 & 0.5 & -5.02 & 4.14 $\pm$ 0.03 & -0.54 $\pm$ 0.021 & -- & 3,4 \\
70 & HD199981 & 4559.0 $\pm$ 149 & 2.893 & 8.093 & 1.1 & -4.68 & 4.3 $\pm$ 0.55 & -0.16 $\pm$ 0.05 & 12.6 $\pm$ 2.2 & 3 \\
71 & HD200133 & 6142.0 $\pm$ 25 & 1.498 & 1.935 & 3.7 & -5.05 & 4.38 $\pm$ 0.04 & 0.3 $\pm$ 0.018 & -- & 3 \\
72 & HD202917 & 5902.0 $\pm$ 109 & 197.725 & 75.061 & 19.7 & -4.09 & 5.0 $\pm$ 0.11 & -0.095 $\pm$ 0.076 & 3.6 $\pm$ 1.0 & w \\
73 & HD203850 & 4793.0 $\pm$ 52 & 1.046 & 1.729 & 0.5 & -4.79 & 4.43 $\pm$ 0.12 & -0.68 $\pm$ 0.021 & -- & 3 \\
74 & HD20407 & 5835.0 $\pm$ 16 & 2.137 & 0.881 & 0.5 & -4.9 & 4.46 $\pm$ 0.02 & -0.44 $\pm$ 0.013 & -- & 2,3 \\
75 & HD205536 & 5426.0 $\pm$ 20 & 1.042 & 0.803 & 1.4 & -5.03 & 4.31 $\pm$ 0.04 & -0.05 $\pm$ 0.015 & 3.2 $\pm$ 0.9 & 4 \\
76 & HD206683 & 5945.0 $\pm$ 21 & 1.362 & 3.065 & 3.5 & -5.05 & 4.35 $\pm$ 0.06 & 0.24 $\pm$ 0.017 & -- & 3 \\
77 & HD206998 & 5775.0 $\pm$ 28 & 2.133 & 2.099 & 0.5 & -5.01 & 4.14 $\pm$ 0.03 & -0.72 $\pm$ 0.021 & -- & 1,2,3 \\
78 & HD207129 & 5919.0 $\pm$ 13 & 1.761 & 6.185 & 1.9 & -4.91 & 4.43 $\pm$ 0.02 & -0.02 $\pm$ 0.01 & -- & 3 \\
79 & HD207869 & 5480.0 $\pm$ 18 & 1.668 & 1.511 & 0.5 & -4.96 & 4.41 $\pm$ 0.03 & -0.45 $\pm$ 0.014 & -- & 24 \\
80 & HD20807 & 5832.0 $\pm$ 15 & 1.896 & 1.666 & 1.1 & -4.9 & 4.49 $\pm$ 0.03 & -0.23 $\pm$ 0.012 & -- & 2,3 \\
81 & HD209100 & 4776.0 $\pm$ 154 & 2.287 & 2.061 & 1.2 & -4.77 & 4.59 $\pm$ 0.39 & -0.105 $\pm$ 0.057 & -- & 3 \\
82 & HD210918 & 5743.0 $\pm$ 10 & 1.487 & 0.557 & 1.2 & -5.02 & 4.3 $\pm$ 0.02 & -0.1 $\pm$ 0.009 & -- & 1,2,3,4 \\
83 & HD211415 & 5839.0 $\pm$ 14 & 2.209 & 4.479 & 1.7 & -4.97 & 4.42 $\pm$ 0.03 & -0.21 $\pm$ 0.011 & -- & 3,4 \\
84 & HD21161 & 5916.0 $\pm$ 13 & 1.985 & 0.886 & 2.7 & -5.06 & 4.26 $\pm$ 0.03 & 0.09 $\pm$ 0.01 & -- & 1 \\
85 & HD212036 & 5683.0 $\pm$ 19 & 6.71 & 1.003 & 1.3 & -5.05 & 4.34 $\pm$ 0.03 & -0.01 $\pm$ 0.014 & 2.5 $\pm$ 1.4 & 2 \\
86 & HD21209A & 4682.0 $\pm$ 46 & 1.987 & 5.407 & 0.5 & -4.84 & 4.5 $\pm$ 0.14 & -0.41 $\pm$ 0.064 & 38.8 $\pm$ 6.9 & 3 \\
87 & HD213042 & 4709.0 $\pm$ 125 & 2.434 & 2.428 & 1.3 & -4.88 & 4.15 $\pm$ 0.4 & 0.165 $\pm$ 0.057 & -- & 3 \\
88 & HD213628 & 5553.0 $\pm$ 22 & 0.624 & 1.66 & 1.5 & -5.0 & 4.38 $\pm$ 0.04 & 0.01 $\pm$ 0.017 & -- & 4 \\
89 & HD215456 & 5797.0 $\pm$ 11 & 3.617 & 0.789 & 2.0 & -5.09 & 4.07 $\pm$ 0.02 & -0.09 $\pm$ 0.009 & -- & 2,3 \\
90 & HD215906 & 6219.0 $\pm$ 29 & 0.889 & 0.666 & 4.0 & -4.91 & 4.37 $\pm$ 0.03 & -0.28 $\pm$ 0.02 & 2.3 $\pm$ 0.6 & 3 \\
91 & HD216054 & 5380.0 $\pm$ 20 & 1.081 & 1.725 & 0.4 & -5.01 & 4.37 $\pm$ 0.04 & -0.12 $\pm$ 0.015 & -- & 1,2 \\
92 & HD218860 & 5557.0 $\pm$ 29 & 26.463 & 23.599 & 6.7 & -4.31 & 4.43 $\pm$ 0.05 & -0.015 $\pm$ 0.022 & 5.0 $\pm$ 2.7 & 4 \\
93 & HD220339 & 4918.0 $\pm$ 40 & 3.02 & 4.789 & 0.6 & -4.84 & 4.31 $\pm$ 0.08 & -0.31 $\pm$ 0.02 & -- & 3 \\
94 & HD220507 & 5672.0 $\pm$ 13 & 1.543 & 0.735 & 2.0 & -5.07 & 4.22 $\pm$ 0.05 & 0.01 $\pm$ 0.011 & -- & 1,2,3 \\
95 & HD221575 & 5065.0 $\pm$ 41 & 4.806 & 11.43 & 3.7 & -4.49 & 4.4 $\pm$ 0.09 & 0.01 $\pm$ 0.022 & 11.0 $\pm$ 3.9 & w \\
96 & HD221638 & 6349.0 $\pm$ 33 & 3.124 & 0.99 & 4.2 & -4.91 & 4.45 $\pm$ 0.04 & -0.225 $\pm$ 0.023 & -- & 1,2 \\
97 & HD222237 & 4688.0 $\pm$ 130 & 2.985 & 1.716 & 0.5 & -4.91 & 4.17 $\pm$ 0.27 & -0.31 $\pm$ 0.188 & -- & 3 \\
98 & HD222335 & 5237.0 $\pm$ 32 & 6.024 & 0.735 & 0.5 & -4.8 & 4.4 $\pm$ 0.07 & -0.16 $\pm$ 0.022 & -- & 1,2,3 \\
99 & HD222480 & 5840.0 $\pm$ 15 & 3.932 & 3.453 & 2.7 & -5.04 & 4.15 $\pm$ 0.04 & 0.19 $\pm$ 0.012 & -- & 1,2,3 \\
100 & HD222669 & 5896.0 $\pm$ 15 & 2.294 & 2.139 & 2.5 & -4.9 & 4.45 $\pm$ 0.04 & 0.05 $\pm$ 0.012 & -- & 2 \\
101 & HD223171 & 5797.0 $\pm$ 14 & 2.559 & 0.883 & 2.5 & -5.04 & 4.09 $\pm$ 0.04 & 0.115 $\pm$ 0.011 & 13.8 $\pm$ 5.9 & 1,2,3 \\
102 & HD223681 & 4995.0 $\pm$ 55 & 62.608 & 48.955 & 6.1 & -4.05 & 4.39 $\pm$ 0.11 & -0.555 $\pm$ 0.025 & 5.6 $\pm$ 1.0 & 2 \\
103 & HD224228 & 4771.0 $\pm$ 59 & 5.262 & 9.418 & 2.4 & -4.44 & 4.19 $\pm$ 0.15 & 0.01 $\pm$ 0.028 & 27.2 $\pm$ 3.9 & w \\
104 & HD224230 & 4786.0 $\pm$ 79 & 6.436 & 3.322 & 1.0 & -4.99 & 4.25 $\pm$ 0.19 & -0.1 $\pm$ 0.037 & -- & 2 \\
105 & HD224619 & 5454.0 $\pm$ 19 & 1.369 & 2.502 & 0.9 & -4.97 & 4.4 $\pm$ 0.04 & -0.2 $\pm$ 0.015 & 15.8 $\pm$ 3.9 & 3 \\
106 & HD23249 & 5120.0 $\pm$ 128 & 2.264 & 3.408 & 2.2 & -5.21 & 4.3 $\pm$ 0.24 & 0.09 $\pm$ 0.074 & -- & 2,3 \\
107 & HD24062 & 5990.0 $\pm$ 19 & 5.086 & 4.646 & 4.6 & -5.06 & 4.22 $\pm$ 0.03 & 0.23 $\pm$ 0.015 & -- & 3 \\
108 & HD25587 & 6207.0 $\pm$ 29 & 5.015 & 2.262 & 7.2 & -5.0 & 4.26 $\pm$ 0.03 & -0.12 $\pm$ 0.022 & 23.1 $\pm$ 4.1 & 2 \\
109 & HD28471 & 5730.0 $\pm$ 13 & 1.814 & 1.775 & 1.6 & -5.0 & 4.32 $\pm$ 0.02 & -0.05 $\pm$ 0.011 & 18.7 $\pm$ 7.2 & 2 \\
110 & HD30306 & 5552.0 $\pm$ 25 & 1.427 & 2.078 & 2.2 & -5.09 & 4.26 $\pm$ 0.05 & 0.18 $\pm$ 0.019 & 44.8 $\pm$ 2.3 & 2 \\
111 & HD3074A & 6100.0 $\pm$ 17 & 1.895 & 0.837 & 4.0 & -5.03 & 4.39 $\pm$ 0.04 & -0.01 $\pm$ 0.013 & 12.2 $\pm$ 0.7 & 2,3 \\
112 & HD30876 & 4985.0 $\pm$ 47 & 9.128 & 10.06 & 1.7 & -4.52 & 4.36 $\pm$ 0.1 & -0.085 $\pm$ 0.024 & 9.6 $\pm$ 0.8 & 2 \\
113 & HD31822 & 6016.0 $\pm$ 19 & 5.119 & 1.059 & 2.5 & -4.86 & 4.5 $\pm$ 0.03 & -0.19 $\pm$ 0.014 & 3.6 $\pm$ 0.6 & 3,4 \\
114 & HD32564 & 5518.0 $\pm$ 16 & 3.944 & 1.884 & 1.5 & -5.02 & 4.36 $\pm$ 0.03 & 0.08 $\pm$ 0.012 & -- & 3 \\
115 & HD35854 & 4788.0 $\pm$ 61 & 1.023 & 1.238 & 0.8 & -4.82 & 4.2 $\pm$ 0.17 & -0.13 $\pm$ 0.03 & -- & 3,4 \\
116 & HD36003 & 4558.0 $\pm$ 278 & 3.732 & 2.226 & 0.6 & -4.89 & 4.13 $\pm$ 0.76 & -0.13 $\pm$ 0.038 & 2.4 $\pm$ 0.7 & 3,4 \\
117 & HD36379 & 6026.0 $\pm$ 15 & 1.441 & 1.726 & 1.6 & -5.0 & 4.26 $\pm$ 0.03 & -0.17 $\pm$ 0.011 & 2.8 $\pm$ 1.0 & 2,3 \\
118 & HD37990 & 6201.0 $\pm$ 24 & 4.961 & 2.348 & 1.7 & -4.75 & 4.48 $\pm$ 0.04 & -0.03 $\pm$ 0.017 & 12.3 $\pm$ 2.2 & 3 \\
119 & HD3823 & 6008.0 $\pm$ 17 & 1.807 & 1.701 & 1.0 & -5.01 & 4.28 $\pm$ 0.04 & -0.28 $\pm$ 0.013 & -- & 2,3 \\
120 & HD38459 & 5320.0 $\pm$ 38 & 20.766 & 12.281 & 0.5 & -4.51 & 4.28 $\pm$ 0.1 & 0.11 $\pm$ 0.026 & 12.0 $\pm$ 2.2 & 2 \\
121 & HD38858 & 5705.0 $\pm$ 13 & 2.098 & 3.607 & 0.5 & -4.94 & 4.46 $\pm$ 0.03 & -0.23 $\pm$ 0.01 & -- & 3 \\
122 & HD39194 & 5188.0 $\pm$ 17 & 2.935 & 1.768 & 0.5 & -4.96 & 4.47 $\pm$ 0.04 & -0.61 $\pm$ 0.012 & 20.4 $\pm$ 6.5 & 3 \\
123 & HD3964 & 5723.0 $\pm$ 15 & 3.26 & 6.874 & 1.7 & -4.9 & 4.43 $\pm$ 0.02 & 0.05 $\pm$ 0.012 & -- & 1,2 \\
124 & HD40397 & 5504.0 $\pm$ 18 & 1.899 & 3.75 & 1.1 & -5.02 & 4.35 $\pm$ 0.04 & -0.13 $\pm$ 0.014 & 2.7 $\pm$ 0.5 & 3,4 \\
125 & HD40865 & 5678.0 $\pm$ 18 & 2.46 & 1.333 & 0.5 & -4.94 & 4.39 $\pm$ 0.03 & -0.38 $\pm$ 0.015 & -- & 2,3 \\
126 & HD42936 & 5075.0 $\pm$ 60 & 2.123 & 2.281 & 0.5 & -5.1 & 4.18 $\pm$ 0.16 & 0.18 $\pm$ 0.039 & -- & 2 \\
127 & HD52449 & 6342.0 $\pm$ 23 & 3.295 & 1.895 & 0.5 & -4.74 & 4.49 $\pm$ 0.03 & 0.09 $\pm$ 0.017 & 27.7 $\pm$ 5.7 & 1,2 \\
128 & HD53705 & 5765.0 $\pm$ 29 & 1.26 & 9.736 & 1.3 & -5.01 & 4.35 $\pm$ 0.05 & -0.227 $\pm$ 0.023 & -- & 4 \\
129 & HD56380 & 5310.0 $\pm$ 25 & 2.124 & 3.119 & 0.5 & -5.0 & 4.3 $\pm$ 0.04 & -0.42 $\pm$ 0.018 & -- & 2,3 \\
130 & HD6107 & 5797.0 $\pm$ 13 & 2.296 & 0.687 & 2.3 & -5.06 & 4.04 $\pm$ 0.02 & -0.055 $\pm$ 0.011 & 13.6 $\pm$ 5.8 & 1,2,3 \\
131 & HD65277 & 4646.0 $\pm$ 69 & 1.132 & 1.311 & 0.5 & -5.04 & 4.33 $\pm$ 0.17 & -0.31 $\pm$ 0.03 & 25.7 $\pm$ 4.6 & 3 \\
132 & HD65907A & 5911.0 $\pm$ 80 & 1.338 & 7.067 & 0.8 & -4.94 & 4.57 $\pm$ 0.09 & -0.31 $\pm$ 0.061 & 13.8 $\pm$ 2.2 & 3 \\
133 & HD67200 & 6090.0 $\pm$ 18 & 1.843 & 2.036 & 3.0 & -5.06 & 4.36 $\pm$ 0.04 & 0.32 $\pm$ 0.014 & 34.1 $\pm$ 6.1 & 2 \\
134 & HD69611 & 5776.0 $\pm$ 22 & 1.69 & 2.087 & 0.5 & -4.98 & 4.3 $\pm$ 0.03 & -0.58 $\pm$ 0.017 & -- & 2,3 \\
135 & HD71835 & 5475.0 $\pm$ 24 & 2.957 & 1.768 & 1.5 & -4.96 & 4.41 $\pm$ 0.04 & -0.04 $\pm$ 0.017 & -- & 2,3 \\
136 & HD72673 & 5201.0 $\pm$ 21 & 1.319 & 0.791 & 0.4 & -4.94 & 4.42 $\pm$ 0.04 & -0.38 $\pm$ 0.015 & -- & 2,3,4 \\
137 & HD73524 & 5979.0 $\pm$ 17 & 1.886 & 0.981 & 2.8 & -5.03 & 4.36 $\pm$ 0.03 & 0.15 $\pm$ 0.013 & -- & 2,3,4 \\
138 & HD74698 & 5797.0 $\pm$ 13 & 6.49 & 1.943 & 2.3 & -5.07 & 4.23 $\pm$ 0.03 & 0.07 $\pm$ 0.01 & -- & 1,24 \\
139 & HD75881 & 6094.0 $\pm$ 28 & 8.375 & 3.554 & 5.9 & -5.15 & 4.06 $\pm$ 0.04 & 0.07 $\pm$ 0.022 & -- & w \\
140 & HD76849 & 5285.0 $\pm$ 36 & 1.272 & 1.75 & 0.5 & -5.04 & 4.34 $\pm$ 0.09 & -- & -- & 4 \\
141 & HD78429 & 5738.0 $\pm$ 12 & 3.698 & 3.201 & 1.5 & -4.93 & 4.25 $\pm$ 0.03 & 0.089 $\pm$ 0.01 & 34.4 $\pm$ 4.2 & 2,3 \\
142 & HD82342 & 4728.0 $\pm$ 607 & 0.989 & 1.592 & 0.5 & -5.0 & 4.23 $\pm$ 1.21 & -0.54 $\pm$ 1.145 & 30.8 $\pm$ 4.9 & 3 \\
143 & HD82516 & 4998.0 $\pm$ 43 & 1.989 & 1.536 & 1.3 & -5.01 & 4.24 $\pm$ 0.12 & 0.01 $\pm$ 0.023 & 10.6 $\pm$ 1.9 & 2,3,4 \\
144 & HD85249 & 6567.0 $\pm$ 51 & 4.04 & 2.693 & 12.6 & -4.92 & 4.63 $\pm$ 0.07 & 0.18 $\pm$ 0.035 & -- & 3 \\
145 & HD85725 & 6000.0 $\pm$ 26 & 4.187 & 1.488 & 6.5 & -5.19 & 4.0 $\pm$ 0.05 & 0.15 $\pm$ 0.02 & -- & w \\
146 & HD87838 & 5993.0 $\pm$ 33 & 1.888 & 1.072 & 0.5 & -4.94 & 4.24 $\pm$ 0.03 & -0.4 $\pm$ 0.023 & 27.4 $\pm$ 4.9 & 2,3,4 \\
147 & HD88218 & 5845.0 $\pm$ 14 & 4.48 & 1.785 & 1.9 & -5.05 & 4.09 $\pm$ 0.02 & -0.14 $\pm$ 0.012 & -- & 3,4 \\
148 & HD8828 & 5386.0 $\pm$ 21 & 1.722 & 1.072 & 0.3 & -5.01 & 4.4 $\pm$ 0.03 & -0.16 $\pm$ 0.016 & -- & 2,3 \\
149 & HD89839 & 6245.0 $\pm$ 25 & 1.973 & 4.409 & 4.2 & -5.01 & 4.4 $\pm$ 0.04 & 0.04 $\pm$ 0.018 & -- & 2 \\
150 & HD92719 & 5794.0 $\pm$ 14 & 4.34 & 2.368 & 1.9 & -4.85 & 4.43 $\pm$ 0.02 & -0.11 $\pm$ 0.011 & 9.5 $\pm$ 1.7 & 4 \\
151 & HD93489 & 5910.0 $\pm$ 13 & 2.21 & 1.456 & 2.2 & -5.0 & 4.36 $\pm$ 0.02 & -0.02 $\pm$ 0.011 & -- & 2,3 \\
152 & HD94151 & 5625.0 $\pm$ 17 & 2.131 & 2.82 & 1.9 & -4.93 & 4.41 $\pm$ 0.03 & 0.04 $\pm$ 0.013 & 10.6 $\pm$ 1.9 & 3 \\
153 & HD94270 & 5983.0 $\pm$ 11 & 2.443 & 3.483 & 1.5 & -5.0 & 4.37 $\pm$ 0.02 & 0.02 $\pm$ 0.009 & 5.9 $\pm$ 2.0 & 3 \\
154 & HD94771 & 5618.0 $\pm$ 16 & 2.93 & 5.22 & 3.3 & -5.18 & 3.93 $\pm$ 0.04 & 0.22 $\pm$ 0.013 & -- & 2,4 \\
155 & HD95456 & 6292.0 $\pm$ 17 & 2.13 & 0.927 & 2.7 & -5.02 & 4.39 $\pm$ 0.04 & 0.16 $\pm$ 0.013 & -- & 2,3,4 \\
156 & HD96700 & 5838.0 $\pm$ 12 & 2.847 & 2.685 & 0.8 & -4.96 & 4.34 $\pm$ 0.02 & -0.18 $\pm$ 0.01 & 6.0 $\pm$ 1.1 & 4 \\
157 & HD97343 & 5390.0 $\pm$ 26 & 3.043 & 1.873 & 1.4 & -5.02 & 4.32 $\pm$ 0.04 & -0.055 $\pm$ 0.019 & -- & 3,4 \\
158 & HIP102152 & 5706.0 $\pm$ 13 & 2.939 & 3.455 & 1.5 & -5.02 & 4.3 $\pm$ 0.03 & -0.03 $\pm$ 0.01 & -- & 2,3 \\
159 & HIP105184 & 5831.0 $\pm$ 13 & 6.028 & 2.552 & 2.5 & -4.65 & 4.46 $\pm$ 0.02 & -0.02 $\pm$ 0.011 & -- & 3 \\
160 & HIP114328 & 5778.0 $\pm$ 13 & 2.046 & 4.235 & 1.5 & -5.03 & 4.34 $\pm$ 0.03 & -0.017 $\pm$ 0.011 & -- & 4 \\
161 & HIP114615 & 5784.0 $\pm$ 16 & 4.824 & 3.251 & 1.6 & -4.78 & 4.4 $\pm$ 0.02 & -0.057 $\pm$ 0.013 & 24.7 $\pm$ 0.4 & 3,4 \\
162 & HIP17157 & 4534.0 $\pm$ 150 & 13.832 & 8.864 & 2.5 & -4.64 & 4.27 $\pm$ 0.48 & -- & 15.0 $\pm$ 2.7 & 1,2 \\
163 & HIP21934 & 4507.0 $\pm$ 247 & 17.715 & 4.513 & 1.3 & -4.93 & 4.07 $\pm$ 0.71 & 0.03 $\pm$ 0.149 & 30.3 $\pm$ 0.9 & 1,2,3,4 \\
164 & HIP22263 & 5836.0 $\pm$ 31 & 10.863 & 4.133 & 3.2 & -4.56 & 4.54 $\pm$ 0.06 & -- & 10.6 $\pm$ 1.9 & 2 \\
165 & HIP25670 & 5768.0 $\pm$ 12 & 1.368 & 1.16 & 1.8 & -4.99 & 4.39 $\pm$ 0.03 & 0.06 $\pm$ 0.009 & 10.1 $\pm$ 1.8 & 2,3,4 \\
166 & HIP28066 & 5695.0 $\pm$ 12 & 2.0 & 6.057 & 1.2 & -5.03 & 4.23 $\pm$ 0.02 & -0.147 $\pm$ 0.01 & -- & 2 \\
167 & HIP36515 & 5801.0 $\pm$ 17 & 13.82 & 7.731 & 3.7 & -4.42 & 4.45 $\pm$ 0.03 & -0.07 $\pm$ 0.013 & 5.3 $\pm$ 2.2 & 2 \\
168 & HIP41317 & 5698.0 $\pm$ 11 & 6.261 & 1.169 & 1.4 & -5.02 & 4.37 $\pm$ 0.02 & -0.081 $\pm$ 0.009 & -- & 3 \\
169 & HIP54597 & 4679.0 $\pm$ 87 & 5.441 & 2.325 & 0.5 & -5.04 & 4.23 $\pm$ 0.23 & -0.22 $\pm$ 0.04 & -- & 2,3,4 \\
170 & HIP68468 & 5848.0 $\pm$ 14 & 0.341 & 1.074 & 2.7 & -5.06 & 4.32 $\pm$ 0.03 & 0.07 $\pm$ 0.012 & -- & 3 \\
171 & HIP79361 & 4490.0 $\pm$ 226 & 2.446 & 7.522 & 0.6 & -4.87 & 4.11 $\pm$ 0.55 & 0.52 $\pm$ 0.467 & -- & 3 \\

\end{longtable}
\footnotetext[1]{using the whole sample}
\footnotetext[2]{mean \vsini~ error: 1.0 [$km \;s^{-1}$]}
\footnotetext[3]{mean $\log(R' _{HK})$ error: 0.01}
\footnotetext[4]{subsample classifications, w: only considered in whole sample}

}
\renewcommand{\thefootnote}{\fnsymbol{footnote}}

\longtab[2]{
\begin{longtable}{ccccccc}
\caption{ \label{list_tic} List of starts in Fig.~\ref{mag_p_tess} with the peak-to-peak of light curve variation < 6.5 ppt and some stellar parameters presents in TIC \citep{2018AJ....156..102S}} \\
\hline\hline
TIC ID & Full TOI id & peak-to-peak light curve & TESS mag & \teff & $R_{star}$  & $\log g$ \\
 &  & [ppt] & & [K] & [$R_{Sun}$] & [cm $s^{-2}$] \\
 \hline
\endfirsthead
\caption{continued.}\\
\hline\hline
 TIC ID & Full TOI id & peak-to-peak light curve & TESS mag & \teff & $R_{star}$ & $\log g$ \\
\hline
\endhead
\hline
\endfoot
9033144 & 367.01 & 6.02 & 9.7 & 5756.9 $\pm$ 195.2 & 1.3 $\pm$ 4.0 & 3.9 $\pm$ 0.1 \\
9725627 & 239.01 & 5.89 & 11.0 & 6362.5 $\pm$ 83.0 & 0.9 $\pm$ 1.2 & 4.6 $\pm$ 0.2 \\
9727392 & 236.01 & 6.5 & 11.2 & 6421.5 $\pm$ 83.0 & 1.5 $\pm$ 5.1 & 4.2 $\pm$ 0.2 \\
13021029 & 439.01 & 5.4 & 11.7 & 6407.0 $\pm$ 105.8 & 1.3 $\pm$ 0.5 & 4.3   \\
14091704 & 445.01 & 2.54 & 9.1 & 6307.9 $\pm$ 124.2 & 1.6 $\pm$ 0.5 & 4.0 $\pm$ 0.1 \\
15445551 & 747.01 & 5.89 & 10.3 & 6271.0 $\pm$ 196.0 & 1.7 $\pm$ 3.0 & 4.1 $\pm$ 0.4 \\
19519368 & 494.01 & 4.58 & 10.0 & 4985.0 $\pm$ 114.0 & 1.1 $\pm$ 3.8 & 4.7 $\pm$ 0.2 \\
22221375 & 652.01 & 1.32 & 7.4 & 5903.0 & 1.0 $\pm$ 1.4 & 4.4 \\
22529346 & 495.01 & 3.84 & 10.0 & 6459.0 & 1.5 $\pm$ 0.1 & 4.2 \\
23434737 & 1203.01 & 1.54 & 8.0 & 5742.2 $\pm$ 30.4 & 1.4 $\pm$ 0.7 & 4.3 $\pm$ 0.1 \\
25155310 & 114.01 & 3.56 & 10.6 & 5800.0 & 1.3 $\pm$ 0.1 & 4.3 \\
29831208 & 124.01 & 3.48 & 11.0 & 5080.0 $\pm$ 181.0 & 0.9 $\pm$ 9.3 & 4.5 $\pm$ 2.0 \\
29960109 & 393.01 & 6.14 & 10.7 & 3856.0 $\pm$ 68.0 & 0.6 $\pm$ 0.3 & 4.7 $\pm$ 0.3 \\
30037565 & 1209.01 & 6.01 & 9.6 & 5914.0 $\pm$ 40.3 & 1.9 $\pm$ 1.9 & 4.0 $\pm$ 0.1 \\
30853470 & 807.01 & 4.9 & 11.2 & 4662.0 $\pm$ 174.0 & 0.8 $\pm$ 2.9 & 4.6 $\pm$ 0.2 \\
31553893 & 1058.01 & 4.08 & 9.3 & 5785.0 $\pm$ 189.0 & 1.1 $\pm$ 3.4 & 4.3 $\pm$ 2.0 \\
33692729 & 469.01 & 1.22 & 8.7 & 6114.0 $\pm$ 194.0 & 1.0 $\pm$ 0.5 & 4.5 $\pm$ 0.3 \\
37575651 & 568.01 & 2.2 & 8.3 & 5780.0 & 1.0 $\pm$ 1.1 & 4.4 \\
37749396 & 260.01 & 5.68 & 8.4 & 4111.0 $\pm$ 171.0 & 0.6 $\pm$ 1.5 & 4.6 $\pm$ 0.4 \\
37770169 & 470.01 & 2.66 & 10.6 & 5418.0 $\pm$ 185.0 & 0.8 $\pm$ 1.1 & 4.6 $\pm$ 0.3 \\
38696105 & 281.01 & 2.28 & 10.5 & 6002.2 $\pm$ 42.2 & 1.5 $\pm$ 3.5 & 4.1 $\pm$ 0.1 \\
41227743 & 804.01 & 3.52 & 10.7 & 5988.0 $\pm$ 192.0 & 1.2 $\pm$ 1.6 & 4.3 $\pm$ 2.0 \\
42054565 & 280.01 & 3.54 & 10.0 & 5454.0 $\pm$ 185.0 & 0.8 $\pm$ 4.8 & 4.6 $\pm$ 0.3 \\
43647325 & 423.01 & 3.63 & 10.4 & 5904.0 $\pm$ 156.8 & 1.1 $\pm$ 0.8 & 4.4   \\
47384844 & 1022.01 & 4.87 & 8.7 & 6084.0 $\pm$ 111.0 & 1.4 $\pm$ 0.8 & 4.1 $\pm$ 0.3 \\
47911178 & 471.01 & 2.73 & 9.8 & 6400.0 & 1.3 $\pm$ 0.1 & 4.3 \\
48476907 & 658.01 & 2.53 & 9.4 & 6521.0 $\pm$ 199.0 & 1.5 $\pm$ 19.9 & 4.2 $\pm$ 2.0 \\
48476908 & 659.01 & 2.49 & 9.5 & 5990.0 $\pm$ 192.0 & 1.7 $\pm$ 17.9 & 4.0 $\pm$ 0.7 \\
49687222 & 254.01 & 5.37 & 9.8 & 6101.0 $\pm$ 193.0 & 1.0 $\pm$ 4.5 & 4.5 $\pm$ 0.4 \\
49899799 & 416.01 & 3.41 & 8.3 & 6084.8 $\pm$ 35.2 & 1.9 $\pm$ 0.2 & 4.0 $\pm$ 0.3 \\
50309953 & 1109.01 & 2.14 & 9.8 & 5317.0 $\pm$ 65.7 & 0.9 $\pm$ 2.7 & 4.5 $\pm$ 0.1 \\
50312495 & 1211.01 & 2.62 & 9.8 & 5572.9 $\pm$ 104.0 & 1.2 $\pm$ 2.2 & 3.8 $\pm$ 0.2 \\
50618703 & 544.01 & 3.84 & 9.6 & 4665.0 $\pm$ 177.0 & 0.7 $\pm$ 0.4 & 4.7 $\pm$ 0.3 \\
52204645 & 209.01 & 6.34 & 10.8 & 4892.0 $\pm$ 179.0 & 0.8 $\pm$ 4.1 & 4.5 $\pm$ 2.0 \\
53189332 & 660.01 & 3.75 & 11.0 & 6055.0 & 1.4 $\pm$ 0.3 & 4.2 \\
53735810 & 661.01 & 4.3 & 11.2 & 6580.0 & 1.8 $\pm$ 0.3 & 4.1 \\
54085154 & 662.01 & 2.56 & 8.2 & 6380.0 $\pm$ 80.0 & 1.3 $\pm$ 0.5 & 4.4 \\
55559618 & 695.01 & 5.45 & 10.7 & 5591.0 $\pm$ 187.0 & 1.1 $\pm$ 3.4 & 4.4 $\pm$ 2.0 \\
61538902 & 752.01 & 6.32 & 11.3 & 5900.0 & 0.9 $\pm$ 0.2 & 4.5 \\
70513361 & 262.01 & 0.76 & 8.1 & 5302.6 $\pm$ 21.0 & 0.8 $\pm$ 0.7 & 4.5 \\
70914192 & 427.01 & 3.34 & 10.3 & 5271.7 $\pm$ 43.7 & 1.1 $\pm$ 4.3 & 4.3 $\pm$ 0.4 \\
73228647 & 755.01 & 2.82 & 9.5 & 6003.0 $\pm$ 192.0 & 1.0 $\pm$ 0.8 & 4.5 $\pm$ 0.3 \\
76989773 & 182.01 & 2.62 & 9.6 & 5569.0 $\pm$ 187.0 & 1.1 $\pm$ 5.7 & 4.3 $\pm$ 0.3 \\
89020549 & 132.01 & 4.81 & 10.8 & 5673.0 $\pm$ 188.0 & 0.9 $\pm$ 1.6 & 4.6 $\pm$ 0.4 \\
94986319 & 421.01 & 4.01 & 9.2 & 5718.0 $\pm$ 189.0 & 1.1 $\pm$ 6.4 & 4.4 $\pm$ 2.0 \\
96097215 & 728.01 & 4.47 & 10.3 & 5272.0 $\pm$ 183.0 & 0.9 $\pm$ 4.7 & 4.4 $\pm$ 2.0 \\
100100827 & 185.01 & 2.44 & 8.8 & 6400.0 & 1.2 & 4.4 \\
101230735 & 1060.01 & 3.83 & 9.6 & 5687.9 $\pm$ 76.0 & 1.3 $\pm$ 6.0 & 4.6 $\pm$ 0.1 \\
101955023 & 667.01 & 4.14 & 10.9 & 3202.0 & 0.3 $\pm$ 0.8 & 4.9 \\
103633434 & 1235.01 & 2.69 & 9.9 & 3912.0 $\pm$ 157.0 & 0.6 $\pm$ 0.2 & 4.6 \\
106402532 & 733.01 & 4.2 & 8.8 & 5969.0 $\pm$ 192.0 & 1.0 $\pm$ 1.1 & 4.5 $\pm$ 0.3 \\
111991770 & 820.01 & 4.8 & 10.4 & 6300.0 & 1.5 $\pm$ 0.1 & 4.2 \\
116483734 & 1412.01 & 6.46 & 11.1 & 4223.4 $\pm$ 133.4 & 0.7 $\pm$ 11.3 & 4.6 $\pm$ 0.1 \\
117979897 & 443.01 & 6.05 & 12.0 & 5928.0 $\pm$ 114.4 & 1.4 $\pm$ 0.5 & 4.3   \\
119700084 & 1413.01 & 4.77 & 10.1 & 5427.0 $\pm$ 141.3 & 0.9 $\pm$ 8.7 & 4.5 $\pm$ 0.1 \\
120602501 & 1366.01 & 3.07 & 8.5 & 5705.0 $\pm$ 185.1 & 1.7 $\pm$ 5.4 & 4.0 $\pm$ 0.1 \\
120610833 & 229.01 & 6.0 & 11.4 & 5100.0 & 0.9 $\pm$ 0.2 & 4.5 \\
120960812 & 1237.01 & 3.49 & 10.3 & 6212.0 $\pm$ 132.8 & 1.5 $\pm$ 4.1 & 4.2 $\pm$ 0.1 \\
122612091 & 264.01 & 1.9 & 10.5 & 6250.0 & 2.0 $\pm$ 0.1 & 4.0 \\
123702439 & 499.01 & 1.9 & 10.1 & 5997.0 $\pm$ 192.0 & 1.0 $\pm$ 1.6 & 4.5 $\pm$ 0.3 \\
124573851 & 669.01 & 2.81 & 10.1 & 5581.9 $\pm$ 76.0 & 1.2 $\pm$ 0.6 & 4.1 $\pm$ 0.1 \\
126733133 & 570.01 & 5.39 & 10.0 & 5973.0 $\pm$ 192.0 & 1.3 $\pm$ 4.4 & 4.3 $\pm$ 0.3 \\
127530399 & 822.01 & 6.41 & 11.1 & 4775.0 & 0.7 $\pm$ 0.2 & 4.6 \\
128790976 & 1124.01 & 1.86 & 7.2 & 6541.0 $\pm$ 94.0 & 1.6 $\pm$ 4.7 & 4.2 $\pm$ 0.1 \\
131081852 & 758.01 & 4.91 & 9.9 & 6072.0 $\pm$ 193.0 & 1.3 $\pm$ 19.4 & 4.3 $\pm$ 2.0 \\
131419878 & 720.01 & 1.15 & 8.8 & 6482.3 $\pm$ 205.2 & 1.5 $\pm$ 0.3 & 4.2 $\pm$ 0.4 \\
134537478 & 501.01 & 4.55 & 10.4 & 5720.0 & 1.5 $\pm$ 0.1 & 4.2 \\
138588540 & 1434.01 & 5.57 & 8.1 & 5393.9 $\pm$ 139.7 & 0.7 $\pm$ 0.9 & 4.7 $\pm$ 0.1 \\
140068425 & 140.01 & 6.47 & 9.8 & 6174.0 $\pm$ 194.0 & 1.3 $\pm$ 5.1 & 4.3 $\pm$ 2.0 \\
140760434 & 1229.01 & 5.38 & 10.7 & 6028.8 $\pm$ 44.1 & 2.0 $\pm$ 3.8 & 4.0 $\pm$ 0.1 \\
141527965 & 1216.01 & 5.87 & 11.4 & 4280.0 $\pm$ 171.0 & 0.7 $\pm$ 0.4 & 4.6 $\pm$ 0.2 \\
143350972 & 440.01 & 1.27 & 7.7 & 5759.2 $\pm$ 38.2 & 1.0 $\pm$ 0.1 & 4.4 \\
144065872 & 105.01 & 4.56 & 9.5 & 5630.0 & 1.1 $\pm$ 0.1 & 4.4 \\
147950620 & 1194.01 & 3.25 & 10.5 & 5339.9 $\pm$ 155.2 & 1.0 $\pm$ 0.5 & 4.4 $\pm$ 0.1 \\
148782377 & 1415.01 & 2.84 & 8.4 & 6383.0 $\pm$ 130.4 & 1.4 $\pm$ 0.6 & 4.2 $\pm$ 0.1 \\
149301575 & 809.01 & 4.56 & 10.3 & 5804.3 $\pm$ 76.0 & 1.3 $\pm$ 1.0 & 3.9 $\pm$ 0.1 \\
149603524 & 102.01 & 2.37 & 9.7 & 6280.0 & 1.3 & 4.3 \\
150030205 & 286.01 & 1.72 & 9.1 & 5245.0 $\pm$ 183.0 & 0.8 $\pm$ 1.7 & 4.6 $\pm$ 0.3 \\
150098860 & 220.01 & 5.38 & 9.7 & 5272.7 $\pm$ 45.6 & 0.9 $\pm$ 1.9 & 4.0 $\pm$ 0.1 \\
153065527 & 406.01 & 4.14 & 11.2 & 3349.0 $\pm$ 63.0 & 0.3 $\pm$ 0.7 & 4.9 $\pm$ 0.4 \\
153976959 & 1435.01 & 4.15 & 10.0 & 5142.0 $\pm$ 122.7 & 0.8 $\pm$ 2.4 & 4.5 $\pm$ 0.1 \\
154089169 & 1174.01 & 5.29 & 10.3 & 5029.5 $\pm$ 59.9 & 0.8 $\pm$ 0.2 & 4.6 $\pm$ 0.1 \\
154383539 & 1436.01 & 3.86 & 11.1 & 5011.2 $\pm$ 102.7 & 0.7 $\pm$ 1.9 & 4.6 $\pm$ 0.1 \\
154840461 & 1153.01 & 2.35 & 8.4 & 6528.3 $\pm$ 179.4 & 1.7 $\pm$ 0.5 & 4.2 $\pm$ 0.1 \\
158002130 & 1180.01 & 2.55 & 10.1 & 4738.1 $\pm$ 139.3 & 0.7 $\pm$ 0.2 & 4.6 $\pm$ 0.1 \\
158025009 & 1416.01 & 2.17 & 9.1 & 4946.0 $\pm$ 128.6 & 0.8 $\pm$ 1.8 & 4.5 $\pm$ 0.1 \\
158623531 & 246.01 & 3.55 & 11.2 & 5070.0 & 0.9 $\pm$ 0.1 & 4.5 \\
158978373 & 823.01 & 2.91 & 10.0 & 6309.0 $\pm$ 196.0 & 1.4 $\pm$ 1.1 & 4.2 $\pm$ 0.3 \\
159510109 & 1141.01 & 6.31 & 8.9 & 5894.7 $\pm$ 90.3 & 1.1 $\pm$ 0.3 & 4.4 $\pm$ 0.1 \\
159951311 & 265.01 & 4.21 & 11.7 & 5310.0 & 0.8 $\pm$ 0.2 & 4.6 \\
160074939 & 230.01 & 1.79 & 8.7 & 6467.0 $\pm$ 198.0 & 1.6 $\pm$ 1.6 & 4.2 $\pm$ 0.3 \\
160148385 & 247.01 & 4.87 & 11.9 & 5540.0 & 1.0 $\pm$ 0.2 & 4.4 \\
166739520 & 190.01 & 1.56 & 9.6 & 6038.0 & 1.2 $\pm$ 0.1 & 4.4 \\
166836920 & 267.01 & 1.27 & 8.9 & 6180.0 & 1.8 $\pm$ 0.1 & 4.1 \\
167303382 & 802.01 & 1.56 & 7.5 & 5235.8 $\pm$ 28.6 & 0.8 $\pm$ 0.4 & 4.4 \\
167342439 & 707.01 & 5.55 & 10.1 & 5409.2 $\pm$ 41.8 & 1.0 $\pm$ 1.7 & 4.5 $\pm$ 0.1 \\
167415965 & 214.01 & 1.73 & 8.0 & 5346.2 $\pm$ 12.9 & 0.9 $\pm$ 0.4 & 4.5 \\
167754523 & 409.01 & 4.84 & 9.1 & 4913.8 $\pm$ 71.5 & 0.8 $\pm$ 0.1 & 4.5 $\pm$ 0.2 \\
170102285 & 477.01 & 5.05 & 11.7 & 5150.0 & 0.8 $\pm$ 0.1 & 4.6 \\
172193428 & 502.01 & 3.76 & 10.3 & 5642.0 $\pm$ 188.0 & 1.4 $\pm$ 1.5 & 4.2 $\pm$ 0.3 \\
175180796 & 816.01 & 4.7 & 11.1 & 4146.6 $\pm$ 148.6 & 0.6 $\pm$ 1.1 & 4.6 $\pm$ 0.1 \\
176778112 & 408.01 & 5.84 & 9.6 & 5600.0 $\pm$ 187.0 & 1.0 $\pm$ 5.9 & 4.4 $\pm$ 0.3 \\
177258735 & 801.01 & 1.34 & 7.5 & 6283.4 $\pm$ 44.4 & 1.6 $\pm$ 0.4 & 4.4 \\
179317684 & 163.01 & 3.55 & 10.8 & 6387.9 $\pm$ 196.4 & 1.7 $\pm$ 1.0 & 4.2 $\pm$ 2.0 \\
183120439 & 169.01 & 6.23 & 11.7 & 5780.6 $\pm$ 401.1 & 1.3 $\pm$ 1.5 & 4.3 $\pm$ 2.0 \\
183537452 & 192.01 & 2.65 & 10.2 & 4800.0 & 0.8 $\pm$ 0.1 & 4.5 \\
183985250 & 193.01 & 4.66 & 9.1 & 5422.0 $\pm$ 185.0 & 1.0 $\pm$ 1.4 & 4.4 $\pm$ 0.3 \\
184952758 & 719.01 & 2.59 & 9.0 & 5761.2 $\pm$ 89.8 & 0.9 $\pm$ 0.1 & 4.5 $\pm$ 0.3 \\
188768068 & 1462.01 & 0.72 & 5.9 & 5801.9 $\pm$ 124.6 & 1.3 $\pm$ 1.3 & 4.3 $\pm$ 0.1 \\
190990336 & 585.01 & 3.74 & 8.9 & 6118.1 $\pm$ 137.4 & 1.7 $\pm$ 0.9 & 4.3 $\pm$ 0.3 \\
198212955 & 1242.01 & 4.35 & 11.6 & 4255.0 $\pm$ 128.2 & 0.7 $\pm$ 2.8 & 4.6 $\pm$ 0.1 \\
198241702 & 1269.01 & 3.21 & 10.9 & 5591.0 $\pm$ 185.0 & 0.8 $\pm$ 0.4 & 4.4   \\
198356533 & 1437.01 & 2.39 & 8.7 & 6093.0 $\pm$ 124.7 & 1.2 $\pm$ 0.7 & 4.3 $\pm$ 0.1 \\
198390247 & 1453.01 & 2.31 & 10.1 & 4920.0 $\pm$ 127.0 & 0.7 $\pm$ 1.6 & 4.6 $\pm$ 0.1 \\
199688472 & 1292.01 & 3.65 & 9.8 & 5672.9 $\pm$ 168.9 & 1.5 $\pm$ 6.6 & 4.1 $\pm$ 0.1 \\
200807066 & 869.01 & 2.93 & 8.9 & 5908.0 $\pm$ 191.0 & 1.2 $\pm$ 2.6 & 4.3 $\pm$ 0.3 \\
201793781 & 248.01 & 2.8 & 8.4 & 5711.7 $\pm$ 57.0 & 1.2 $\pm$ 0.7 & 3.8 $\pm$ 0.1 \\
206466531 & 410.01 & 2.47 & 10.0 & 6000.2 $\pm$ 130.8 & 1.5 $\pm$ 0.2 & 4.4 $\pm$ 0.1 \\
207081058 & 121.01 & 4.51 & 9.9 & 6053.0 $\pm$ 193.0 & 1.2 $\pm$ 2.2 & 4.3 $\pm$ 0.3 \\
207084429 & 381.01 & 0.78 & 7.2 & 5883.0 $\pm$ 52.4 & 1.9 $\pm$ 0.2 &  --   \\
211438925 & 194.01 & 3.69 & 10.2 & 5940.0 & 1.4 $\pm$ 0.1 & 4.2 \\
219239945 & 701.01 & 2.74 & 10.9 & 5837.0 $\pm$ 190.0 & 1.1 $\pm$ 0.7 & 4.4 $\pm$ 0.3 \\
219338557 & 133.01 & 4.54 & 9.9 & 4028.0 $\pm$ 170.0 & 0.7 $\pm$ 1.9 & 4.6 $\pm$ 0.2 \\
219379012 & 397.01 & 4.52 & 10.2 & 5973.5 $\pm$ 185.9 & 1.0 $\pm$ 2.5 & 3.9 $\pm$ 0.1 \\
219388773 & 399.01 & 5.22 & 10.6 & 4752.8 $\pm$ 64.0 & 1.0 $\pm$ 0.5 & 4.4 $\pm$ 0.1 \\
219698950 & 766.01 & 5.49 & 11.5 & 5870.0 & 1.0 $\pm$ 0.1 & 4.4 \\
219850915 & 1244.01 & 5.37 & 11.0 & 4599.0 $\pm$ 123.0 & 0.7 $\pm$ 2.1 & 4.6 $\pm$ 0.1 \\
220396259 & 379.01 & 2.29 & 9.9 & 5825.0 $\pm$ 97.8 & 1.7 $\pm$ 0.2 & 4.0 $\pm$ 0.1 \\
224297258 & 1279.01 & 2.68 & 10.0 & 5477.0 $\pm$ 120.1 & 0.8 $\pm$ 2.0 & 4.6 $\pm$ 0.1 \\
229650439 & 1438.01 & 4.09 & 10.3 & 5374.0 $\pm$ 183.0 & 0.8 $\pm$ 0.3 & 6.0 $\pm$ 0.3 \\
229770036 & 1348.01 & 5.24 & 10.7 & 4567.0 $\pm$ 143.3 & 0.8 $\pm$ 0.2 & 4.4 $\pm$ 0.1 \\
230017324 & 1280.01 & 4.07 & 10.6 & 4665.0 $\pm$ 128.1 & 0.7 $\pm$ 2.5 & 4.6 $\pm$ 0.1 \\
230088370 & 1176.01 & 2.2 & 10.2 & 6557.7 $\pm$ 83.3 & 1.6 $\pm$ 0.4 & 4.2 $\pm$ 0.1 \\
230982885 & 195.01 & 1.58 & 10.0 & 5640.0 & 1.1 $\pm$ 0.1 & 4.4 \\
231912935 & 215.01 & 4.17 & 10.5 & 5821.0 $\pm$ 190.0 & 1.0 $\pm$ 1.7 & 4.4 $\pm$ 0.3 \\
232025086 & 874.01 & 2.15 & 8.8 & 6437.0 $\pm$ 198.0 & 1.4 $\pm$ 1.3 & 4.2 $\pm$ 0.3 \\
232540264 & 1247.01 & 1.49 & 8.5 & 5711.8 $\pm$ 106.3 & 1.1 $\pm$ 0.6 & 4.4 $\pm$ 0.1 \\
232612416 & 1248.01 & 3.79 & 11.1 & 5227.0 $\pm$ 119.5 & 0.9 $\pm$ 0.6 & 4.5 $\pm$ 0.1 \\
232967440 & 1173.01 & 2.39 & 10.3 & 5429.6 $\pm$ 292.5 & 1.0 $\pm$ 0.5 & 4.4 $\pm$ 0.1 \\
232971294 & 1281.01 & 2.46 & 9.7 & 6191.0 $\pm$ 127.5 & 1.3 $\pm$ 1.4 & 4.3 $\pm$ 0.1 \\
232976128 & 1249.01 & 2.67 & 10.2 & 5453.0 $\pm$ 117.9 & 1.0 $\pm$ 2.3 & 4.4 $\pm$ 0.1 \\
232982558 & 1439.01 & 1.94 & 10.0 & 5873.0 $\pm$ 123.8 & 1.6 $\pm$ 0.9 & 4.0 $\pm$ 0.1 \\
233087860 & 1184.01 & 5.62 & 9.9 & 4681.4 $\pm$ 80.5 & 0.7 $\pm$ 1.1 & 4.6 $\pm$ 0.1 \\
233390838 & 1341.01 & 2.25 & 9.5 & 6477.6 $\pm$ 101.6 & 1.7 $\pm$ 5.4 & 4.1 $\pm$ 0.1 \\
233541860 & 1486.01 & 2.44 & 10.0 & 4154.0 $\pm$ 173.4 & 0.6 $\pm$ 0.1 & 4.5 $\pm$ 0.1 \\
233617847 & 1440.01 & 4.07 & 9.6 & 5780.0 & 1.0 $\pm$ 0.9 & 4.4 \\
233681149 & 1340.01 & 2.96 & 10.8 & 5929.0 $\pm$ 124.8 & 1.0 $\pm$ 0.4 & 4.5 $\pm$ 0.1 \\
233684822 & 1253.01 & 2.08 & 9.9 & 5779.0 $\pm$ 188.0 & 1.5 $\pm$ 3.2 & 4.9 $\pm$ 0.4 \\
233951353 & 1441.01 & 2.02 & 9.5 & 5885.0 $\pm$ 201.6 & 1.2 $\pm$ 1.4 & 4.3 $\pm$ 0.1 \\
234345288 & 213.01 & 3.34 & 9.7 & 4566.4 $\pm$ 64.0 & 0.7 $\pm$ 1.1 & 4.9 $\pm$ 0.1 \\
236714379 & 1254.01 & 3.13 & 10.8 & 5451.0 $\pm$ 125.9 & 1.1 $\pm$ 0.9 & 4.3 $\pm$ 0.1 \\
237222864 & 1255.01 & 3.04 & 9.1 & 5126.0 $\pm$ 124.4 & 0.9 $\pm$ 0.5 & 4.5 $\pm$ 0.1 \\
237232044 & 1443.01 & 2.52 & 9.9 & 5236.0 $\pm$ 125.3 & 0.7 $\pm$ 0.5 & 4.6 $\pm$ 0.1 \\
238176110 & 116.01 & 3.61 & 11.0 & 4920.0 & 0.9 $\pm$ 0.1 & 4.5 \\
238197638 & 903.01 & 5.37 & 11.7 & 6266.0 $\pm$ 41.0 & 1.4 $\pm$ 2.0 & 4.3 $\pm$ 0.2 \\
240968774 & 1467.01 & 6.01 & 10.6 & 3834.0 $\pm$ 157.0 & 0.5 $\pm$ 2.4 & 4.7 \\
243200602 & 826.01 & 4.95 & 10.5 & 5625.0 & 1.0 $\pm$ 0.1 & 4.5 \\
248075138 & 769.01 & 5.9 & 11.3 & 5315.0 & 0.9 $\pm$ 0.2 & 4.5 \\
250386181 & 390.01 & 4.45 & 9.9 & 6321.5 $\pm$ 53.5 & 1.1 $\pm$ 6.5 & 4.2 $\pm$ 0.1 \\
253990973 & 1061.01 & 4.12 & 9.7 & 5525.0 $\pm$ 186.0 & 1.2 $\pm$ 0.6 & 4.3 $\pm$ 0.3 \\
255685030 & 799.01 & 4.42 & 10.5 & 5643.0 $\pm$ 188.0 & 1.0 $\pm$ 3.5 & 4.5 $\pm$ 0.3 \\
257567854 & 403.01 & 6.04 & 11.2 & 6153.0 & 1.2 $\pm$ 0.2 & 4.3 \\
258514800 & 1444.01 & 4.17 & 10.2 & 5466.0 $\pm$ 121.5 & 0.9 $\pm$ 3.4 & 4.5 $\pm$ 0.1 \\
258777137 & 1183.01 & 4.93 & 9.9 & 5095.6 $\pm$ 1241.5 & 1.1 $\pm$ 0.2 &  --   \\
259511357 & 271.01 & 2.53 & 8.4 & 6025.0 $\pm$ 192.0 & 1.2 $\pm$ 1.5 & 4.3 $\pm$ 0.4 \\
259592689 & 429.01 & 2.74 & 10.2 & 5160.7 $\pm$ 90.5 & 1.0 $\pm$ 0.4 & 4.4 $\pm$ 2.0 \\
259701242 & 401.01 & 3.79 & 8.2 & 5928.0 $\pm$ 85.6 & 1.0 $\pm$ 0.1 & 4.5 $\pm$ 0.2 \\
260043723 & 217.01 & 4.65 & 8.7 & 6278.0 $\pm$ 196.0 & 1.7 $\pm$ 1.3 & 4.1 $\pm$ 0.3 \\
260708537 & 486.01 & 3.47 & 9.3 & 3445.0 $\pm$ 87.0 & 0.4 $\pm$ 1.0 & 4.9 $\pm$ 0.1 \\
261257684 & 904.01 & 5.75 & 10.8 & 3992.0 $\pm$ 170.0 & 0.6 $\pm$ 1.0 & 4.6 $\pm$ 0.2 \\
261867566 & 905.01 & 5.65 & 10.5 & 5565.0 $\pm$ 187.0 & 1.0 $\pm$ 1.8 & 4.5 $\pm$ 0.3 \\
263003176 & 130.01 & 1.79 & 7.4 & 6295.7 $\pm$ 56.5 & 1.1 $\pm$ 1.1 & 4.4 \\
264979636 & 518.01 & 5.27 & 10.1 & 5596.0 $\pm$ 187.0 & 1.1 $\pm$ 4.2 & 4.3 $\pm$ 0.3 \\
266980320 & 118.01 & 1.68 & 9.2 & 5586.0 $\pm$ 187.0 & 1.1 $\pm$ 0.6 & 4.4 $\pm$ 0.3 \\
268644785 & 505.01 & 2.85 & 10.7 & 6003.0 & 1.5 $\pm$ 0.1 & 4.2 \\
270341214 & 173.01 & 1.52 & 8.9 & 6409.0 $\pm$ 197.0 & 1.5 $\pm$ 1.3 & 4.2 $\pm$ 0.3 \\
271168962 & 828.01 & 6.22 & 9.4 & 6030.0 & 1.5 $\pm$ 0.1 & 4.1 \\
271893367 & 150.01 & 3.26 & 10.9 & 5766.0 $\pm$ 37.3 & 1.6 $\pm$ 0.3 & 4.1 $\pm$ 0.1 \\
271900960 & 389.01 & 2.91 & 8.3 & 6457.7 $\pm$ 115.5 & 1.4 $\pm$ 2.6 & 4.4 $\pm$ 0.3 \\
272635712 & 1349.01 & 2.25 & 9.3 & 5711.1 $\pm$ 365.0 & 1.3 $\pm$ 0.4 & 4.2 $\pm$ 0.1 \\
273985865 & 1208.01 & 3.83 & 10.5 & 5626.0 $\pm$ 188.0 & 0.9 $\pm$ 2.3 & 4.6 $\pm$ 0.3 \\
274138511 & 506.01 & 3.15 & 10.4 & 5280.0 $\pm$ 183.0 & 0.9 $\pm$ 3.1 & 4.4 $\pm$ 2.0 \\
277103955 & 284.01 & 3.21 & 9.2 & 4779.8 $\pm$ 128.6 & 0.7 $\pm$ 0.1 & 4.6 $\pm$ 0.3 \\
277507814 & 1286.01 & 4.42 & 10.2 & 6588.0 $\pm$ 247.6 & 1.6 $\pm$ 3.1 & 4.2 $\pm$ 0.1 \\
277683130 & 138.01 & 5.77 & 9.5 & 5722.0 $\pm$ 189.0 & 1.1 $\pm$ 8.3 & 4.4 $\pm$ 2.0 \\
279425357 & 739.01 & 6.5 & 11.5 & 6083.6 $\pm$ 125.0 & 1.2 $\pm$ 2.6 & 4.3 $\pm$ 0.2 \\
280095254 & 235.01 & 2.79 & 9.3 & 5454.4 $\pm$ 77.3 & 1.0 $\pm$ 5.8 & 3.4 $\pm$ 0.1 \\
280206394 & 677.01 & 4.18 & 9.2 & 6446.0 $\pm$ 198.0 & 1.2 $\pm$ 1.5 & 4.4 $\pm$ 0.3 \\
280830734 & 188.01 & 2.48 & 9.4 & 6340.3 $\pm$ 204.0 & 1.6 $\pm$ 7.5 & 4.2 $\pm$ 0.3 \\
281575427 & 205.01 & 2.2 & 9.4 & 6222.5 $\pm$ 83.0 & 1.5 $\pm$ 2.7 & 4.1 $\pm$ 0.2 \\
286132427 & 635.01 & 4.94 & 8.1 & 5914.0 $\pm$ 191.0 & 1.0 $\pm$ 0.9 & 4.5 $\pm$ 0.3 \\
287156968 & 1230.01 & 1.69 & 9.6 & 5696.0 $\pm$ 188.0 & 1.2 $\pm$ 1.1 & 4.3 $\pm$ 0.3 \\
288132261 & 1258.01 & 5.71 & 9.1 & 6124.0 $\pm$ 126.6 & 1.5 $\pm$ 0.4 & 4.2 $\pm$ 0.1 \\
290131778 & 123.01 & 1.22 & 8.8 & 6188.0 $\pm$ 194.0 & 1.3 $\pm$ 7.7 & 4.3 $\pm$ 2.0 \\
290348383 & 1099.01 & 4.49 & 7.4 & 4867.0 $\pm$ 179.0 & 1.0 $\pm$ 0.6 & 4.4 \\
294301883 & 774.01 & 6.17 & 11.3 & 6070.0 & 1.1 $\pm$ 0.1 & 4.4 \\
294471966 & 1446.01 & 5.04 & 10.8 & 5180.0 $\pm$ 101.3 & 0.8 $\pm$ 2.1 & 4.5 $\pm$ 0.1 \\
294781547 & 1218.01 & 3.68 & 10.3 & 5456.0 $\pm$ 48.9 & 1.0 $\pm$ 2.9 & 4.5 $\pm$ 0.1 \\
294981566 & 1219.01 & 6.24 & 10.1 & 6734.0 $\pm$ 202.0 & 1.6 $\pm$ 2.1 & 4.2 $\pm$ 0.4 \\
299799658 & 1062.01 & 4.04 & 9.4 & 5394.0 $\pm$ 185.0 & 0.9 $\pm$ 0.5 & 4.5 $\pm$ 0.3 \\
300293197 & 211.01 & 5.7 & 9.4 & 5874.0 $\pm$ 102.6 & 1.1 $\pm$ 6.4 & 4.4 $\pm$ 0.1 \\
304042899 & 775.01 & 2.71 & 9.5 & 6746.0 $\pm$ 202.0 & 1.2 $\pm$ 8.5 & 4.4 $\pm$ 0.4 \\
306362738 & 479.01 & 2.9 & 10.8 & 5600.0 & 1.0 $\pm$ 0.1 & 4.4 \\
309791156 & 533.01 & 5.76 & 10.9 & 4666.1 $\pm$ 64.0 & 1.1 $\pm$ 1.7 & 4.2 $\pm$ 0.1 \\
314865962 & 208.01 & 3.96 & 10.3 & 5493.0 $\pm$ 186.0 & 1.0 $\pm$ 1.3 & 4.4 $\pm$ 0.3 \\
317060587 & 1052.01 & 2.74 & 9.0 & 5958.2 $\pm$ 76.0 & 1.6 $\pm$ 4.5 & 4.3 $\pm$ 0.1 \\
317548889 & 480.01 & 1.05 & 6.8 & 6212.6 $\pm$ 49.0 & 1.6 $\pm$ 1.8 & 4.4 \\
320004517 & 1055.01 & 2.73 & 8.1 & 5783.5 $\pm$ 7.8 & 1.1 $\pm$ 0.6 & 4.5 \\
321857016 & 1420.01 & 4.48 & 11.2 & 5387.0 $\pm$ 136.6 & 0.9 $\pm$ 0.7 & 4.5 $\pm$ 0.1 \\
322063810 & 253.01 & 1.78 & 9.3 & 4020.0 $\pm$ 170.0 & 0.6 $\pm$ 1.6 & 4.4 \\
325680697 & 414.01 & 3.1 & 9.1 & 5773.7 $\pm$ 76.0 & 1.3 $\pm$ 5.3 & 3.5 $\pm$ 0.1 \\
327301957 & 1074.01 & 2.9 & 10.6 & 5206.0 $\pm$ 183.0 & 0.8 $\pm$ 0.5 & 4.6 $\pm$ 0.3 \\
335499997 & 680.01 & 5.44 & 9.0 & 5967.0 $\pm$ 192.0 & 1.8 $\pm$ 1.4 & 3.9 $\pm$ 0.4 \\
339672028 & 481.01 & 4.4 & 9.4 & 5760.0 $\pm$ 189.0 & 1.7 $\pm$ 1.6 & 4.0 $\pm$ 0.3 \\
339733013 & 417.01 & 4.24 & 10.0 & 5834.2 $\pm$ 179.5 & 1.3 $\pm$ 0.3 & 4.2 $\pm$ 0.1 \\
339857675 & 686.01 & 2.26 & 9.9 & 6279.8 $\pm$ 127.2 & 1.6 $\pm$ 0.3 & 4.3 $\pm$ 0.4 \\
346418409 & 1423.01 & 5.63 & 10.9 & 4993.0 $\pm$ 127.1 & 0.7 $\pm$ 7.7 & 4.7 $\pm$ 0.1 \\
348844154 & 916.01 & 5.23 & 11.2 & 5493.2 $\pm$ 126.6 & 1.9 $\pm$ 1.2 & 3.9 $\pm$ 0.5 \\
349827430 & 1148.01 & 1.51 & 7.9 & 6403.9 $\pm$ 184.1 & 1.5 $\pm$ 0.9 & 4.2 $\pm$ 0.1 \\
350153977 & 908.01 & 3.68 & 10.6 & 5500.0 $\pm$ 76.0 & 1.1 $\pm$ 3.9 & 5.0 $\pm$ 0.1 \\
350584963 & 787.01 & 3.81 & 9.5 & 6623.0 $\pm$ 200.0 & 1.3 $\pm$ 2.3 & 4.4 $\pm$ 0.3 \\
350743714 & 165.01 & 2.67 & 9.8 & 6038.0 $\pm$ 54.0 & 1.9 $\pm$ 2.1 & 4.0 $\pm$ 0.1 \\
352413427 & 1473.01 & 2.73 & 8.3 & 5958.0 $\pm$ 121.3 & 1.0 $\pm$ 0.5 & 4.5 $\pm$ 0.1 \\
352764091 & 1287.01 & 1.81 & 8.6 & 5891.0 $\pm$ 124.6 & 1.2 $\pm$ 1.7 & 4.3 $\pm$ 0.1 \\
354594208 & 1492.01 & 3.45 & 9.8 & 5904.0 $\pm$ 249.1 & 1.2 $\pm$ 0.4 & 4.3 $\pm$ 0.1 \\
358460246 & 867.01 & 4.07 & 10.4 & 5777.4 $\pm$ 113.6 & 1.2 $\pm$ 1.0 & 4.3 $\pm$ 0.2 \\
359271092 & 741.01 & 5.51 & 7.7 & 3766.0 $\pm$ 95.0 & 0.5 $\pm$ 0.7 & 4.8 $\pm$ 0.1 \\
364107753 & 909.01 & 1.0 & 7.0 & 6059.1 $\pm$ 50.4 & 1.6 $\pm$ 0.8 & 4.3 $\pm$ 0.1 \\
364186197 & 1408.01 & 2.4 & 8.8 & 6594.0 $\pm$ 90.9 & 1.6 $\pm$ 0.6 & 4.2 $\pm$ 0.2 \\
364393429 & 1207.01 & 2.92 & 9.2 & 6282.0 $\pm$ 196.0 & 1.2 $\pm$ 1.0 & 4.4 $\pm$ 0.3 \\
365733349 & 1288.01 & 3.15 & 9.9 & 6180.0 $\pm$ 362.0 & 0.9 $\pm$ 0.4 & 4.6 $\pm$ 0.1 \\
366989877 & 1054.01 & 1.64 & 8.4 & 6122.0 $\pm$ 194.0 & 1.2 $\pm$ 0.9 & 4.3 $\pm$ 0.3 \\
369327947 & 910.01 & 4.38 & 10.3 & 3194.0 $\pm$ 101.0 & 0.3 $\pm$ 1.3 & 5.0 $\pm$ 0.1 \\
373844472 & 275.01 & 4.27 & 11.0 & 5616.0 $\pm$ 187.0 & 1.1 $\pm$ 18.7 & 4.4 $\pm$ 2.0 \\
374829238 & 785.01 & 6.3 & 11.5 & 3783.0 & 1.0 $\pm$ 0.2 & 4.4 \\
374997123 & 1222.01 & 3.37 & 10.3 & 6082.0 $\pm$ 193.0 & 1.2 $\pm$ 2.4 & 4.3 $\pm$ 0.3 \\
375059587 & 786.01 & 2.54 & 9.8 & 5853.8 $\pm$ 45.2 & 1.4 $\pm$ 1.2 & 4.2 $\pm$ 0.1 \\
377293776 & 1450.01 & 3.14 & 10.0 & 3407.0 $\pm$ 157.0 & 0.5 $\pm$ 2.6 & 4.8 \\
382188882 & 276.01 & 3.75 & 10.6 & 6366.0 $\pm$ 197.0 & 1.5 $\pm$ 5.3 & 4.2 $\pm$ 0.5 \\
382437043 & 1223.01 & 4.81 & 9.3 & 6516.0 $\pm$ 199.0 & 1.9 $\pm$ 3.0 & 4.0 $\pm$ 0.4 \\
382474101 & 349.01 & 4.08 & 10.2 & 6054.0 $\pm$ 193.0 & 1.2 $\pm$ 14.0 & 4.3 $\pm$ 2.0 \\
382626661 & 283.01 & 5.09 & 9.6 & 5250.0 $\pm$ 183.0 & 0.7 $\pm$ 0.9 & 4.6 $\pm$ 0.3 \\
386435344 & 575.01 & 2.4 & 8.7 & 6529.0 $\pm$ 199.0 & 1.9 $\pm$ 13.5 & 4.0 $\pm$ 0.3 \\
387259626 & 1455.01 & 3.87 & 10.2 & 6458.8 $\pm$ 157.5 & 1.4 $\pm$ 0.9 & 4.3 $\pm$ 0.1 \\
391949880 & 128.01 & 1.79 & 7.9 & 6086.0 $\pm$ 193.0 & 1.3 $\pm$ 0.9 & 4.3 $\pm$ 0.3 \\
393414358 & 483.01 & 2.35 & 10.4 & 5570.0 & 1.9 $\pm$ 0.2 & 4.0 \\
394657039 & 159.01 & 3.44 & 10.5 & 6404.0 $\pm$ 463.3 & 1.6 $\pm$ 3.1 & 4.4 $\pm$ 0.2 \\
394698182 & 170.01 & 6.4 & 11.9 & 5768.2 $\pm$ 56.4 & 1.2 $\pm$ 0.4 & 4.3 $\pm$ 2.0 \\
399144800 & 1213.01 & 3.59 & 10.6 & 6270.0 $\pm$ 196.0 & 1.0 $\pm$ 1.9 & 4.5 $\pm$ 0.4 \\
402026209 & 232.01 & 5.69 & 11.8 & 5436.0 & 0.9 $\pm$ 0.1 & 4.5 \\
403224672 & 141.01 & 1.19 & 7.4 & 5795.3 $\pm$ 44.6 & 1.2 $\pm$ 0.4 & 4.4 \\
406941612 & 912.01 & 3.81 & 10.4 & 3566.0 $\pm$ 66.0 & 0.4 $\pm$ 0.4 & 4.8 $\pm$ 0.4 \\
407126408 & 913.01 & 5.4 & 9.6 & 4948.0 $\pm$ 180.0 & 0.7 $\pm$ 1.0 & 4.6 $\pm$ 0.3 \\
407966340 & 554.01 & 1.7 & 6.4 & 6337.9 $\pm$ 44.7 & 1.8 $\pm$ 0.9 & 4.4 \\
408159788 & 420.01 & 3.67 & 10.6 & 5747.3 $\pm$ 518.8 & 0.8 $\pm$ 0.4 & 4.6 $\pm$ 0.3 \\
408310006 & 576.01 & 5.23 & 8.8 & 6313.0 $\pm$ 196.0 & 1.3 $\pm$ 0.8 & 4.3 $\pm$ 0.3 \\
410245915 & 725.01 & 5.31 & 10.2 & 6173.3 $\pm$ 152.9 & 0.8 $\pm$ 0.2 & 4.2 $\pm$ 2.0 \\
415969908 & 233.01 & 5.1 & 11.3 & 3644.0 $\pm$ 64.0 & 0.5 $\pm$ 0.6 & 4.8 $\pm$ 0.4 \\
417676622 & 1290.01 & 2.17 & 9.5 & 5875.2 $\pm$ 137.9 & 1.3 $\pm$ 3.0 & 4.3 $\pm$ 0.1 \\
417931607 & 1451.01 & 3.53 & 9.0 & 5781.0 $\pm$ 130.5 & 1.0 $\pm$ 2.1 & 4.4 $\pm$ 0.1 \\
418959198 & 1424.01 & 5.14 & 10.5 & 4529.0 $\pm$ 104.6 & 0.8 $\pm$ 3.3 & 4.5 $\pm$ 0.1 \\
421894914 & 1056.01 & 4.96 & 9.7 & 6124.3 $\pm$ 83.0 & 1.4 $\pm$ 1.0 & 4.4 $\pm$ 0.2 \\
424865156 & 1265.01 & 2.82 & 10.0 & 6532.2 $\pm$ 109.2 & 2.0 $\pm$ 0.7 & 4.0 $\pm$ 0.1 \\
427348923 & 484.01 & 5.54 & 11.6 & 4425.0 $\pm$ 90.0 & 0.7 $\pm$ 1.2 & 4.6 $\pm$ 0.1 \\
432549364 & 1476.01 & 3.99 & 10.2 & 6596.0 $\pm$ 156.6 & 1.5 $\pm$ 0.6 & 4.2 $\pm$ 0.1 \\
437242640 & 744.01 & 2.7 & 9.7 & 5700.0 & 0.9 $\pm$ 0.3 & 4.5 \\
437248515 & 683.01 & 4.82 & 11.4 & 6302.0 & 1.2 $\pm$ 0.1 & 4.3 \\
440100539 & 1548.01 & 2.9 & 8.7 & 5891.9 $\pm$ 92.3 & 1.2 $\pm$ 0.2 & 4.3 $\pm$ 0.1 \\
445859771 & 1273.01 & 5.9 & 10.4 & 5736.0 $\pm$ 164.8 & 1.1 $\pm$ 4.6 & 4.4 $\pm$ 0.1 \\
451606970 & 1214.01 & 2.03 & 10.0 & 5262.4 $\pm$ 76.0 & 1.1 $\pm$ 2.0 & 4.0 $\pm$ 0.1 \\
453211454 & 509.01 & 1.69 & 7.9 & 5560.3 $\pm$ 43.4 & 1.0 $\pm$ 0.5 & 4.4 \\
453260209 & 1215.01 & 6.03 & 9.6 & 3751.0 $\pm$ 68.0 & 0.5 $\pm$ 1.0 & 4.8 $\pm$ 0.3 \\
455135327 & 490.01 & 4.15 & 9.9 & 6304.0 & 1.2 $\pm$ 0.1 & 4.4 \\
458478250 & 1165.01 & 2.79 & 9.8 & 5787.0 $\pm$ 109.9 & 1.0 $\pm$ 0.6 & 4.5 $\pm$ 0.1 \\
459942762 & 430.01 & 1.86 & 8.2 & 5997.7 $\pm$ 78.3 & 1.0 $\pm$ 0.1 &  --   \\
459969957 & 1274.01 & 6.28 & 11.9 & 4968.1 $\pm$ 172.4 & 0.8 $\pm$ 0.5 & 4.5 $\pm$ 0.1 \\
467666275 & 1204.01 & 5.16 & 8.0 & 6711.0 $\pm$ 201.0 & 1.8 $\pm$ 2.6 & 4.1 $\pm$ 0.3 \\
468148930 & 1086.01 & 5.26 & 12.2 & 5818.0 $\pm$ 190.0 & 1.1 $\pm$ 8.3 & 4.3 $\pm$ 2.0 \\
1400770435 & 1344.01 & 1.64 & 8.9 & 5941.5 $\pm$ 161.9 & 1.4 $\pm$ 1.3 & 4.2 $\pm$ 0.1 \\

\end{longtable}
}

\end{appendix}

\end{document}